\documentclass{emulateapj}

\usepackage{longtable,lscape}
\usepackage{natbib}
\newcommand{\flux}{ergs\,s$^{-1}$\,cm$^{-2}$}
\begin{document}
\title{X-ray Spectral Properties of the BAT AGN Sample}
\author{Lisa M. Winter\altaffilmark{1}, 
Richard F. Mushotzky\altaffilmark{2}, 
Christopher S. Reynolds\altaffilmark{1},
Jack Tueller\altaffilmark{2}}
\email{lwinter@astro.umd.edu}

\altaffiltext{1}{University of Maryland, College Park, MD}
\altaffiltext{2}{ NASA Goddard Space Flight Center, Greenbelt, MD}

\begin{abstract}
The 9-month SWIFT Burst Alert Telescope (BAT) catalog provides the first unbiased ($N_H < 10^{24}$\,cm$^{-2}$) look at local ($<z> = 0.03$) AGN.  In this paper, we present the collected X-ray properties (0.3 -- 12\,keV) for the 153 AGN detected.  In addition, we examine the X-ray properties for a complete sample of non-beamed sources, above the Galactic plane (b$\ge 15^{\circ}$).  Of these, 45\% are best fit by simple power law models while 55\% require the more complex partial covering model.  One of our goals was to determine the fraction of ``hidden'' AGN, which we define as sources with scattering fractions $\le 0.03$ and ratios of soft to hard X-ray flux $\le 0.04$.  We found that ``hidden'' AGN constitute a high percentage of the sample (24\%), proving that they are a very significant portion of local AGN.  Further, we find that the fraction of absorbed sources does increase at lower unabsorbed 2--10\,keV luminosities, as well as accretion rates.  This suggests that the unified model requires modification to include luminosity dependence, as suggested by models such as the 'receding torus' model
(Lawrence 1991).  Some of the most interesting results for the BAT AGN sample involve the host galaxy properties.  We found that 33\% are hosted in peculiar/irregular galaxies and only 5/74 hosted in ellipticals.  Further, 54\% are hosted in interacting/merger galaxies.  Finally, we present both the average X-ray spectrum (0.1--10\,keV) and $\log N$-$\log S$ in the 2-10\,keV band.  With our average spectrum, we have the remarkable result of reproducing the measured CXB X-ray power law slope of $\Gamma \approx 1.4$ (Marshall {\it et al.} 1980).  From the $\log N$-$\log S$ relationship, we show that we are complete to $\log S \ge -11$ in the 2--10\,keV band.  Below this value, we are missing as many as 3000 sources at $\log S = -12$.  Both the collected X-ray properties of our uniform sample and the $\log N$-$\log S$ relationship will now provide valuable input to X-ray background models for $z \approx 0$. 
\end{abstract}

\section{Introduction}
Active galactic nuclei (AGN) are among the most powerful sources of energy in the Universe, where the brightest quasars can outshine all of the stars in their host galaxy by 100 times.  Optical studies of AGN reveal strong narrow and broad emission lines indicative of AGN activity in the nearby Universe (z$ < 1$).  However, X-ray and optical surveys fail to select the same AGN samples \citep{2004ASSL..308...53M}.  X-ray surveys identify more sources whose 2 -- 10\,keV emission is obscured by high column density absorbing material in the line of sight.  Still, even the X-ray surveys are affected by heavy obscuration, making it difficult to detect sources with column densities $> 10^{24}$\,cm$^{-2}$ (Compton thick sources).

This provides a major question that AGN surveys need to address: how many heavily obscured or Compton thick sources exist?  \citet{2000MNRAS.318..173M} estimated that the number could be as high as an order of magnitude more than the unobscured sources, which are easily detected in optical and soft X-ray surveys.  In order to account for these additional heavily obscured sources and determine their contribution to the cosmic X-ray background, very hard X-ray ($> 10$\,keV) surveys are needed.  Only at these wavelengths is the AGN emission penetrating enough to pass through much of the dust and gas in the line of sight.

With the Burst Alert Telescope (BAT) on board SWIFT, we now have the first sensitive unbiased AGN survey, towards all but the most heavily obscured sources ($N_H > 10^{24}$\,cm$^{-2}$).  This is due to BAT's sensitivity at very high X-ray energies (14 -- 195\,keV).  With a much larger sample than previous (such as HEAO-1) and contemporary (Integral is most sensitive along the Galactic plane) very hard X-ray missions, BAT is the first sensitive all-sky very hard X-ray survey in 28 years.

From the 9-month catalog of BAT AGN \citep{2007arXiv0711.4130T}, a large enough sample has been obtained (153 sources) to determine the statistical properties.  In the catalog paper, only the X-ray derived column density and a complexity flag were indicated.  However, in this paper we provide a more detailed look at the X-ray properties, including archival data, analyses from the literature, and previously unanalyzed SWIFT X-ray Telescope (XRT) observations.  This work follows upon an XMM-Newton follow-up study of 22 BAT AGN, in which we reported that the ``hidden''/buried AGN described in \citet{2007ApJ...664L..79U} may be a significant fraction of the BAT AGN \citep{2008ApJ...674..686W}.

Our goals are two-fold.  First, we present the X-ray properties for the entire 9-month catalog.  Second, we examine in more detail the collective properties of a uniform sample.  This sample consists of non-beamed sources with Galactic latitudes $\ge 15^{\circ}$.
In Section~\ref{bat-data}, we describe the observations and the spectral fits, including data from ASCA, XMM-Newton, Chandra, Suzaku, and SWIFT XRT.  In Section~\ref{bat-spectra}, we describe the general properties of the entire BAT 9-month AGN sample.  In Section~\ref{bat-unbiased}, we describe in more depth the properties of our uniform sample.  In Section~\ref{bat-cxb}, we present the average X-ray spectrum as well as the 2--10\,keV $\log N$-$\log S$ relation. These X-ray properties can now be used as input to X-ray background models for $z \approx 0$. We then discuss the properties of the host galaxies in Section~\ref{bat-host}.  Finally, we summarize our results in Section~\ref{bat-summary}.

\section{The Data}
\label{bat-data}
\subsection{Source Selection}
The sources in the 9-month BAT AGN catalog were selected based on a detection at 4.8$\sigma$ or higher.  This corresponds to fluxes in the 14--195\,keV band of $\ga 2 \times 10^{-11}$\flux.  The survey and the method of selection are described in \citep{2007arXiv0711.4130T}.  To summarize, the BAT positions for the detected sources have an error of $\le 6$\arcmin at 4.8$\sigma$.  Therefore, the positions of identified sources were compared with available optical, radio, and X-ray observations.  Where the BAT source could not be identified with a previously known AGN source, SWIFT XRT observations were obtained.  With an identified XRT counterpart, the position is narrowed to an error of $\sim 4$\arcsec.  The source identification was further constrained by the requirement that each of the BAT AGN sources have a clear optical/IR counterpart in the Digital Sky Survey/2MASS.   Many of the sources in this entire 9-month sample have identifications based on optical spectra either from archived data, the literature, or our own follow-up data.  The sample consists of 17 blazars/BL Lacs,  49 Sy 1--1.2s, 34 Sy 1.5--1.9s, and 45 Sy 2s.  Of the remaining sources, 7 are unidentified, 2 have optical spectra showing a normal galaxy (i.e. no AGN emission lines; NGC 612 and NGC 4992), and 2MASX J09180027+0425066 is identified as a type 2 quasar.

In order to study the properties of the local BAT-detected AGN, we identify a uniform sample from the 9-month catalog.  This requires the exclusion of beamed sources as well as sources within the Galactic plane ($b < 15^{\circ}$).  We exclude beamed sources (the 17 blazars/BL Lacs), since they make up only $\approx 10$\% of AGN and are at higher redshifts than our sample of local sources.  Further, the physics behind the spectra are different between beamed and unbeamed sources, since these sources are jet dominated.  Only high Galactic latitude sources are included ($b > 15^{\circ}$), since the identification of AGN in the Galactic plane by BAT are less certain due to the higher background and large number of unidentified sources (Galactic and extragalactic). This uniform sample consists of 102 sources.  Among these, 34 are Sy 1--1.2s, 28 are Sy 1.5--1.9s, and 36 are Sy2s, with none of the sources without an identification.  Since the normal galaxy spectra sources and the type 2 quasar show no broad lines, we include them as Sy2s in the following discussions.

Since the 9-month survey sources are moderately bright, many of them were well-known AGN sources with archival data/published papers detailing the X-ray properties.  Thus, in compiling the X-ray properties of this sample, we first searched the literature for analyses of the X-ray spectra of our sources.
In the following section, we describe the X-ray data and analysis for the entire sample.

\subsection{X-ray Data and Analysis}
 The X-ray emission from AGN primarily takes the form of an absorbed power law \citep{1993ARAA..31..717M}.  However, there are additional features which are present in the 0.3--10\,keV spectra of many AGN, primarily a soft excess and an Fe K-$\alpha$ line.  The soft excess is often modeled as a blackbody component ($kT \approx 0.1$\,keV) and its origin is believed to be either thermal emission from star formation, photoionized gas, blurred reflection  
\citep{2003AA...412..317C,2005MNRAS.358..211R}, or blurred absorption \citep{2004MNRAS.349L...7G}.  The Fe K-$\alpha$ line is a fluorescent line from lowly ionized iron at 6.41\,keV.  Finally, many AGN exhibit a complex spectrum indicating emission components absorbed with different column densities.  This type of spectrum may be the result of scattering of direct AGN emission, a dusty environment where the AGN emission is partially covered by absorbing material, or contamination of the AGN spectrum by less-absorbed X-ray binaries.  In our effort to uniformly compare the properties of our sources, we searched the literature for simple models fit to the X-ray data, based on these components (power law, absorption (simple or complex), soft excess, and Fe K line).  We did not include models for reflection since higher signal-to-noise spectra would be required for the entire sample.

\begin{figure}
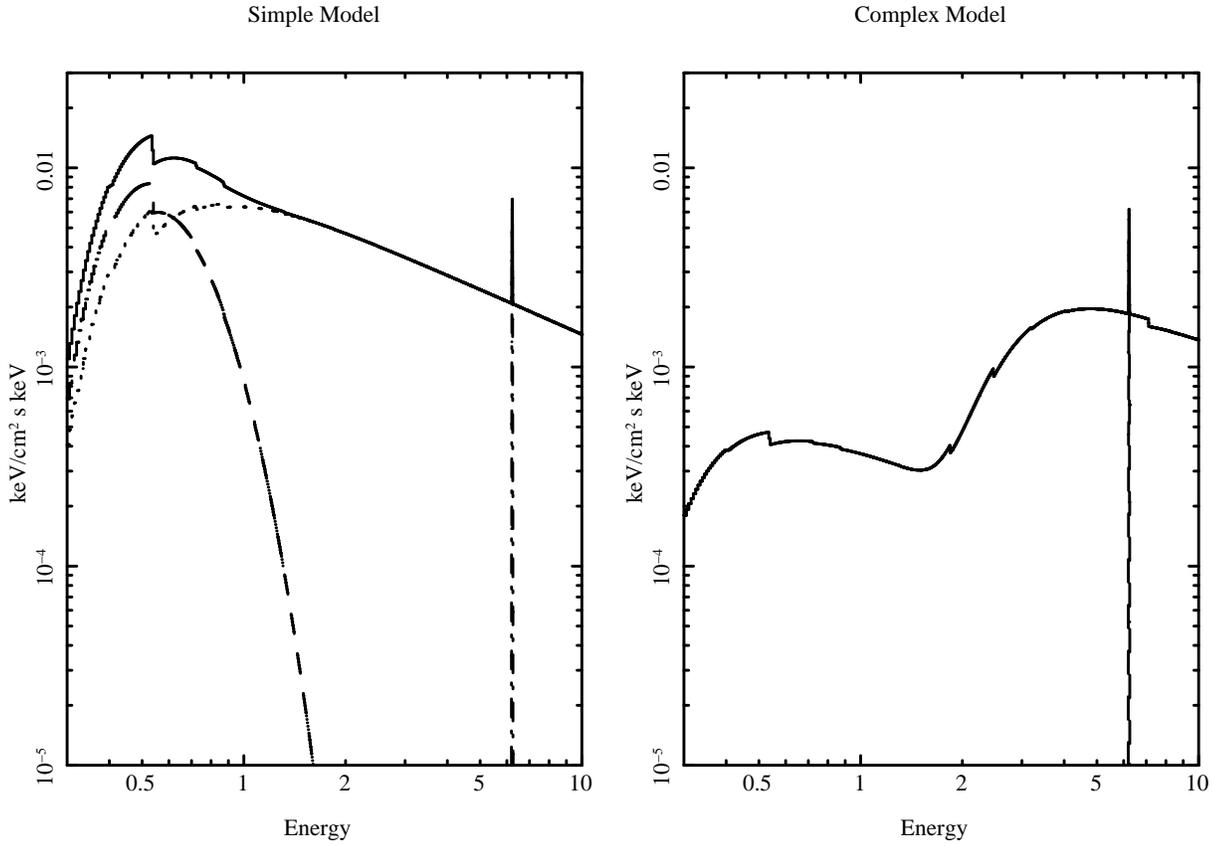

\includegraphics[width=8cm]{f1a.eps}
\hspace{0.1cm}
\includegraphics[width=8cm]{f1b.eps}
\caption[Simple and Complex Models]{The above figures are examples of the complex and simple models used in the paper.  The simple model (left) is an absorbed power law model ($N_H = 8 \times 10^{20}$\,cm$^{-2}$ and $\Gamma = 1.75$).  In addition to the power law, we include a blackbody component ($kT = 0.1$\,keV, dashed line) and a 60\,EW Fe K-$\alpha$ line at 6.41\,keV.  The complex model (right) is an absorbed partial covering model (or a double power law model, which gives similar results).  Here, the same power law and gaussian components are used as for the simple model, with $N_H = 1 \times 10^{23}$\,cm$^{-2}$ and a covering fraction of 0.95.
\label{fig-models}}
\end{figure}

In our study of 22 {\it XMM-Newton} follow-ups of BAT AGN \citep{2008ApJ...674..686W}, we had used the same components to classify the AGN X-ray spectra.  We chose a simple and complex model that is well-used throughout the literature, allowing for easy comparison with other AGN studies.  In this study, we adopt these same models. Our simple model is an absorbed power law model (Figure~\ref{fig-models}a), with an additional soft blackbody and/or gaussian (for the Fe-K$\alpha$ line) model where required.  Using the standard X-ray software for fitting X-ray spectral models, XSPEC \citep{ADASS_96_A}, our simple model is represented as {\tt tbabs}*({\tt pow} + {\tt bbody} + {\tt zgauss}).  Here, {\tt tbabs} is a standard neutral absorption model \citep{2000ApJ...542..914W}.
We categorize a complex spectrum as one well-fitted with either a partial covering model (Figure~\ref{fig-models}b) or a double power law model, both of which give similar results (in $\chi^2$ and spectral parameters), along with an Fe-K$\alpha$ line.  Our complex/partial covering model is formally implemented in XSPEC as {\tt pcfabs}*({\tt pow} + {\tt zgauss}) for partial covering or {\tt tbabs}*({\tt pow} + {\tt tbabs}*({\tt pow} + {\tt zgauss})) for a double power law model.  Further discussions of our choice of these models can be found in \citet{2008ApJ...674..686W}

For all of the 9-month BAT AGN, we list the source, position, optical type, host galaxy type, and details of the observation in Table~\ref{tbl-1}.  We include the X-ray satellite used, references, exposure time, and count rate, where available.  Where count rates and exposure times are quoted, they correspond to the pn detector for {\it XMM-Newton} and SIS0 for ASCA.  For many sources, spectra were available from many different satellites.  Our preference was to choose ASCA analyses first, followed by {\it XMM-Newton} and Chandra.  Mostly, this is due to the uniform way the spectral properties of Seyfert 1s and Seyfert 2s are presented in papers analyzing ASCA data such as: 
\citet{1997ApJ...476...70N,1997ApJS..113...23T,1997MNRAS.286..513R}.  Also, especially with the increased resolution of the grating spectrometers and higher signal-to-noise CCD data on {\it XMM-Newton} and Chandra, more accurate/complicated models are often used to analyze spectra from these satellites, particularly for observations with a large number of counts.  Where multiple observations were available, we chose the longest observation below 50\,ks, to be comparable to the quality of spectra collected for other sources.    

Where references were not available, or comparable models were not used, we analyzed either SWIFT XRT, ASCA (downloaded from the Tartarus database\footnote{\url{http://astro.ic.ac.uk/Research/Tartarus/}}), or Suzaku spectra (we obtained data for $\approx 10$ sources through AO-1/AO-2 proposals).  For the XRT spectra, we analyzed these data following the same procedure as in \citet{2008ApJ...674..686W}.  In XRT observations with few counts, if there were multiple observations available, we combined the spectra from the longest observations with the FTOOL {\tt mathpha}.  The analysis of the Suzaku spectra is contained in Ueda et al. (in prep), which includes detailed analyses of the spectra.  The observation and source properties for the SWIFT XRT, ASCA, and Suzaku observed sources are also listed in Table~\ref{tbl-1}.

\begin{figure}
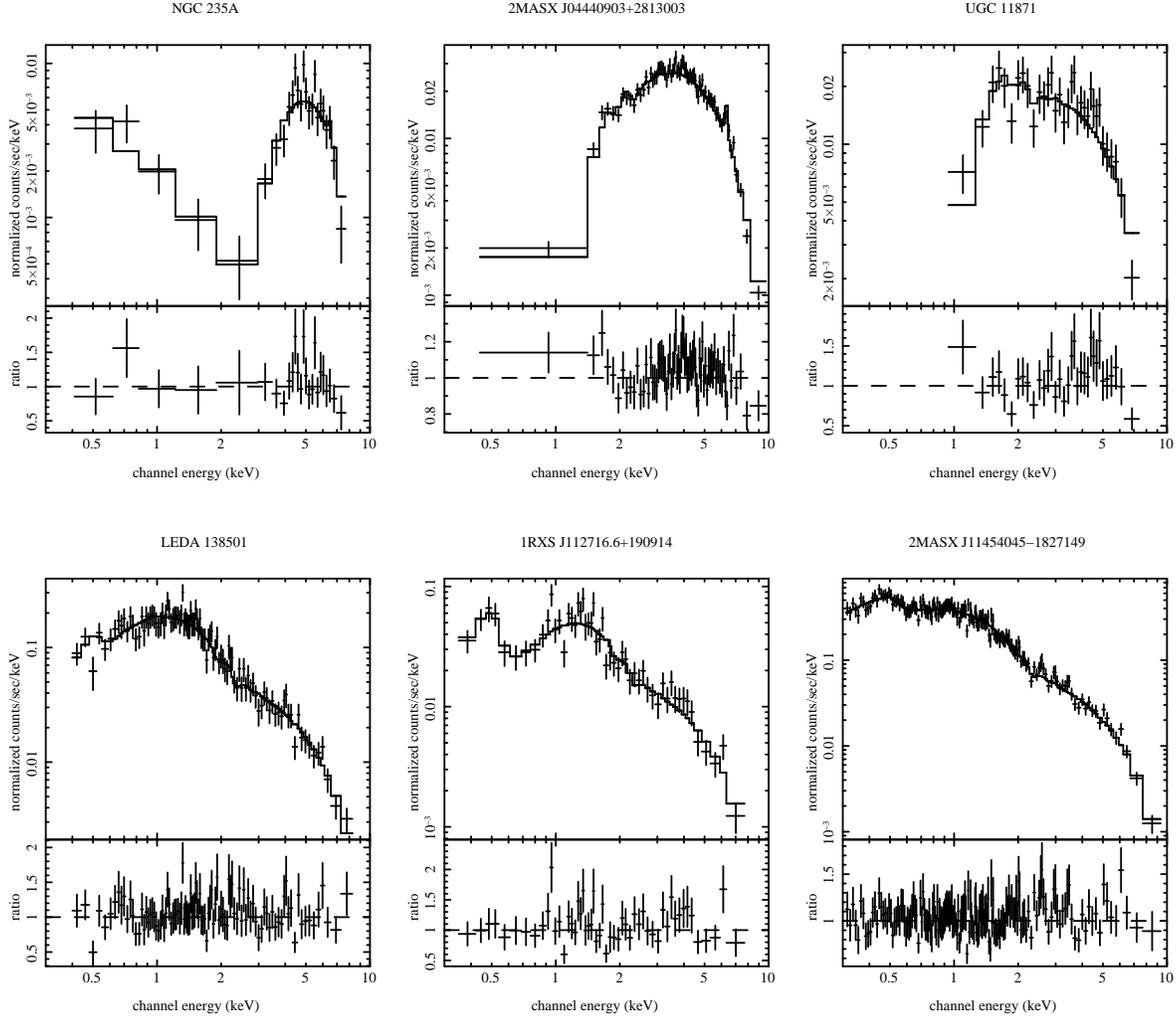

\includegraphics[width=5cm]{f2a.eps}
\hspace{0.1cm}
\includegraphics[width=5cm]{f2b.eps}
\hspace{0.1cm}
\includegraphics[width=5cm]{f2c.eps} \\

\vspace{0.3cm}

\includegraphics[width=5cm]{f2d.eps}
\hspace{0.1cm}
\includegraphics[width=5cm]{f2e.eps}
\hspace{0.1cm}
\includegraphics[width=5cm]{f2f.eps} \\

\caption[Example X-ray Spectra]{Examples of XRT spectra for sources fit with simple and complex models, along with residuals.  The spectra
of  2MASX J04440903+2813003 and 2MASX J11454045-1827149 were rebinned to better show the residuals from the model.
Details of each of the fits are recorded in Table~\ref{tbl-2}.  The sources in the top row are all well-fit with the complex model, while the lower row sources have simple spectra (see Figure~\ref{fig-models}).
\label{fig-1}}
\end{figure}

In Tables~\ref{tbl-2} and~\ref{tbl-3}, we list the spectral properties from the observations.  Table~\ref{tbl-2} includes our XSPEC fits to the XRT data using an absorbed power law ({\tt tbabs}*{\tt pow}) or an absorbed partial covering model ({\tt pcfabs}*{\tt pow}).  For each of these fits, an additional neutral absorber ({\tt tbabs}) was fixed to the Galactic column density.  For spectra with clear residuals from these models indicating a soft component or an Fe K-$\alpha$ line, we added the necessary components and included the details in the appendix.  Table~\ref{tbl-3} includes the compiled properties from the literature as well as fits to ASCA spectra downloaded from the Tartarus database.  For these spectra, the model used is clearly specified in the table.  In both tables, we categorize the spectra as being simple (S: well-fit by an absorbed power law or power law and blackbody) or complex (C), as in \citet{2008ApJ...674..686W}.  In Figure~\ref{fig-1}, we plot some examples of the XRT spectra from each category.  Notice that the spectrum of UGC 11871 has no data points below 1\,keV.  For this source, as well as 3 other sources in Table~\ref{tbl-2}, we classify the source as S*.  This indicates that it was best fit with a simple power law, however, with longer exposure times, and thus more data points below 1\,keV, it would likely be a complex source, showing the characteristic complex/double power law shape.

We have collected the X-ray properties for 145/154 AGN.  Though there are 153 BAT sources, one BAT source (NGC 6921/MCG +04-48-002) is the combination of two interacting galaxies, each hosting AGN \citep{2008ApJ...674..686W}.  In the appendix, we include details on each of the 9 sources excluded from this analysis.  The sources are excluded due to a lack of data available (2) or complex spectra (7). For instance, Cen A, an AGN hosted in a merger galaxy, is a complex source that is excluded.  In the following sections, we will describe the X-ray properties for the remaining 145 sources.

\section{Properties of the Spectra}
\label{bat-spectra}
In Tables~\ref{tbl-1}--~\ref{tbl-3}, we provide information on the 9-month BAT AGN sources, including the X-ray spectral parameters and fluxes, all derived in a uniform way.  In order to study the properties of the local BAT-detected AGN, we need to look at a uniform sample.  However, it is first worth noting some of the general properties of the sources detected in the 9 month catalog.   

As a means to summarize these properties, we plot a color-color plot of F$_{0.5 - 2 keV}$/F$_{2 - 10 keV}$ vs. F$_{14 - 195 keV}$/F$_{2 - 10 keV}$ (all observed fluxes), first used in \citet{2008ApJ...674..686W}.  This plot includes all sources excepting 3 AGN (the unabsorbed sources ESO 416-G002, MCG -01-13-025, and Mrk 79) which had only broad band fluxes available, the 4 sources which had no high-quality or publicly available X-ray spectra, and the 4 very complex spectra sources (NGC 1275, Cen A, NGC 6240, and GRS 1734-292).  As in our previous paper, the unabsorbed/low absorption sources (triangles, $N_H \le 10^{22}$\,cm$^{-2}$) occupy the upper left portion, with one exception.  Cyg A is a complex source with a complex spectrum, the product of an AGN in a recent merger galaxy.  A strong thermal component, mainly from the hot thermal emission of the rich galaxy cluster Cyg A inhabits, is seen at low energies, modeled by a Raymond-Smith thermal model \citep{1999ApJ...526...60S}.  This thermal emission gives Cyg A its unique position on the diagram (as indicated in the plot).

For the unabsorbed sources, the soft band mean ($\mu$) color is 0.55 with standard deviation ($\sigma$) of 0.39.  The hard color is much higher, with $\mu = 6.67$ and $\sigma = 17.50$.  Many of the heavily absorbed sources (circles, $N_H >10^{23}$\,cm$^{-2}$) are located towards the lower right corner.  These sources clearly have much less flux in the soft band (with the exception of Cyg A).  The mean and standard deviations are: $\mu = 0.06$, $\sigma = 0.13$ (soft) and $\mu = 20.43$, $\sigma = 16.16$ (hard).  Finally, the squares represent the intermediary sources, with $10^{22}$\,cm$^{-2}$\,$ < N_H \le 10^{23}$\,cm$^{-2}$.  These sources have colors clearly in the middle, with soft colors ($\mu = 0.07$, $\sigma = 0.07$) similar to the more absorbed sources and hard colors ($\mu = 6.68$, $\sigma = 8.48$) closer to the less absorbed sources.

In our color-color plot, there are some clear outliers towards the upper right (shown within a box on the graph).  The high column density sources include the complex `changing-look' AGN, NGC 1365, and NGC 5728, a strongly barred galaxy whose host is likely the main contributer to the soft thermal emission.  The remaining sources include Mrk 3 (square) and both NGC 4945 and NGC 6814 (triangles).  Among these, Mrk 3 and NGC 4945 have complex spectra.  NGC 6814 is well fit by an unabsorbed power law but has a strong Fe-K$\alpha$ line (EW$ = 545$\,eV).  There is very little flux in the ASCA observation below 1\,keV.  The strong emission in Fe K$\alpha$ and weak flux at soft energies contribute to give it an unusual location in the plot.  In short, all the sources at this location have complex or atypical spectra.  However, another factor which can lead to unusual positions in the color-color plot is time variability.  This variability could be
over the softer X-ray band ($< 10$\,keV) or in the BAT observations.  The complex source NGC 6814, for instance, is known to show X-ray variability by at least 10 times over time scales of years
 \citep{2003ApJ...597..479M}.  In the 14--195\,keV band, the corresponding BAT fluxes are time averaged over months.  Therefore, sources that are variable in the X-ray or BAT bands may have unusual hard colors.  Analysis of the X-ray light curves, however, is beyond the scope of this paper. 

We next summarize the X-ray spectral shape of the BAT AGNs.  In Tables~\ref{tbl-2} and~\ref{tbl-3}, we included a column with the type of model used (simple or complex).  Excluding the 17 blazars, 63 sources are best fit with a simple model while 65 require a more complex model.  As stated earlier, our complex model referred to a better fit (over the simple power law model) with either a partial covering or double power law model.  Among the complex sources, 5 corresponded to low column density sources ($N_H \le 10^{22}$\,cm$^{-2}$), 20 to sources with intermediate columns ($10^{22}$\,cm$^{-2}$\,$ < N_H \le 10^{23}$\,cm$^{-2}$), and 40 to the heavily absorbed sources ($N_H >10^{23}$\,cm$^{-2}$).  All of the heavily absorbed sources were better fit with a complex model.  This shows that our color-color plot is particularly good for selecting complex sources, which lie to the right of the constant $\Gamma = 1.5$ line (indicating colors for a fixed spectral index with an absorbed power law model).  

Among the complex absorbed sources, we classify 28 AGN as ``hidden''/buried sources.  ``Hidden'' AGN were first distinguished as a new class discovered by BAT in \citet{2007ApJ...664L..79U}.  These sources are extremely obscured, possibly by a geometrically thick torus.  Since the hidden sources were identified in the X-ray and only two sources were identified in the discovery paper, the multi-wavelength properties are as of yet not fully classified.  The X-ray spectra are characterized by a very small scattering component ($\la 3$\%) in the soft band.  In the partial covering model, the model is a multiplicative model defined as $M(E) = f \times e^{-N_H \sigma(E)} + (1 - f)$.  Here, $f$ is the covering fraction, $N_H$ is the neutral hydrogen column density, and $\sigma(E)$ is the photo-electric cross-section.  The scattering fraction is then the value (1 - f).  For our spectral fits, the partial covering model is applied to a simple power law spectrum.

We adopted the criteria that a hidden source is one where the scattering component is $\le 0.03$ and the ratio of soft (0.5 -- 2\,keV) to hard (2 -- 10\,keV) flux is $\le 0.04$.  Of the 28 identified, 9 are from \citet{2008ApJ...674..686W}.  The additional sources include 5 sources with ASCA spectra (NGC 788, NGC 3081, NGC 3281, NGC 4507, and IC 5063) and 4 with Suzaku spectra (ESO 297-018, ESO 005-G004,  and 2MASX J09180027+0425066, and 3C 105).  The remaining 10 sources have XRT spectra (listed in Table~\ref{tbl-2}) which are identified as complex (C) or as well-fit with a simple model but with little/no soft flux (S*).  The median column density of the hidden sources is $\log N_H = 23.57$ and the median soft/hard flux is 0.013, consistent with the properties of the previously identified hidden sources.  

\section{The `Uniform' Sample}
\label{bat-unbiased}
Having summarized the general properties of the entire 9-month BAT AGN sample, we now will focus on a uniform sample.  Our uniform sample includes 102 Seyferts at $|b| \ge 15$\degr, with 46 simple power law model sources and 56 complex model (partial covering or double power law model) sources.  In this section, we present general properties of this uniform sample including the distribution of $N_H$, $\Gamma$, $L^{corr}_{2-10 keV}$, and L$^{corr}_{2- 10 keV}$/L$_{Edd}$.  We also discuss the existence of correlations between these properties, particularly with the Fe-K$\alpha$ equivalent width.  Finally, we discuss the properties of our classified simple and complex sources in more detail.  We note that our comparisons throughout, for instance the fraction of complex sources, have not been corrected for their relative numbers in the luminosity function. Therefore, our conclusions are based solely on the sources in our BAT AGN sample and not in the entire phase space of AGN.  

\subsection{Distribution of $N_H$ and $\Gamma$}
In Figure~\ref{fig-nh}, we plot the normalized column density distribution of the simple (left) and complex (right) sources.  The column densities are columns above the Galactic value, measured from the X-ray spectral fits.  Where no additional absorption was necessary, we set $N_H = 10^{20}$\,cm$^{-2}$.  From the plots, it is clear that simple model sources have much lower column densities than sources with spectra modeled with a complex model.  The mean and standard deviation for these $\log N_H$ distributions are: $\mu = 20.58$, $\sigma = 0.74$ (simple) and  $\mu = 23.03$, $\sigma = 0.71$ (complex).  Notice that none of the complex model sources have columns of $\log N_H \le 21$.  As already noted above, none of the simple power law model sources have columns of $\log N_H \ge 23$.        

With our simple model fits, we have determined the photon spectral index, $\Gamma$, best fit to the continua of our sources.  We excluded a treatment of reflection, using a simpler model which is recorded more uniformly throughout the literature (absorbed power law/ partially covering absorbed power law models).  This allows us to use results from the literature for most of our sources. If we had used the reflection model ({\tt pexrav} in XSPEC), many fewer literature results could be included. In Figure~\ref{fig-pow}, we plot the distributions for the simple power law model (left) and the more complex model (right).  For the complex sources, we exclude 3C 452 whose flat spectrum is best characterized by reflection \citep{2006ApJ...642...96E}.  The mean and standard deviation for the simple model is $\mu = 1.78$ and $\sigma = 0.24$.  This is consistent with the $<\Gamma> = 1.75$ we obtained from our representative sample  \citep{2008ApJ...674..686W}.   For the complex sources, 10 had been best fit by a double power law model while the rest were fit with a partial covering model.  Of these 10, all but 2 had details of the corresponding best-fit partial covering model in \citet{1997ApJS..113...23T}.  We computed the average difference between the two model fits ($<\Gamma_{pcfabs}> = <\Gamma_{dbl}>/1.18$) and applied this correction to estimate $\Gamma_{pcfabs}$ for the 10 applicable sources.  Thus, our distribution is computed using the partial covering model spectral index.  The mean and standard deviation for the complex partial covering model is $\mu = 1.73$ and $\sigma = 0.45$.  There is a larger spread of values for the complex sources, compared to the simple power law model sources.  However, there is only a slight difference in the average spectral index (0.05).  This result is consistent with those of our representative sample in  \citet{2008ApJ...674..686W}, which agrees with the average AGN photon indices reported in \citet{1982ApJ...256...92M} of $\approx 1.8$ from HEAO-1 observations.


\subsection{Distribution of $L^{corr}_{2 - 10 keV}$ and $L^{corr}_{2 - 10 keV}$/L$_{Edd}$}
In this subsection, we test whether Sy 1 and Sy 2 sources have the same or different distributions of both hard band X-ray luminosity (2--10\,keV) and accretion rate.  Towards this end, we needed to compute an absorption independent measure of 2--10\,keV luminosity as well as the Eddington luminosity.  In terms of 2--10\,keV luminosity, absorption has little to no affect on the unabsorbed luminosity for sources with $N_H < 10^{22}$\,cm$^{-2}$.  Therefore, for these sources we used the quoted 2--10\,keV flux to compute 
L$^{corr}_{2 - 10 keV}$ (absorption corrected 2--10\,keV luminosity).  For all luminosity calculations, we used $\Lambda = 0.7$ and H$_0 = 75$\,km\,s$^{-1}$\,Mpc$^{-1}$.  Above $N_H = 10^{22}$\,cm$^{-2}$, absorption has a significant affect on the 2--10\,keV flux.  For these sources, we used {\tt XSPEC} to calculate the unabsorbed 2--10\,keV flux, which we then used to compute L$^{corr}_{2 - 10 keV}$.  

In Figure~\ref{fig-l210lbat}a, we plot our absorption corrected 2--10\,keV luminosities versus the 14--195\,keV luminosities.  We find that the data is well fit ($R^2 = 0.85$) by a line: \begin{math}  \log L^{corr}_{2-10 keV} = (1.06 \pm 0.05) \times \log L_{14-195 keV} + (-3.08 \pm 2.16)\end{math}.  Therefore, the relationship is linear ($L^{corr}_{2-10 keV} \propto L_{14-195 keV}$) showing the validity of our absorption corrected 2--10\,keV luminosities.  On the plot, we label the 5 sources which deviate the most from this relationship (sources with $1.7 \ge L_{14-195 keV}/L^{corr}_{2-10 keV} \ge 1.2$).  
A likely explanation for these sources (NGC 931, 2MASX J09180027+0425066, NGC 2992, NGC 5728, and NGC 6814) deviating from the linear fit is variability in the X-ray spectra, either in the BAT band or the 2--10\,keV band.  Indeed, the 64 day 14--195\,keV light curves of each of these sources \citep{2008ATel.1429....1B} show variability by at least a factor of 2.  See section 3, where we discuss the effects of time variability on hardness ratios.  In Figure~\ref{fig-l210lbat}b, we plot the ratio of $L^{corr}_{2-10 keV}/L_{14-195 keV}$ versus the total luminosity in both bands.  As shown, there is no relationship ($R^2 << 0.1$).  We also indicate on the plot values of spectral index ($\Gamma$) for a constant ratio of $L^{corr}_{2-10 keV}/L_{14-195 keV}$, assuming that the $\Gamma$ is constant between both bands ($\Gamma^{corr}_{2-10 keV} = \Gamma_{14-195 keV}$).

To calculate the Eddington luminosity, we used the black hole mass computed from stellar 2MASS K magnitudes.  In Mushotzky {\it et al.} 2008 (submitted), we calculated the stellar magnitude from the total 2MASS magnitude (obtained from NED) and a nuclear magnitude, calculated by using the IRAF task {\tt qphot} to extract circular photometry from an aperture equal to the PSF FWHM of the 2MASS images.
The stellar absolute magnitude was computed as: \begin{math}  M_{stellar} = 2.5 \log(\chi/(\chi - 1)) + M_{tot} \end{math}, where \begin{math} \chi = F_{tot}/F_{nuc} = 10^{-0.4(M_{tot} - M_{nuc})}\end{math}.  We then transformed the measured stellar 2MASS K magnitudes to mass using the relation set forward in \citet{2006ApJ...637...96N}:
\begin{math} \log M_{BH} = 8.19 + 0.524 \times (M_{K (stellar)} - 23)\end{math}.  The assumption made, in computing the mass, is that the K band stellar light is dominated by the bulge.  Further details can be found in Mushotzky {\it et al.} 2008 (submitted), where we show that the derived stellar luminosity is not correlated with the X-ray luminosity (14--195\,keV) but that the nuclear luminosities are.

Comparing these
black hole masses to the results from reverberation mapping  \citep{2004ApJ...613..682P}, we find that the values agree within the quoted margin of error for the reverberation study (a factor of 3). 
L$_{Edd}$ is then computed as $M_{BH} \times 1.3 \times 10^{38}$\,ergs\,s$^{-1}$.  As an estimate of accretion rate (L/L$_{Edd}$), we use the ratio of L$^{corr}_{2 - 10 keV}$/L$_{Edd}$.  Both quantities are related by the bolometric correction, which could be as large as 100.  Assuming a constant bolometric correction, L$^{corr}_{2 - 10 keV}$ is proportional to L.  We list both black hole mass and L$^{corr}_{2 - 10 keV}$/L$_{Edd}$ in Table~\ref{tbl-4} for the uniform sample.

In Figure~\ref{fig-lledd}, we plot the distribution of our accretion rate proxy, L$^{corr}_{2 - 10 keV}$/L$_{Edd}$ , for the 34 Sy 1--1.2s (red) and 32 Sy 2s (blue).  We find that there is a clear difference in the distributions, with the Sy 2s having lower ratios of L$^{corr}_{2 - 10 keV}$/L$_{Edd}$.  To quantify this, the mean and standard deviations for the logarithm of accretion rate are $\mu = -2.92$ and $\sigma = 0.57$ for Sy 1--1.2s and $\mu = -3.53$ and $\sigma = 0.75$ for Sy 2s.  Since we used unabsorbed luminosities to compute L$^{corr}_{2 - 10 keV}$, this effect is not simply due to the Sy 2s being absorbed and therefore less luminous.  We find that the difference in distributions also holds for luminosity alone, where $\mu = -2.92$ and $\sigma = 0.57$ for Sy 1--1.2s and $\mu = -3.53$ and $\sigma = 0.75$ for Sy 2s.  Using the Kolmogorov-Smirnov test, we find that the differences in distributions are significant, both having a very small P value ($< 0.001$) or probability that the two samples are drawn from the same parent population.

The result of Sy 2s having lower hard X-ray luminosities than Sy 1s is not new, but the Eddington ratio result is.  A lack of X-ray obscured sources at high luminosities was noted by \citet{1986ApJ...303...87R} and
\citet{2002AA...394..835P}.  The results of \citet{2003ApJ...596L..23S} and \citet{2003ApJ...598..886U} showed that the fraction of obscured AGN is lower at high luminosities.  Our luminosity distribution results are new, however, 
in that the BAT AGN include the ``hidden'' sources which had not previously been identified as a class. 
Therefore, our study is  much more unbiased with respect to absorption than the previous studies.

 Also, more importantly, we show that it is not simply the 2 -- 10\,keV luminosities which are different but also the accretion rates, estimated by L$^{corr}_{2 - 10 keV}$/L$_{Edd}$.  These results provide a challenge to the unified AGN model.  If all AGN were essentially the same but viewed at different angles with respect to obscuring material, there would be no difference in the distributions of accretion rate or intrinsic luminosity.  The fact that we are seeing these differences suggests that there is something fundamentally different besides the amount of obscuration.  However, if the bolometric correction for hard X-ray luminosities varies between absorbed (correction of $\approx 85$) and unabsorbed sources (correction of $\approx 35$), as \citet{2005AJ....129..578B} suggest, then our distributions of accretion rates would be the same.  However, the unabsorbed X-ray luminosities would still differ and thus there is still a discrepancy with the unified model.

We are aware of at least three modifications to the unified model that can explain this difference in distributions between L$^{corr}_{2 - 10 keV}$/L$_{Edd}$ in Sy 1s and Sy 2s.  The first possibility is a luminosity
dependent opening angle for the molecular torus, such as the
receding torus model of \citet{1991MNRAS.252..586L}.  A second model is presented in \citet{2000ApJ...530L..65N}, where the broad line region is produced from a wind from the accretion disk.  The wind is produced at a boundary between radiation and gas dominated regions of the disk and its existence is dependent on reaching (or exceeding) a critical accretion rate.  Our observation of a lower distribution of accretion rates in Sy 2s supports this claim.  Their model also predicts that no hidden broad line regions should exist for low accretion rate sources.  Such a test of the existence/non-existence of polarized broad line regions in the lowest accretion rate sources ($L_{bol}/L_{Edd} \le 10^{-3}$ \citep{2003ApJ...589L..13N}), is beyond the scope of this paper.   

Another possibility is that, at lower luminosities, the host galaxy light is so much brighter than the emission from the broad line region that it is completely masked \citep{2002ApJ...579L..71M}.  Therefore, broad line galaxies should be brighter, as we found.  In \citet{2006AJ....131..133P}, it is shown that the properties of high z objects may be modified by dilution of X-ray spectral features 
by star formation.  Given the arc min resolution of ASCA, used for most of our spectra, this allows the blending of emission from star formation and other non-nuclear X-ray features.  We expect this effect to be most pronounced for low luminosity sources, where the ratio of AGN emission to star formation/galaxy emission is lowest.  We quantify the possible effects of this dilution in section 4.6.

\subsection{Correlations of $\Gamma$ with L$^{corr}_{2- 10 keV}$ and L$^{corr}_{2- 10 keV}$/L$_{Edd}$}
Based on recent studies \citep{2004AA...422...85P, 2005AA...432...15P, 2006ApJ...646L..29S}, we wanted to test whether the X-ray power law index is correlated with the 2--10\,keV luminosity or accretion rate.  In comparing our results from a very hard X-ray selected sample with the results from these soft X-ray selected samples, we aim to highlight differences in the spectral properties between the samples.  In Figure~\ref{fig-lleddgamma}, we plot the photon index versus L$^{corr}_{2- 10 keV}$ (top left) and the ratio of L$^{corr}_{2- 10 keV}$/L$_{Edd}$ (bottom left).  We see no evidence of a correlation between $\Gamma$ and 2 -- 10\,keV luminosity or Eddington ratio, even among the Seyfert 1 sources which have smaller associated error bars on $\Gamma$.  Based on previous studies, correlations between $\Gamma$ and L$^{corr}_{2- 10 keV}$ have been seen in high-redshift samples \citep{2004ApJ...605...45D, 2008AJ....135.1505S}, but not among the low-redshift sample ($z \le 0.1$) of  \citet{2000ApJ...531...52G}.  Thus, with our low redshift sample ($<z> = 0.03$), we confirm earlier results showing no correlation.

While we did not find a correlation between $\Gamma$ and our Eddington ratio proxy on average, earlier in \citet{2008ApJ...674..686W} we had found a correlation between flux and spectral index for individual sources.  This correlation was found when we compared multiple X-ray observations from {\it XMM-Newton} and XRT.  Similar results had been seen before for individual AGN \citep{1993ARAA..31..717M}.  Thus, these results show that while on average there is no relationship between $\Gamma$ and a given Eddington rate, on a source by source basis, the photon index becomes steeper with higher accretion rate.  We interpret the fact that we do not see a correlation in the average plot as a result of each individual source having a broad range of luminosities and accretion rates which they vary between.  Since the individual ranges overlap between the sources,  the scatter is a natural result.  

Comparing further with \citet{2006ApJ...646L..29S}, we see at least two possibilities for our different result.  First, the 30 objects selected by \citet{2006ApJ...646L..29S} are moderate to high luminosity RQQs.  Our sample, however, includes lower luminosity sources not present in their sample.  It may be that the correlation is only there for the most luminous AGNs.  In this case, the more luminous AGN may be more similar, perhaps having the same trigger (like a large-scale merger event \citep{2005Natur.433..604D}) which dictates a specific range of accretion rates.  Our sample is more heterogeneous with a larger range in properties and possibly multiple triggers for AGN activity.  Second, the Shemmer black hole masses are estimated from the width of the optical H-$\beta$ line (\citet{2004AA...422...85P} and \citet{2005AA...432...15P} use the luminosity at 5100\AA) while we used the stellar K-band flux.  It is possible that there is a bias in the Shemmer black hole mass determinations relative to ours.  The H-$\beta$ line width may have an explicit dependence on $\Gamma$ \citep{1997ApJ...477...93L}, in addition to black hole mass and luminosity. 

\subsection{The Fe-K$\alpha$ Feature}
Yet another important feature of AGN spectra is the Fe K$\alpha$ line.  For 83/102 sources (81\%), a measurement of the strength of this line, via the equivalent width (EW), was available.  The missing objects include most of the XRT observed sources as well as some from the literature.  For these sources, the data were not of sufficient quality to accurately measure the Fe K line strength.  In Figure~\ref{fig-fek}, we plot the Fe-K$\alpha$ (6.4\,keV) EW versus absorption corrected 2 -- 10\,keV luminosity and our proxy for the Eddington ratio. We note that for three bright Seyfert 1 sources (Fairall 9, 3C 120, and 3C 382), the ASCA data show a strong, broad, Fe K-$\alpha$ line that is not seen in further observations.  While we include the ASCA fit parameters, the high EWs for these sources cause these three points to be clear outliers.

 The X-ray Baldwin/ ``Iwasawa-Taniguchi'' effect \citep{1993ApJ...413L..15I} is the anti-correlation between Fe K-$\alpha$ and luminosity, quantified as $EW \propto L^{-0.17 \pm 0.08}$ by \citet{2004MNRAS.347..316P}.  Recently, this effect has been reported in radio quiet samples of \citet{2006ApJ...644..725J} and \citet{2007AA...467L..19B}.  In the top left panel of Figure~\ref{fig-fek}, we see no evidence of the X-ray Baldwin effect.  However, when we bin the sources by luminosity (excluding the 3 questionable measurements) we do find a correlation (top right panel).  This correlation is seen when we choose the average Fe K EW in each luminosity bin, with $\log EW = (-0.23 \pm 0.03) \times \log L^{corr}_{2- 10 keV} + (12.11 \pm 1.17)$, and is similar to the  \citet{2004MNRAS.347..316P} measured slope.  The significance of the anti-correlation, measured by a correlation co-efficient of R$^2 = 0.93$ (99\% significance), seems to confirm the X-ray Baldwin effect.  However, the results are deceiving.  We find that when we alternatively use the median EW, the correlation becomes much weaker (R$^2 = 0.63$).  This shows that the more absorbed, lower luminosity sources -- among them, the hidden AGN missed in other surveys -- are skewing the results.  Therefore, our data does not confirm the inverse Baldwin effect.
 
 \citet{2006ApJ...644..725J} suggested that the X-ray Baldwin effect is driven by a correlation of Fe K-$\alpha$ EW and Eddington ratio.  Such an anti-correlation was found by \citet{2007AA...467L..19B}, who found that $EW \propto L_{bol}/L_{Edd}^{-0.19 \pm 0.05}$.  In the bottom left panel of Figure~\ref{fig-fek}, we plot the EW versus our proxy for Eddington ratio.  As with the luminosity, we do not immediately find any correlation.  However, when we again bin the values (excluding the 3 questionable measurements), we find an anti-correlation.  Using the mean values of Fe K-$\alpha$ EW, we find that $\log EW = (-0.26 \pm 0.03) \times \log L^{corr}_{2- 10 keV}/L_{Edd} + (1.40 \pm 0.10)$, with R$^2 = 0.89$.  Further, unlike in the luminosity plots, this relation also holds true when we use the median EW.  Thus, our data shows no correlation with unabsorbed 2--10\,keV luminosity but does show an anti-correlation with Eddington rate.

This suggests that the primary relationship causing the X-ray Baldwin effect is the relationship between Fe K  EW and accretion rate.  From Figure~\ref{fig-lledd}, we found that the distribution of $L^{corr}_{2- 10 keV}/L_{Edd}$ is different between absorbed and unabsorbed sources.  The Sy 2s have lower accretion rates than Sy 1s.  We also find this in Figure~\ref{fig-LLEddnh}, where we plot the Eddington rate proxy and the Fe K EW versus hydrogen column density ($N_H$).  In Figure~\ref{fig-LLEddnh} (left), we see that while there is a large range of accretion rates for a given column density, the higher column sources tend to have lower accretion rates.  In Figure~\ref{fig-LLEddnh} (right), we find that the higher $N_H$ sources also have higher Fe K EWs.  This result is expected, particularly if some of the sources are Compton thick.  However, since there are sources with higher EWs and lower $N_H$, there is no direct correlation between EW and column density.

Supporting the case that EW is correlated with accretion rate, \citet{2007ApJ...664..101M} reported a correlation between $\Gamma$ and Fe K EW.  Their data sample consisted of 350 RXTE spectra of 12 Sy 1 -- 1.2 sources, thus, as shown in their Figure 3, it appears that the anti-correlation (except for jet dominated 3C 273, where there is a correlation) is seen mostly for multiple observations of the same source.  Since $\Gamma$ steepens with increasing accretion rate, we claim that the primary relationship is again accretion rate and EW.   


\subsection{Simple Power Law Model Sources}
Nearly half, 46/102, of the uniform sample were well-fit by a simple absorbed power law model.  As mentioned, none of these sources had X-ray column densities $\ge 10^{23}$\,cm$^{-2}$.  In this subsection, we detail additional properties of these sources.  In particular, from the data available we can compare optical Seyfert type with the X-ray column density and examine the sources more closely for soft excesses.  While we hoped to examine sources with warm absorbers in detail, given the non-uniform nature of analyses of warm absorbers in the literature, we defer this topic to a later study.

First, we discuss the optical versus X-ray type.  For each of the simple model sources, an optical Seyfert type is listed in Table~\ref{tbl-1}.  Most of the sources, 30/46 (65\%), have optical classifications of Sy\,1 -- 1.2.  The mean X-ray column density for these sources corresponds to a low column density, log\,$N_H = 20.7$.  As expected, the Sy 1.5 -- 1.9 sources (13/46 or $\approx 28$\%) have a higher mean column density of log\,$N_H = 21.8$.  The 3 Sy\,2 sources have a mean column density of log\,$N_H = 22.8$.  Therefore, for the simple model sources, there is no large discrepancy between the X-ray and optical classifications.  In the unified model of AGN, the presence/absence of optical broad emission lines is explained as an effect of viewing angle.  Therefore, sources whose optical spectra show broad emission lines (Sy1s) would have little obscuring material blocking the central AGN emission while sources with no broad lines (Sy2s) would be more heavily obscured.  Thus, our results agree, in general, with the unified model.

Next, we will examine the spectral fits in more detail.  As a first step, we look at the goodness of fit for our sources through the $\chi^2$ parameter.  While a simple power law or power law and blackbody model was a better choice for our sources than a more complex model (like the partial covering model), 4 of the simple model sources have $\chi^2$/dof values of 1.3 or higher (in Tables~\ref{tbl-2} and~\ref{tbl-3}).  Of these, XSS J05054-2348 only had XRT data available.  The number of data points for this source is small, so the high $\chi^2$/dof value may be a product of poor statistics.  Of the remaining sources, the source with the highest $\chi^2$/dof value (2.2) is Mrk 841.  For this source, XMM-Newton observations analyzed by  \citet{2007AA...470..889P} showed an Fe line complex. In addition, this source is known to have a strong soft excess.  These features were not well fitted by the simple models employed in this study.  Additionally, Chandra observations of Mrk 279 reveal the presence of a weak absorbing outflow \citep{2007ApJ...666..828F}.  Similarly, a high-quality XMM-Newton spectrum of NGC 4593 reveals an ionized warm absorber in addition to the soft excess, as well as cold and ionized Fe K lines \citep{2007ApJ...666..817B}.  

In the spectral fits for our simple model sources, we allowed for the addition of a blackbody component to model the presence of a soft excess.  In our uniform sample, 19/46 ($\approx 41$\%) of the sources required a blackbody for a statistically improved fit.  This is slightly less than the 50\% found in 
\citet{2008ApJ...674..686W}.  The mean blackbody temperature for our sources is $kT = 0.10$\,keV with $\sigma = 0.07$\,keV.  This value agrees with the blackbody temperatures found for the PG quasars \citep{2004AA...422...85P, 2005AA...432...15P} as well as those for type 1 AGN in the Lockman Hole \citep{2005AA...444...79M}.  We note that unlike the results of \citet{2004MNRAS.349L...7G}, who found $<kT> = 0.12$\,keV with $\sigma = 0.02$\,keV for the PG quasars, we do see scatter in the values of $kT$ (which becomes apparent in Figure~\ref{fig-simple}).  However, since they used two Comptonization models instead of a blackbody and power law model, it is hard to make a direct comparison.

For the sources with measured soft excesses, we wanted to test whether there was a relationship between the blackbody temperature ($kT$) and our proxy for Eddington rate, black hole mass, and the photon index ($\Gamma$). In Figure~\ref{fig-simple}, we plot the results of these comparisons.  As seen in the figure, there is no correlation of any of these values.  We tested this by calculating the coefficient of determination, R$^2$, which was $< 0.10$ for each comparison.  Based on current understanding of the soft excess, we did not expect to find correlations for these parameters.  If the soft excess were the result of a thermal process, blackbody emission from a disk surrounding the black hole, we would see a correlation between black hole mass and the blackbody temperature ($T \propto M^{-1/4}$ ${L/L_{Edd}}^{1/4}$).  We do not find a correlation between kT and mass, kT and $L^{corr}_{2-10 keV}/L_{Edd}$, or kT and ${L^{corr}_{2-10 keV}/L_{Edd}}^{1/4} \times (M/M_{\sun})^{-1/4}$ (not shown in Figure!\ref{fig-simple}).

In Figure~\ref{fig-simple}, we also plot the luminosity in the power law component versus the luminosity in the blackbody component.  Combining spectra from our own analysis with downloaded spectra of sources with soft excesses previously analyzed in the literature, we calculated unabsorbed fluxes for 17/19 sources in the 0.3--10\,keV band for each component (power law and blackbody).  As shown in the plot, we did find a correlation between the two values.  The correlation is signficant, R$^2 = 0.48$, with $L_{pow} = (0.79 \pm 0.14) \times L_{kT} +  (9.34 \pm 6.04)$.  Thus, we find that $L_{pow} \propto L_{kT}$.  Such a relationship may provide a challenge to soft excess models where the excess is the result of an absorbing model, unless there is an explicit luminosity dependence between the absorbing wind and source emission.  In terms of a reflection origin, where the soft excess is the result of emission from reprocessed disk emission, there is no implicit contradiction.  However, in this case, our results indicate that since the reprocessed emission ($L_{kT}$) is the same order as the input spectrum ($L_{pow}$) the process would need to be highly efficient, with the reprocessor occupying a large solid angle and a very high reprocessing efficiency.

\subsection{Complex Sources}
Slightly more than half of our sample ($\approx 55$\%) consists of sources whose X-ray spectra were not well-fit by a simple absorbed power law model.  Here, we discuss the optical types of these sources.  We also discuss the fraction of complex sources and the nature of the complexity.

As expected, the optical types for the complex sources are opposite those of the simple model sources.  Here, only 4 sources are identified as Sy\,1 -- 1.5, where the complexity in their spectra is a result of complex absorption (e.g. IC 4329A is known to have at least 7 separate absorbers in the X-ray spectrum \citep{2005AA...432..453S}).  The majority of sources are Sy\,2s (36/56 or 64\%), with the remaining 16 sources Sy\,1.5 -- 1.9s (29\%).  As with the simple model sources, the optical types roughly matched the X-ray column densities.  The mean values of log\,$N_H$ corresponded to 22.0 (Sy\,1 -- 1.2), 23.2 (Sy\,1.5 -- 1.9), and 23.6 (Sy\,2).  Thus, we find no substantial discrepancy between our data and the standard unified AGN model for our entire uniform sample, with respect to X-ray/optical classifications.

One question remaining is the cause of the complexity.  We fit the spectra in this category with a partial covering model, but this model is flexible and can be used to describe more than just a cloud or clouds of material blocking some of the AGN light.  As already mentioned, one cause of complexity is complex absorption, e.g. warm absorption.  The Seyfert 1--1.2 sources from our sample with complex spectra are known to fit in this category.  Another possible cause of complexity is scattering of direct emission from an obscured region into our line of sight, accounting for the soft emission ($< 2$\,keV).  We can not easily test either of these theories with our data.  Yet another cause of complexity could be that the soft emission is not from the AGN, but rather from X-ray binaries, star formation, or hot ionized gas in the host galaxy.  Since we do not expect this emission to exceed a few $\times 10^{41}$\,ergs\,s$^{-1}$ \citep{2003AA...399...39R}, we can automatically rule out this scenario for sources with higher luminosities in the soft emission.

In Figure~\ref{fig-lsoft} (left), we plot the distribution of soft X-ray luminosity in the 0.5--2.0\,keV band for our complex sources.  The mean luminosity is $\log L_{0.5-2.0 keV} = 41$ with $\sigma = 0.94$.  We find that only 13/54 sources have soft emission  high enough to exclude a simple explanation of galactic emission ($\log L_{0.5-2.0 keV} \ge 41.5$).  For these sources, which include all of the 4 Sy 1s as well as 3 ``hidden'' AGN, it is unlikely that the soft emission is from the host galaxy.  The fact that a few of the hidden sources can not be explained by this model suggests that an alternative model, like scattering or partial covering, is more favorable.  However, for 75\% of the complex sources, the luminosities are too low to exclude galactic emission.  In Figure~\ref{fig-lsoft} (right), we plot the observed soft luminosity versus the observed hard luminosity.  We would naively expect the two luminosities to be directly correlated if they are related and not due to galactic emission.  Of course, the effects of obscuration in the 0.5--2\,keV band have not been considered to make this plot, particularly since the nature of the soft emission is ill-determined.  A strong correlation is not seen, but this does not rule out any of the possibilities.  Unfortunately, the present data set has too low an angular resolution, on average, to distinguish galactic sources of soft emission from the AGN.  We know that in some cases, e.g. Circinus, NGC 1365, and NGC 4151, the soft emission is due to X-ray binaries, hot gas from star formation, and extended emission from AGN cores, respectively.  However, higher quality data with the superior spatial resolution of Chandra is needed to solve the problem of the origin of the soft emission.  Even with the low fluxes of many of our sources, not very long ($\approx 10$\,ks) Chandra exposures would be required to obtain images of the soft emission.  

Earlier, we had shown that the distribution of 2--10\,keV luminosities and our Eddington ratio proxy ($L^{corr}_{2-10 keV}/L_{Edd}$) was lower for Sy 2s than Sy 1s.  Another important investigation that we can now make is the fraction of obscured AGN as a function of luminosity and accretion rate.  In Figure~\ref{fig-fraction}, we show the results of the fraction of sources with column densities above $\log N_H = 22$ in each indicated 2--10\,keV unabsorbed luminosity bin and  $\log L^{corr}_{2- 10 keV}/L_{Edd}$ bin.  We also show the subset fraction of sources with $\log N_H \ge 23$ (in black).  These plots show us that there are clearly less obscured sources at high luminosity.  The highest luminosity bin is composed entirely of unabsorbed sources.  One interesting thing to note, however, is that the most absorbed sources ($\log N_H \ge 23$) are not more numerous at lower luminosity.  Rather, they are merely a subset of the absorbed sources.  Instead, the sources with $23 > \log N_H \ge 22$ dominate in the lowest luminosity bins.  It is unclear what this result implies.  However, our results clearly support previous studies which found the fraction of obscured sources low at high luminosities and higher at lower luminosities 
\citep{2003ApJ...598..886U, 2003ApJ...596L..23S, 2005AJ....129..578B}.  Our results argue even more strongly that there must be a modification to the unified model which includes dependence on luminosity.  

Our plot of the fraction of absorbed sources by binned accretion rate (Figure~\ref{fig-fraction} (right)), is more difficult to interpret.  While absorbed sources do dominate at the lowest values of $\log L^{corr}_{2- 10 keV}/L_{Edd}$, the highest Eddington rate bin also shows a large fraction of absorbed sources.  However, we find that this bin includes fewer sources (8), and the result could be a product of poor statistics.  We do note that, as with luminosity, the fraction of the most heavily absorbed sources ($\log N_H \ge 23$) does not appear to increase with decreasing Eddington rate.  Rather, it appears to remain nearly constant.  Still, on the whole, the absorbed sources make up the largest fraction of sources at low Eddington rate and a lower fraction at high Eddington rates. 

Another important conclusion drawn from our analysis is that the percent of ``hidden'' AGN, sources with low scattering fractions ($\le 0.03$), is significant.  These sources comprise 45\% of the complex sources and 24\% of our uniform sample.  This highlights the importance of the BAT survey and its ability to find obscured sources, since these objects have no indication of AGN activity in the soft X-ray band.  Now that these sources are firmly established as an important subset of local AGN, it is important to understand their properties.  In \citet{2008ApJ...674..686W}, we had noted that fitting the sources with a reflection model is problematic.  We were unable to constrain the reflection parameter or the cutoff energy, even with the addition of the BAT spectrum (14--195\,keV).  Further, while the partial covering model provides a decent fit, this model is very flexible.  One probable explanation for the hidden/buried AGN, is that they are embedded in a very geometrically thick torus \citep{2007ApJ...664L..79U}.  

 \citet{1998MNRAS.297L..11F} proposed a model in which low-luminosity AGN were obscured by nuclear starbursts.  This is one possible origin for a geometrically thick torus.  Following \citet{2007ApJ...664L..79U}, we used the 60\,$\mu$m and 100\,$\mu$m fluxes from the {\it Infrared Astronomical Satellite}, obtained from NED, to estimate the far infrared luminosity of the hosts of the hidden sources.   We found these values for 16/24.  The mean value of $\log L_{FIR} = 43.76$ with $\sigma = 0.42$.  We also computed the ratio of $L^{corr}_{2-10 keV}/L_{FIR}$.  Here, we find that  the mean value is 0.26 with $\sigma = 0.26$.  This value is consistent with  ratios for AGNs in the local universe, as pointed out by \citet{2007ApJ...664L..79U}.  It is the sources with very small ratios, $<< 0.1$, which indicate the possibility of significant starburst activity.  In our sample, only 6 sources fall into this category, with the lowest ratio (0.007) corresponding to NGC 7582, an object whose H and K band nuclear light is dominated by young supergiants \citep{1995AA...301...55O}.  However, without higher quality IR observations of the nuclear region and an analysis which can separate out any AGN emission, little conclusions can be drawn from this data.  We (Weaver {\it et al.} in prep) have obtained Spitzer data for a significant fraction of our objects to examine this issue.

Yet another question remaining is how many sources are Compton-thick.  This is not an easy question to answer, especially since different authors use different definitions.  In general, the term has been used to apply to: (1) heavily obscured AGN ($N_H > 1.4 \times 10^{24}$\,cm$^{-2}$, (2) spectra with a high EW Fe K line, (3) spectra with a flat power law continuum, and (4) spectra with a reflection hump.  The last three criteria are all indications of a reflection dominated spectrum.  If we take the Compton-thick definition to apply to sources whose column densities are $> 1.4 \times 10^{24}$\,cm$^{-2}$, none of the BAT-detected sources are Compton-thick.  There are, however, sources which come close (NGC 612, NGC 3281, NGC 1365, NGC 5728, NGC 6921, and NGC 7319), with $N_H \approx 10^{24}$\,cm$^{-2}$.  Without simultaneous data above 10\,keV, it is extremely difficult to definitively discriminate between reflection models and partial covering models.  Even with simultaneous data from Suzaku, we are finding it impossible for some sources (Winter {\it et al.}, in prep).  However, this is not the case for all sources, for instance, NGC 5728 strongly prefers a reflection model and exhibits a strong iron K EW ($\approx 800$\,eV).  Therefore, it is clear that if other criteria are used we do find Compton-thick sources.  For instance, we find that 6 sources exhibit a very flat spectrum ($\Gamma \la 1.0$).  Alternatively,  6 sources have strong Fe K equivalent widths ($EW \ga 600$\,eV).  From our data, we cannot test for the presence of a reflection hump, since the feature is predominately above 10\,keV.  However, with the BAT spectra such studies are forthcoming.  Combining criteria, only 1 source shows both a flat continuum and a strong Fe K EW, NGC 5728.  However, the column density is still below the strict Compton-thick limit.  In this discussion, we have not considered the complicated spectra of Cen A, NGC 1275, and the double nucleus NGC 6240, who may also have consistent properties with the various Compton thick definitions.


\section{The Cosmic X-ray Background}
\label{bat-cxb}
The most recent synthesis models for the cosmic X-ray background (CXB) find that heavily obscured and Compton thick AGN are as important as unabsorbed AGN \citep{2007AA...463...79G}.  With a uniform sample of AGN from the BAT, we can begin to understand the contributions of both types of AGN in the local Universe.  In the previous section, we have presented the distribution of column densities and power law indices from the 0.1 -- 12\,keV X-ray bands for our unique uniform 14 -- 195\,keV X-ray sample.  These properties will provide a valuable input to the CXB models for low redshift ($z \approx 0$) AGN. In addition, in this section we provide the 2 -- 10\,keV $\log N$-$\log S$ relationship.  With this relationship, we will comment on the completeness of a very hard X-ray selected sample in the 2--10\,keV band and its implications.

\subsection{The Average X-ray Spectrum}
Having compiled all of the properties of the 9-month BAT AGN sources, we present the average 0.6--10\,keV X-ray spectrum for our uniform sample of 102 sources in Figure~\ref{fig-avg}.  Since the normalization values are not uniformly recorded in the literature, we could not simply add the spectra together.  Instead, we construct the average spectrum using the shape of the spectrum (from the absorbed power law models) with 2--10\,keV flux used to weight the contribution to the average spectrum.  

To construct the average spectra, we have excluded three important AGN features: blackbody components/soft excess, Fe K lines, and reflection.  We do not include blackbody components since the $<kT> \approx 0.1$\,keV components have very little spectral effects at $E > 0.6$\,keV, the energy above which the CXB is well measured
\citep{2002ApJ...576..188M}.  It is almost impossible to measure the diffuse background below this level because the Galaxy's soft X-ray background is very bright at $E < 0.7$\,keV.  A significant fraction of our spectra are from XRT observations, where one cannot constrain Fe K line emission or reflection. Also many of the ASCA, XMM, and other data sets were analyzed in many different ways with respect to reflection and without re-analyzing the entire data set in a uniform fashion one cannot model reflection correctly.  In particular, we do not believe that the exclusion of reflection has a significant effect on the average spectrum $< 10$\,keV, since the effect of reflection is typically rather small at these energies (for all but the largest reflection fractions).  In fact, as  \citet{1994MNRAS.268..405N} show, a $\Gamma = 1.7$ power law is almost a perfect match to a $\Gamma = 1.9$ power law with reflection in the 2--10\,keV band.

The average spectrum was constructed as: \begin{math} I(E) = \sum A_{i} \times F_{i} \times E^{1 - \Gamma_i}
\end{math}.  For each source, $F_i$ is the observed 2--10\,keV flux.  The measured spectral index was used as the main component of the AGN emission.  Finally, $A_i$ represents the absorbing column density.  To apply the absorption term, we constructed a grid of {\tt XSPEC} absorption models using an input power law with a spectral index of 1.73 (the average of the complex model sources).  The models included simple absorption using {\tt tbabs} with $\log N_H = 20, 20.5, 21, 21.5, 22, 22.5, 23$ and complex absorption using {\tt pcfabs} with a partial covering fraction of 0.95 and $\log N_H = 21.5, 22, 22.5, 23, 23.5, 24$.  The flux of the output {\tt XSPEC} models corresponds to normalized photons\,cm$^{-2}$\,s$^{-1}$\,keV$^{-1}$.  In order to compute $A$ for each model, we divided the model by the power law contribution ($E^{1-1.73}$) and multiplied by energy.  We then applied the appropriate model for each of the sources based on $N_H$ and the complexity flag.

In Figure~\ref{fig-avg}, the contributions of the simple model and complex model sources are shown.  We note that we chose to neglect additional soft emission and Fe K line signatures, as well as reflection.
Fitting a line to our spectrum for energies above 0.6\,keV, we find \begin{math} \log I(E) = (-0.369 \pm 0.004) \log E + (-12.989 \pm 0.003) \end{math}, with $R^2 = 0.97$.  Thus, we find $\Gamma \approx 1.37$.  This result is quite remarkable.  In 1980, HEAO-1 found that the CXB could be modeled as a power law with $\Gamma = 1.4$, below 15\,keV \citep{1980ApJ...235....4M}.  However, the narrow distribution of power law indices ($\Gamma \approx 1.7$) for AGN led to a ``spectral paradox'', where it was unclear how averaging over these dominant bright sources resulted in the flatter power law index fit to the CXB
\citep{1987PhR...146..215B}.  The paradox was resolved theoretically by assuming that many AGN are absorbed, such that $\approx 85$\% of their light is obscured \citep{1999MNRAS.303L..34F}.  With our simple estimate of the average X-ray spectrum from the SWIFT BAT-detected AGN, we have now observationally reproduced the measured CXB slope.  The flux from these bright sources, with a total flux of $1.79 \times 10^{-9}$\flux over 74\% of the sky, accounts for only 0.29\% of the CXB ($2.0 \times 10^{-11}$\flux\,deg$^{-2}$ \citet{2008AA...483..425R}).  However, if the distribution of source properties at $z \approx 1$, where most of the CXB originates, is similar to that of the BAT sources, the spectral paradox is resolved.  A similar conclusion was reached by \citet{2008AA...482..517S}, who calculated the 3--300\,keV SEDs of local AGN using Integral and RXTE data.

\subsection{The 2-10\,keV Log N - Log S Relationship}
Towards determining how complete a 14--195\,keV survey is in the 2--10\,keV band, we plotted the $\log N$-$\log S$ relationship for our uniform sample in the 2--10\,keV band  in Figure~\ref{fig-lognlogs}.  Here, $\log N$ is the logarithm of the number of galaxies with a 2--10\,keV flux above the associated $\log S$ value.  We corrected this value for the BAT sky coverage by using the BAT sky coverage map from \citet{2007arXiv0711.4130T} and the 14--195\,keV flux for each source.  We found that the sample is complete in the 2--10\,keV band to $\log S \approx -11$ ($F_{2-10keV} \approx 1 \times 10^{-11}$\,erg\,cm$^{-2}$\,s$^{-1}$).  To show this, we fit a line to the points above this threshold (shown in the plot).  The fit is very significant, with $R^2 = 0.97$ and \begin{math} \log N(S) = (-1.53 \pm 0.12)\times \log S - (14.93 \pm 1.24). \end{math}
Thus, the measured slope of $-1.53 \pm 0.12$ is consistent with the value $-1.5$ expected for a uniform density of objects.  Our normalization, which corresponds to $1.17 \times 10^{-15}$\,(ergs\,cm$^{-2}$\,s$^{-1}$)$^{-1}$\,sr$^{-1}$, agrees very well with the normalization from the 2--10\,keV $\log N$-$\log S$ relation of HEAO-1 ($2.2^{+0.3}_{-0.2} \times 10^{-15}$ \citep{1982ApJ...253..485P} for AGN).

This shows that above $\approx 1.0 \times 10^{-11}$\,erg\,cm$^{-2}$\,s$^{-1}$ (2 -- 10\,keV) we have a complete sample, consisting of 51 sources.  Only 9 of these AGN are Sy\,2s while 28 are Sy\,1 -- 1.2s, showing that the brightest sources correspond mostly to less absorbed sources even in a very hard X-ray selected sample.  Re-examining the average spectrum of this complete sample (with $F_{2-10 keV} \ga 10^{-11}$\flux), we find that the shape is the same as the larger sample (\begin{math} \log I(E) = (-0.41 \pm 0.005) \log E + (-13.019 \pm 0.003)
\end{math} with $R^2 = 0.973$).  Thus, the average spectrum of these bright sources also replicate the measured CXB slope of $\Gamma \approx 1.4$.  Of this complete sample, all have measured $N_H < 5 \times 10^{23}$\,cm$^{-2}$ and none are Compton-thick.  In fact, most of the sources have much lower column densities (only 6 have $N_H \ge  10^{23}$\,cm$^{-2}$).  This suggests that there are no local Compton-thick objects with 2--10\,keV fluxes above $1.0 \times 10^{-11}$\,erg\,cm$^{-2}$\,s$^{-1}$.  

However, we also find that below this flux threshold we are far from complete.  Particularly, towards $\log S = -12$ the plot suggests that we are missing close to 3000 sources.  Could some of these missing sources be Compton-thick?  From $\log N$-$\log S$ in \citet{2007arXiv0711.4130T}, we know that the 9-month BAT AGN survey is complete in the 14 -- 195\,keV band above $2 \times 10^{-11}$\,erg\,cm$^{-2}$\,s$^{-1}$.  Therefore,  the missing sources in our sample must have 14 -- 195\,keV fluxes below $2 \times 10^{-11}$\,erg\,cm$^{-2}$\,s$^{-1}$ and 2 -- 10\,keV fluxes from $\approx 1$ -- $10 \times 10^{-12}$\,erg\,cm$^{-2}$\,s$^{-1}$.  Since the ratio of $F_{14-195 keV}/F_{2-10 keV}$ for the majority of missing sources (near $\log S = -12$) is $\ge 10$, it is likely that these sources are heavily absorbed.  They may be an extension of the ``hidden'' AGN with higher column densities ($\log N_H \ge 24$).  We can not determine whether or not some of these sources are Compton-thick.


\section{Host Galaxy Properties}
\label{bat-host}
In this section, we present a simple analysis of the host galaxy properties of the uniform sample of BAT-detected AGN.  Since the BAT AGN offer the first unbiased AGN sample with respect to absorption ($N_H \la 10^{24}$\,cm$^{-2}$), studying their host properties can provide major insight into the relationship between the AGN and host.  This is particularly important since the exact trigger of activity around the black hole is not known.

Since the BAT AGN hosts are located nearby ($<z> = 0.03$), archived images are easily available from the Digital Sky Survey (DSS), Sloan Digital Sky Survey (SDSS), and 2MASS.  For our study, we use these data as well as publicly available information from NED. First, we provide a simple comparison of the galaxy major axis/minor axis to the X-ray column density.  We then discuss the host galaxy morphology types as well as the percentage in interacting galaxies.

\subsection{Host Inclination}
One of the questions left to be answered about AGNs is the nature of the obscuring material.  Likely, there are many different answers to this question, depending on the host.  In some cases, the obscuring material may largely be a product of looking through much of the gas and dust in the host galaxy.  In such case, we can use the inclination of the host galaxy to determine how much of the obscuration can be simply attributed to the host.  To test this, we found measurements of the major axis (a) and minor axis (b) in NED.  These data were available for 85/102 sources in our uniform sample.


In Figure~\ref{fig-ba}, we plot the ratio of minor to major axes ($b/a$) versus the X-ray derived column density.  From the plot, we find that sources of all types (Sy\,1 -- 2) have similar distributions of $b/a$.  We find that a face-on galaxy ($b/a = 1$) could host either a Sy\,1 or Sy\,2 source.  However, the edge-on galaxies ($b/a \la 0.4$) only host more absorbed AGNs.  Among the 11 edge-on sources, only one is associated with a Sy\,1.  However, while IC 4329A is an optical Sy\,1.2, the X-ray column density is high for a Sy\,1 ($6 \times 10^{21}$\,cm$^{-2}$) and the X-ray spectrum is complex.  In fact, the XMM-Newton spectrum of this source reveals signatures from 7 different absorbing systems \citep{2005AA...432..453S}.  We note that \citet{1990AJ.....99.1435K} found a deficiency of edge-on galaxies in their optically selected AGN sample, which they explained as a lack of obscured sources in the optical survey.  Our results support this claim, since the unobscured sources are not associated with edge-on galaxies.

For the edge-on sources, it is possible that much of the X-ray obscuration is from the ISM in the host galaxy.   In the Milky Way, we know that the column density looking towards the Galactic Center is a few $\times 10^{22}$\,cm$^{-2}$.  Therefore, this can be a plausible explanation for the 6 edge-on sources with X-ray columns below $10^{23}$\,cm$^{-2}$.  For the other 5 sources, however, the columns are simply too large to be attributed solely to the host galaxy.  

Excluding the edge-on sources, we find the mean values for $b/a$ as $\mu = 0.74$ and $\sigma = 0.13$ for Sy 1 -- 1.2s, $\mu = 0.69$ and $\sigma = 0.14$ for Sy 1.5 -- 1.9s, and $\mu = 0.70$ and $\sigma = 0.15$ for Sy 2s.  This shows that the distributions are effectively the same, with a difference of only 0.04 in $b/a$ between obscured and unobscured AGNs.  Thus, we can not explain differences in obscuration simply through the inclination of the host galaxy for these sources.  The fact that we see a range of $b/a$ values causes us to disagree with \citet{2006ApJ...645..115R}, whose data on 9 optically active AGN (AGN showing optical emission lines) at $0.5 \la z \la 0.8$ led them to conclude that optically active AGN are only in the most face-on or spheroidal host galaxies.  However, they also conclude that optically dull AGN (from a sample of 22 AGN) inhabit a range of $b/a$ ratios, similar to our result for the entire sample.  As for optically dull AGN in the 9-month BAT AGN sample, there are only two sources (NGC 612 and NGC 4992), both of which are intermediate between face-on and edge-on with $b/a = 0.64$ and 0.58, respectively.

 \subsection{Host Morphology}
For our uniform sample, 74/102 sources had morphologies available from NED or LEDA (listed in Table~\ref{tbl-1}).  For all of these sources, we expressed the morphology as a T-type, following the 2MASS large galaxy atlas \citep{2003AJ....125..525J} and the de Vaucouleurs method.  Here, each morphology is assigned a numerical value (E = -1, S0 = 0, Sa = 1, Sab = 2, ..., Irr/Peculiars = 9.5).  In Figure~\ref{fig-hosts} (left), we plot the distribution of morphologies.  For all of these galaxies, we also looked at the accompanying DSS images as confirmation of the NED classification.  Additionally, these images are available for all of the 9-month surveys online at {\url http://swift.gsfc.nasa.gov/docs/swift/results/bs9mon/} \citep{2008ATel.1429....1B}.

From the plot of the distribution of morphologies, we find that our sample includes very few ellipticals ($\approx 8$\%).  This is in direct contrast to deeper AGN surveys where the hosts are predominately in elliptical or red hosts.  For instance, the $z \approx 1$ X-ray selected sample of \citet{2007ApJ...660L..11N} mostly have red colors consistent with early type galaxies (ellipticals and lenticulars).  Meanwhile, our $<z> = 0.03$ sample includes a larger fraction of bluer hosts (spirals and peculiars).  This suggests that there may be an evolutionary change in the host properties between local AGN and those at higher redshift.  However, the SDSS results show a connection between Seyfert 2s and early type hosts but also young stellar populations at $0.02 < z < 0.3$ \citep{2003MNRAS.346.1055K}.  In order to compare more directly with both the SDSS and X-ray selected samples, we need more robust data including spectra and photometry.  Thus, our results, based on NED classifications and DSS images, are preliminary and we are in the process of analyzing higher quality images and photometry (Koss {\it et al.} in prep) as well as spectra (Winter {\it et al.} in prep) to substantiate these findings.  

In the morphology distribution plot, it is clear that a large fraction of the hosts (33\%) are irregular/peculiars, with an equal percentage of Sy 1-1.2s, Sy 1.5-1.9s, and Sy2s with peculiar hosts.  In some sense, the classification as peculiar in NED can be somewhat subjective.  However, we looked through all of the DSS images to confirm the NED classifications.  In a study of 256 nearby ($z \la 0.035$), optically selected galaxies with HST,  \citet{1998ApJS..117...25M} found only $\approx 20$\% of the hosts ``normal'' (i.e. axi-symmetric, a bulge with regular elliptical isophotes, a thin, planar disk).  This suggests that even more than the 33\% identified would fit within this category.   In Figure~\ref{fig-peculiar}, we show optical images of four representative cases.  Some of these sources are quite famous  and have high quality HST images available showing large dust lanes and other disturbances (i.e. Cen A, NGC 1275).  In fact, many of these sources appear to be interacting with nearby galaxies (i.e. ESO 490-G026, ESO 511-G030) or are identified as the result of a merger (as for Cen A).  More images can be seen at the indicated BAT 9-month survey website.  

This leads to the most important result we have found in our analysis of the host galaxies.  Namely, that a large fraction of the galaxies are interacting.  In Table~\ref{tbl-1}, we included a note of Int or Int? to indicate sources that we classify as interacting or possibly interacting.  Here, we use the term loosely to describe sources which, from the available DSS images, have a nearby companion galaxy or show a heavily distorted morphology, which has previously been identified as the result of a merger in the literature.  We supplement the list with sources known to be the product of a merger (like Cen A and Cyg A).  In addition to ESO 490-G026, Cen A, and ESO 511-G030, shown in Figure~\ref{fig-peculiar}, more images of interacting hosts are found in \citet{2007arXiv0711.4130T}.  All six of the sources shown in Figure 8 of that paper are classified as interacting.  Additionally, NGC 454, NGC 6240, and NGC 7469 are interacting sources with HST images released in the recent ``Cosmic Collisions Galore!'' news release.  Just from these available images, we account for 28\% of the sources classified as interacting.  Images of the additional sources are available online at the BAT 9-month survey website.

Of the sources with a known morphology (including those listed as S or S?), we classify 45 sources (54\%) as interacting.  The distinction between peculiars and interacting sources is somewhat arbitrary, since a merger galaxy would be expected to have a peculiar shape.  However, not all of the classified peculiars are also classified as interacting, though a large fraction are (66\%).  If interactions cause disruptions to the galaxy, thereby making it peculiar, the percentage of interacting galaxies is even higher -- including all of the peculiars.  It is also important to note that not all interacting galaxies are peculiar.  The distinction for these sources may be due to a companion that is much smaller than the AGN host or at a greater distance.  Further investigations of these effects will be presented in Koss {\it et al.}, in prep. 

The important thing to note is that a large fraction of local AGN have companion galaxies or are the result of a merger (54\%).  This is significant considering that 12-33\% of optically selected Seyferts \citep{1995AJ....109.1546R} and 15\% of more distant X-ray selected AGN ($0.2 \le z \le 1.2$) \citep{2007ApJ...660L..19P} are hosted in an interacting galaxy.   It is possible that our high percentage of interacting galaxies indicates an evolutionary difference in AGN hosts.  Another possibility is that our value is higher simply because we can see smaller companions at low redshift.  Thus, the sources at $z \ge 0.1$ may have the same number of companions or merger galaxies like Cen A and Cyg A but the available images may not be of sufficient quality to distinguish this.  However, differences in classifying a source as ``interacting'' can also cause our higher value.  For instance, \citet{1998ApJS..117...25M} find less than 0.5\% of their optically-selected Seyferts as interacting based on the requirement of a clear double nucleus and 5\% if they include collisions/mergers. Their study misses a number of the clear mergers in our sample (e.g., NGC 1275, Cen A, Cyg A), suggesting that optical samples select less mergers due to the difficulty in detecting optical AGN signatures in heavily absorbed AGN.

Since such a large percentage of hosts are interacting, we wanted to test whether the distribution of luminosities and Eddington ratios were the same or different between AGN with interacting and non-interacting hosts.  In Figure~\ref{fig-int_lum}, we plot the results.  We find no difference in the AGN unabsorbed 2--10\,keV luminosity or the Eddington rate proxy ($L^{corr}_{2-10keV}/L_{Edd}$) between the two distributions.  For interacting systems, $\mu = 42.97$ and $\sigma = 0.80$ for $\log L^{corr}_{2-10keV}$ while $\mu = -3.38$ and $\sigma = 0.76$ for $L^{corr}_{2-10keV}/L_{Edd}$.  The non-interacting systems have a very similar distribution, with $\mu = 43.12$ and $\sigma = 0.81$ for $\log L^{corr}_{2-10keV}$ and $\mu = -3.29$ and $\sigma = 0.70$ for $L^{corr}_{2-10keV}/L_{Edd}$.  Using the Kolmogorov-Smirnov test, the P values (0.65 for luminosity and 0.91 for Eddington ratio) are not small and therefore do not suggest a difference between the distributions.

Further, when we examine the morphologies of the interacting host galaxies, we find that they follow the same distribution as the larger sample (Figure~\ref{fig-hosts} (right)).  When we look at the distribution of optical Seyfert type within interacting galaxies, we find no preference for absorbed (Sy 2) systems.  Broken down by optical type, 14 correspond to Sy 1 -- 1.2, 13 in Sy 1.5 -- 1.9, and 13 in Sy 2 sources.  This is similar to the results of 
\citet{1995AJ....109.1546R}, who also found no difference in the percentage of interacting sources between Sy 1s and Sy 2s.  However, this result is somewhat confusing since we would expect optically-selected samples to find less of the interacting Sy 2s due to heavy absorption.  Since the morphologies are a nearly even mix of Sy\,1 -- Sy\,2 sources, it is not surprising that the mean 2--10\,keV luminosities and Eddington ratio proxies are intermediate between the Sy\,1 and Sy\,2 values quoted in Section 4 (for both interacting and non-interacting hosts). 

\vspace{0.5cm}

Based on simple analyses of publicly available data on the host galaxies, we have found a few interesting results.  We have found that the host inclination (approximated by $b/a$) does have an effect on the amount of obscuration we see in the X-ray band, but that the effect is slight.  In particular, there are no unabsorbed sources hosted in an edge-on galaxy.  However, there are heavily obscured AGN in face-on galaxies.  This shows that for the bulk of the obscuring medium, the origin is most likely more local to the accreting black hole.  This is further supported by HST observations of nearby AGN, which find Sy2s to be more likely to have dust lanes or irregular or disturbed dust absorption through the galactic nucleus \citep{1998ApJS..117...25M}.

Also, we found that many of the hosts are in spirals and pecular/irregular galaxies.  We find very few (5/74) in ellipticals, contrasting with the results of \citet{2007ApJ...660L..11N} who find red hosts for the majority of $z \approx 1$, X-ray selected AGN.  This suggests an evolutionary change in AGN host properties, from red at $z \approx 1$ to blue at $z \approx 0.03$.  However, high quality photometry is necessary to make a direct comparison with \citet{2007ApJ...660L..11N}.
 
More than half of our sample is associated with a close companion or recent/ongoing merger.  We find that Sy 1 and Sy 2 sources are equally likely to be hosted in interacting galaxies.  Considering that optical surveys at similar redshift found between 12-33\% \citep{1995AJ....109.1546R}, it is tempting to attribute the difference to selection effect, i.e. the heavily obscured sources missed in the optical.  However, we find that the hosts of the interacting systems are not dominated by absorbed sources.  

We also find that the distribution of host morphologies are the same between interacting/non-interacting hosts.  Further, the distributions of L$^{corr}_{2-10 keV}$ and $L^{corr}_{2-10keV}/L_{Edd}$ are also the same.  Likely, mergers are not the sole trigger for our sample.  The next step in understanding the differences between the interacting and non-interacting systems is an in-depth source by source look at the local environments, with particular attention to star-formation and dust.  This, however, is beyond the scope of our paper. 

Currently, our team is compiling and analyzing data from SDSS images and our own ground-based observations for the 22-month BAT AGN sample.  With this data, we will obtain morphologies for dimmer optical sources and colors for a complete sample.  In addition, we will revisit the nature of the interacting galaxies, quantifying the type of interaction involved as in \citet{1985ApJS...57..643D}.  These results will be presented in Koss {\it et al.} (in prep).

\section{Summary}
\label{bat-summary}
In this paper, we present the X-ray properties of a uniform sample of very hard X-ray (14 -- 195\,keV) selected AGN.  We present a number of interesting results that highlight the many uses of a uniform very hard X-ray survey.  This paper is complimentary to the 9-month AGN survey paper \citep{2007arXiv0711.4130T}, which presents the 14--195\,keV properties of the sources.  Additionally, this paper confirms the results of our earlier study on {\it XMM-Newton} observations of a representative sample of the BAT AGN  \citep{2008ApJ...674..686W}.  Among these, we show that: (1) the X-ray and optical classifications agree, i.e. Sy 1s have low X-ray column densities while Sy 2s are more obscured, (2) the average power law index, $\Gamma \approx 1.8$, agrees with the results from HEAO-1 \citep{1982ApJ...256...92M}, (3) ``hidden'' AGN are a significant fraction of local AGN, where we can now quantify this value as $\approx 24$\%, and (4) nearly half  (45\%) of local AGN are well-fit by a simple model (all with $\log N_H < 23$) while the remaining sources (55\%) require a more complex model.  In addition, this paper presents a number of additional, important results. 

From examining the host galaxy properties, we found that the majority of the X-ray obscuration is not simply from the host galaxy (by comparing host inclination ($b/a$) to X-ray column density).  The most surprising results, however, were that many of the host galaxies are peculiar/irregular galaxies (33\%).  Further, an even larger fraction (54\%) have either a close companion galaxy or are known mergers. This is observational proof that galaxy interactions may be driving activity in local supermassive black holes.  However, we also find that the distribution of AGN 2--10\,keV luminosities and accretion rates, as well as morphologies, are the same between interacting and non-interacting hosts.  While, it is unclear what these results mean, however, there appears to be more than one trigger besides mergers for local AGN activity.  Our team is currently compiling higher quality images and photometry to better quantify these results.

From our uniform sample (102 sources with $|b| \ge 15^{\circ}$), we found that the distributions of both unabsorbed 2 -- 10\,keV luminosity and accretion rate are significantly lower for Sy 2s than Sy 1s.  While earlier studies found this connection in 2 -- 10\,keV luminosity \citep{2003ApJ...596L..23S, 2003ApJ...598..886U}, this is the first time it has been reported in accretion rate.    We also showed that the fraction of obscured AGN is indeed larger for lower luminosities (absorption corrected $L_{2-10 keV}$) and accretion rates.  However, we note that the most heavily obscured sources ($\log N_H \ge 23$) do not dominate this relationship. Since the unified model predicts differences between absorbed and unabsorbed sources are a product of viewing angle alone, our results provide a challenge, arguing in favor of a luminosity-dependent AGN model.

Another result involves the correlation between accretion rate and $\Gamma$.  In \citet{2008ApJ...674..686W}, we had found indication of a connection between Eddington ratio (or 2 -- 10\,keV luminosity) and $\Gamma$ using the spectral fits of multiple observations for individual sources.  The fact that we did not observe a correlation in our larger sample seems to be a result of our sources having a larger range of Eddington ratios (or 2 -- 10\,keV luminosities).  We suggest that previous studies, for instance by Shemmer et al. 2006, see this correlation because their samples have a narrower range of properties (being mid- to high luminosity AGNs).  The primary correlation appears to be with accretion rate and not hard band luminosity. Such a correlation should appear when comparing $\Gamma$ to  $L^{corr}_{2-10 keV}/L_{Edd}$ for multiple observations of individual sources or for a sample of sources with a narrow range of accretion rates.  

In a similar manner, we found that while our sample did not immediately confirm the X-ray Baldwin effect, binning the sources by luminosity, we were able to reproduce the anti-correlation between unabsorbed 2 -- 10\,keV luminosity and Fe K EW.  The primary  anti-correlation, however, again appears to be with Eddington rate.  When we binned the values by our Eddington ratio proxy, we found that $EW \propto {L^{corr}_{2-10 keV}/L_{Edd}}^{0.26 \pm 0.03}$ (agreeing with the results of \citet{2007AA...467L..19B}).  Since both $\Gamma$ and Fe K EW are dependent on accretion rate, this suggests that the $\Gamma$-EW correlation found by  \citet{2007ApJ...664..101M} is a result of the accretion rate dependences.

Having classified the X-ray spectra of our sample into simple and complex categories, we were able to examine the properties of the two sub-samples in more detail.  For the simple model sources, we found that 41\% of the sources exhibited a soft excess.  Having modeled this parameter with a simple blackbody model, we found the average temperature to be $kT = 0.10$\,keV.  We also found that there was a significant amount of scatter in this value ($\sigma = 0.07$\,keV), contrasting with the \citet{2004MNRAS.349L...7G} results for PG quasars.  We found no correlation between the blackbody temperature and Eddington ratio, black hole mass, or photon index.  However, we did find a correlation between the luminosity of the blackbody component and the luminosity in the power law.  This relationship is linear ($L_{pow} \propto L_{kT}$) and may provide a challenge to the current soft excess models. 

Examining the complex model sources, we found that the majority of these sources included absorbed AGN.  Of the 4 Sy 1s in this category, all have complex absorption features in their X-ray spectra.   For these sources, we showed that the nature of the soft emission ($L_{0.5-2 keV}$) for these sources is unclear.  Over half have soft band luminosities low enough to be the result of galactic emission from star formation/X-ray binaries.  However, of the sources with higher soft luminosities, 3 are ``hidden''/buried AGN.  This argues that the soft emission may be scattered AGN emission ($\le 0.03$).

An important result we found is that the ``hidden''/buried AGN, sources with a high covering fraction, are a significant fraction of local AGN.  Among the complex sources, 45\% are ``hidden''.  For these sources, we found that the FIR luminosity is not consistent with an increased star formation rate, as suggested by \citet{2007ApJ...664L..79U}.  However, without higher quality X-ray spectra and multi-wavelength observations, we are unable to further explore the nature of these sources.

While BAT is quite good at finding heavily obscured sources, we found that none of the 9\,month sources in our uniform sample have spectra consistent with heavily obscured Compton-thick objects ($N_H > 1.4 \times 10^{24}$\,cm$^{-2}$).   However, we do detect sources classified as Compton-thick in other studies based on a reflection dominated spectrum or strong Fe K EW (for instance 3C 452 and NGC 4945).  Additionally, we did not include an analysis of the very complex source NGC 6240, which may also be classified as Compton thick.  Since the Compton hump lies above 10\,keV, spectral fits with and without reflection can be degenerate in the 0.1--10\,keV band.  Therefore, a full analysis of the Compton thick nature of the BAT sources must be deferred to future studies.

One remarkable result we found came from the average spectrum we constructed in the 0.1--10\,keV band with the measured spectral properties of our uniform sample.  Here, our data reproduce the measured slope of the CXB ($\approx 1.4$).  This highlights the importance of the BAT survey in selecting heavily absorbed sources.  More importantly, this is observational proof that the combination of BAT-detected absorbed and unabsorbed local AGN replicate the shape of the CXB.  If the distribution of source properties at $z \approx 1$, where much of the CXB originates, is the same as that of the BAT-detected AGN, the spectral paradox is resolved.

To test our completeness in the 2--10\,keV band, we plotted the distribution of $\log N$-$\log S$ for the entire uniform sample.  This showed that while the sample is complete in the 14--195\,keV band \citep{2007arXiv0711.4130T}, we are only complete above 
$\log S = -11$ in the 2--10\,keV band.  Further, this distribution suggests that we are missing as many as 3000 sources at $\log S = -12$, requiring that these sources have 14--195\,keV fluxes below the current flux limit of the BAT survey.  Possibly these sources are ``hidden'' AGN with even higher X-ray columns ($\log N_H \ge 24$).  Such sources must have a high ratio of $F_{2-10 keV}/F_{14-195 keV}$, like NGC 1068.  Also, they may or may not contain Compton-thick sources, an answer to which our data can not supply.  These results, in addition to the X-ray properties (including column densities and spectral indices) will provide important input for CXB models at low redshift ($z \approx 0$).
 
Overall, our analysis of the X-ray properties (and some host galaxy properties) show the interesting nature of very hard X-ray selected AGN.  In order to understand the properties further, we are continuing to collect and analyze the properties in the optical through spectra (Winter et al. in prep) and imaging (Koss et al. in prep), the IR through {\it Spitzer} observations (Weaver et al. in prep), and radio (Sambruna et al. in prep).  Additionally, BAT is continuing to discover more sources at fainter fluxes, with sensitivity increasing as $\sqrt t$.  With the additional sources in future BAT catalogs, we will obtain an even better understanding of local AGN.

\section*{Appendix}
Here, we present notes on individual sources.  These notes include complexities associated with the sources.  Particularly, in notes to Table 1, we include details on why sources needed to be excluded or why XRT observations were used instead of data available from other missions.  Also, we include details on extra components added in order to model the X-ray spectra, corresponding to the XRT analysis (Notes to Table 2) or the ASCA/{\it XMM-Newton}/Chandra/Suzaku analyses (Notes to Table 3).  Throughout, the sources are ordered in RA.

\subsection*{Notes to Table 1}
\par{\bf NGC 1275: } This source is located within the Perseus cluster.  As such, emission from the cluster is difficult to separate from the AGN emission.  Therefore, we do not include an analysis of this source.

\par{\bf PKS 0326-288: } The XRT observation of this source is too short to extract an X-ray spectrum.

\par{\bf NGC 1365: } This source is a ``changing-look'' AGN, observed in both Compton-thin and Compton-thick states.  The source has been observed extensively by XMM-Newton and Chandra with relevant papers -- for instance \citet{2007ASPC..373..458R} and \citet{2005ApJ...623L..93R}.  For simplicity, we averaged the XRT observations, taken within a day of each other.  However, the spectral parameters of this source, particularly column density, are quite variable.

\par{\bf 4U 0517+17: } The {\it XMM-Newton} observation of this source is not yet publicly available.

\par{\bf Mrk 6: } This source is known to have a complex absorber, with ASCA and XMM-Newton spectra well modeled by a double partial covering model \citep{1999ApJ...510..167F,2003AJ....126..153I}.  Due to its complex nature, we only include our XRT analysis with a simple model, for easy comparison with the other sources.

\par{\bf IGR J12026-5349: } The Chandra observation of this source is not yet publicly available.

\par{\bf NGC 4102: } The Chandra observation is too short to compare with the spectra of other sources in this survey.

\par{\bf Cen A: } Cen A recently underwent a galaxy merger.  The X-ray spectrum is extremely complicated, as presented in \citep{2004ApJ...612..786E}.  Due to the complex form, we do not include an analysis of this source.

\par{\bf NGC 6240: } This peculiar source hosts two active nuclei, as found by Chandra \citep{2003ApJ...582L..15K}.  Therefore, we do not include an analysis of this source.

\par{\bf GRS 1734-292: } This source is a Sy 1 seen through the Galactic plane in a heavily obscured region.  Therefore, the exact hydrogen column is not known.  An analysis of the spectrum of this source is not included.

\subsection*{Notes to Table 2}
\par{\bf NGC 1365: }{The spectral fit to this source also includes the addition of an {\tt apec} model with kT of 0.74$^{+0.07}_{-0.10}$\,keV and an Fe K line at 6.54$^{+0.08}_{-0.09}$\,keV with EW $\ga 780$\,eV.  The fitted column density has large errors.}

\par{\bf 2MASX J04440903+2813003: }{The residuals to this fit show clear evidence of an Fe K-$\alpha$ line ($\Delta \chi^2 = 20.9$).  Adding a gaussian for this component, with $\sigma$ fixed at 0.01\,keV, we find E $= 6.41^{+0.05}_{-0.03}$\,keV and EW$= 140^{+70}_{-54.6}$\,eV.}

\par{\bf 1RXS J045205.0+493248: }{The model includes a significant ($\Delta\chi^2 = 23$) soft excess modeled with a blackbody (kT$ = 0.10^{+0.05}_{-0.03}$).  The residuals indicate the presence of possible additional lines (including Fe K-$\alpha$) in the spectrum, but a higher quality spectrum is necessary to fully resolve these features. }

\par{\bf Mrk 6: }{The residuals to this fit show clear evidence of an Fe K-$\alpha$ line ($\Delta \chi^2 = 10.7$).  Adding a gaussian for this component, we find E $= 6.41^{+0.07}_{-0.06}$\,keV, $\sigma = 0.09^{+0.10}_{-0.05}$\,keV, and EW$= 498^{+264}_{-250}$\,eV. }

\par{\bf SDSS J074625.87+254902.2: }{\citet{2006ApJ...646...23S} provide an analysis of the BAT + XRT spectra, however, we re-analyze an observation for comparison of the parameters from a fit without the BAT spectrum. 
}

\par{\bf IRAS 09149-6206: }{An unabsorbed blackbody component was required ($\Delta\chi^2 = 100$) at soft energies, probably a foreground Galactic object given the AGN's position in the Galactic plane, with kT $= 0.13 \pm 0.02$\,keV.}

\par{\bf LEDA 093974: }{The residuals to this fit show evidence of an Fe K-$\alpha$ line ($\Delta \chi^2 = 5.4$).  Adding a gaussian for this component, with $\sigma$ fixed at 0.01\,keV, we find E $= 6.46^{+0.09}_{-0.14}$\,keV and EW$= 297^{+326}_{-187}$\,eV. }

\par{\bf 1RXS J112716.6+190914: }{A blackbody component was also added ($\Delta\chi^2 \approx 20$) with kT$= 52^{+13}_{-12}$\,eV.  There are some residuals in the fit but a higher quality spectrum is necessary to investigate these features.}

\par{\bf 1RXS J174538.1+290823: }{We model a significant soft excess ($\Delta\chi^2 = 20$) is added with kT$= 36.1^{+8.2}_{-9.6}$\,eV.}

\par{\bf NGC 7319: }{We also include an Fe K line in this fit with E fixed at 6.4\,keV with $\sigma = 0.01$\,keV.  The resultant EW $ = 319^{+313}_{-296}$\,eV.}

\subsection*{Notes to Table 3}

\par{\bf RBS 476: }{We found that the ASCA observation of RBS 476 had the wrong position in the 
ASCA catalog and that a large part of the emission was outside the observation's field of view.  Using archived {\it XMM-Newton} data, we found that a broken power law is the best fit to the data, with the indicated fluxes recorded in the table.}

\par{\bf EXO 055620-3820.2: }{We note that the authors (see Table 1 for the reference) suggest that an ionized partial covering absorber is a better physical explanation than neutral absorption.}

\par{\bf ESO 005-G004: }{In addition to the partial covering model, a thermal model is also required ({\tt apec}) at soft energies with kT $= 0.57^{+0.09}_{-0.12}$\,keV.}

\par{\bf Mrk 110: }{The spectrum is noted to be more complex in the XMM-Newton spectra analyzed by
\citet{2007AA...465...87B}.  However, for our purposes a simple power law is a good fit.}

\par{\bf NGC 3516: }{NGC 3516 has a very complex spectrum.  A very significant ($\Delta\chi^2 \approx 600$) blackbody component was also added with kT$= 52.2^{+ 1.4}_{-2.0}$\,eV and a normalization of 0.086.  This source is extensively studied, particularly because of the complex absorption present.}

\par{\bf NGC 3728: }{NGC 3783 has a complex spectrum, which has been observed with a 900\,ks Chandra observation.  A very significant ($\Delta\chi^2 = 214$) blackbody component was also added with kT$= 63.4^{+ 0.6}_{-0.7}$\,eV and a normalization of 0.008.}

\par{\bf NGC 4138: }{Alternatively, this source can be fit with a partial covering model with a high amount of scattering ($> 99$\%). Further, the blackbody component can alternatively be fit with an {\tt apec} model of similar temperature with no difference in $\chi^2$.}

\par{\bf MCG -03-34-064: }{The spectrum also has a significant ($\Delta\chi^2 = 34$) soft component, which we fit with an {\tt apec} model with kT$ = 0.82^{+0.11}_{-0.09}$\,keV.  A more complete analysis of this source is presented in \citet{2007MNRAS.375..227M}, where they model the spectra as reflection-dominated with thermal plasmas, a broad Fe K line, and additional Fe emission lines.}

\par{\bf IC 4329A: }{This source has complex absorption, studied most recently in the {\it XMM-Newton} spectrum analyzed in \citet{2005AA...432..453S}.  They find signatures from 7 different absorbing systems.}

\par{\bf NGC 5506: }{A full description of the X-ray history and properties of NGC 5506 is found in
\citet{2003AA...402..141B} where BeppoSax observations are also presented (a strong reflection component is evident).}

\par{\bf NGC 5728: }{Also included is a thermal model ({\tt mekal}) with kT $= 0.15 \pm 0.01$\,keV.  A reflection model is also used, which is a worse fit to the data.  We performed an independent analysis of the spectrum confirming these results.}

\par{\bf Mrk 841: }{This source has complex iron features.  The authors (see Table 1 for reference) get an acceptable fit when a relativistic disk line is included along with the narrow gaussian fit to Fe-K$\alpha$.  The blackbody component is also an unacceptable approximation for the soft excess.}

\par{\bf NGC 6300: }{In the analysis, the authors (see Table 1 for reference) model the soft and hard spectrum separately.  The additional parameter is the ratio of the hard to soft power law flux.}

\par{\bf Cyg A: }{For Cyg A, the authors include a Raymond-Smith model to describe the soft emission in the spectrum with $kT = 0.88^{+0.07}_{-0.11}$\,keV, $Y = 0.53^{+4.57}_{-0.19}$.}

\par{\bf 3C 452: }{3C 452 has an extremely flat spectrum which is much better fit with the addition of a reflection model \citep{2006ApJ...642...96E}.  The values quoted are from our own analysis of the data with a simplistic model.}

\acknowledgements
This research has made use of the Tartarus (Version 3.2) database, created by Paul O'Neill and Kirpal Nandra at Imperial College London, and Jane Turner at NASA/GSFC. Tartarus is supported by funding from PPARC, and NASA grants NAG5-7385 and NAG5-7067.  Also, this research has made use of the NASA/IPAC Extragalactic Database (NED) which is operated by the Jet Propulsion Laboratory, California Institute of Technology, under contract with the National Aeronautics and Space Administration.

\bibliography{ms.bib}

\clearpage

\scriptsize
\begin{landscape}
\begin{center}

\begin{longtable}{llllllllll}

\caption{X-ray Observation Details}\label{tbl-1} \\

\hline \hline \\[-2ex]

\multicolumn{1}{c}{No.} & \multicolumn{1}{c}{Source} &  \multicolumn{1}{c}{RA (h\,m\,s)} &  \multicolumn{1}{c}{Dec (\degr\,\arcmin\,\arcsec)} &
\multicolumn{1}{c}{Type\tablenotemark{1}} & \multicolumn{1}{l}{Host Galaxy\tablenotemark{1}} & 
\multicolumn{1}{l}{Obs.\tablenotemark{2}} & \multicolumn{1}{c}{Obs ID/ref} & \multicolumn{1}{l}{Ct Rate} & \multicolumn{1}{l}{Exp Time}
\\[0.5ex] \hline \\[-1.8ex]
\endfirsthead

\multicolumn{10}{c}{{\tablename} \thetable{} -- Continued} \\[0.5ex]
\hline \hline \\[-2ex]
\multicolumn{1}{c}{No.} & \multicolumn{1}{c}{Source} & \multicolumn{1}{c}{RA (h\,m\,s)} &  \multicolumn{1}{c}{Dec (\degr\,\arcmin\,\arcsec)} &
\multicolumn{1}{c}{Type\tablenotemark{1}} & \multicolumn{1}{l}{Host Galaxy\tablenotemark{1}} & 
\multicolumn{1}{l}{Obs.\tablenotemark{2}} & \multicolumn{1}{c}{Obs ID/ref} & \multicolumn{1}{l}{Ct Rate} & \multicolumn{1}{l}{Exp Time}
\\[0.5ex] \hline \\[-1.8ex]
\endhead

\multicolumn{10}{l}{{Continued on Next Page\ldots}} \\
\endfoot

\\[-1.8ex] \hline \hline
\endlastfoot

1 & NGC 235A & 00 42 52.8 & -23 32 27.6 & Sy2 & S0 pec, Int & XRT & 00035648001, 02 &  0.02196  &  21402\\
2 & Mrk 348 & 00 48 47.1 & 31 57 25.2 & Sy2 & SA(s)0/a, Int & ASCA & \citet{2000ApJ...542..175A} &  0.027  &  28000\\
3 & Mrk 352 & 00 59 53.3 & 31 49 36.8 & Sy1 & SA0 & XMM & \citet{2008ApJ...674..686W} &  7.3  &  9773\\
4 & NGC 454 & 01 14 22.7 & -55 23 55.0 & Sy2 & Pec, Int & XRT & 00035244001, 02, 03 &  0.017  &  20290\\
5 & Fairall 9 & 01 23 45.8 & -58 48 20.5 & Sy1 & S, Int? & ASCA & \citet{1997MNRAS.286..513R} &  1.001  &  22000 \\
6 & NGC 526A & 01 23 54.4 & -35 03 55.4 & Sy1.5 & S0pec?, Int & ASCA & \citet{1997ApJS..113...23T} &  0.641  &  43000\\
7 & NGC 612 & 01 33 57.7 & -36 29 35.9 & Gal & SA0+ pec, Int & XMM & \citet{2008ApJ...674..686W} &  0.12  &  9744\\
8 & ESO 297-018 & 01 38 37.2 & -40 00 41.0 & Sy2 & Sa sp, Int & Suzaku &  &   &  \\
9 & NGC 788 & 02 01 06.5 & -06 48 55.8 & Sy2 & SA(s)0/a & ASCA & 77040000 &  0.03  &  1921\\
10 & Mrk 1018 & 02 06 16.0 & 00 17 29.0 & Sy1.5 & S0, Int & XRT & 00035166001 &  0.6109  &   4256 \\
11 & LEDA 138501 & 02 09 34.3 & 52 26 33.0 & Sy1 & ? & XRT & 00035246002 &  0.365  &  6435.5 \\
12 & Mrk 590 & 02 14 33.6 & 00 46 00.1 & Sy1.2 & SA(s)a & XMM & \citet{2007AA...470...73L} &    &  71000\\
13 & 2MASX J02162987+5126246 & 02 16 29.8 & 51 26 24.7 & ? & -- & XMM & \citet{2008ApJ...674..686W} &  1.62  &  8921\\
14 & $[$HB89$]$ 0212+735 & 02 17 30.8 & 73 49 32.5 & BL Lac & ? & XRT & 00035241002 &  0.08307  &   6485.4 \\
15 & NGC 931 & 02 28 14.5 & 31 18 42.1 & Sy1.5 & Sbc, Int & ASCA & \citet{1995MNRAS.276.1311R} &  0.133  &  18900 \\
16 & NGC 985 & 02 34 37.8 & -08 47 15.4 & Sy1 & SBbc? p (Ring) & Chandra & \citet{2005ApJ...620..165K} &  0.18  &  77000 \\
17 & ESO 416-G002 & 02 35 13.4 & -29 36 16.9 & Sy1.9 & Sa & XMM & \citet{2006MNRAS.365..688G} &    &  4269 \\
18 & ESO 198-024 & 02 38 19.7 & -52 11 32.3 & Sy1 & N, Int & XMM & \citet{2004AA...413..913P} &   &  22900\\
19 & QSO B0241+622 & 02 44 57.7 & 62 28 06.6 & Sy1 & ? & XRT & 00036305001 &   0.4485  &   10618 \\
20 & NGC 1142 & 02 55 12.2 & 00 11 01.0 & Sy2 & S pec (Ring B), Int & XMM & \citet{2008ApJ...674..686W} &  0.28  &  8921 \\
21 & 2MASX J03181899+6829322 & 03 18 19.0 & 68 29 31.6 & Sy1.9 & ? & XMM & \citet{2008ApJ...674..686W} & 1.47  &  6578 \\
22 & NGC 1275 & 03 19 48.2 & 41 30 42.1 & Sy2 & cD;pec &  &  &   &   \\
23 &  PKS 0326-288 & 03 28 36.5 & -28 41 48.5 & Sy1.9 & ? & none &  &   &    \\
24 & NGC 1365 & 03 33 36.4 & -36 08 25.4 & Sy1.8 & (R')SBb(s)b, Int & XRT & 00035458001, 02 &  0.027  &  19434 \\
25 & ESO 548-G081 & 03 42 03.7 & -21 14 39.8 & Sy1 & SB(rs)a pec?, Int? & XMM & \citet{2008ApJ...674..686W} & 11.95  &  8924 \\
26 & RBS 476 & 03 49 23.2 & -11 59 26.9 & BL Lac & -- & ASCA & \citet{2000MNRAS.316..234R} &   &  109000\\
27 & PGC 13946 & 03 50 23.8 & -50 18 35.6 & ? & ?, Int & XRT & 00035251003 &  0.02  &  8808 \\
28 & 2MASX J03565655-4041453 & 03 56 56.5 & -40 41 45.6 & Sy1.9 & ? & XRT & 00035600001, 02 &  0.0628  &  12579 \\
29 & 3C 105 & 04 07 16.5 & 03 42 25.6 & Sy2 & ? & XRT & 00035625001, 02, 03, 04 &  0.007  &  22398\\
30 & 3C 111 & 04 18 21.3 & 38 01 35.8 & Sy1 & N & ASCA & \citet{1998MNRAS.299..410R} &  0.702  &  32800 \\
31 & 1H 0419-577 & 04 26 00.8 & -57 12 00.4 & Sy1 & ? & ASCA & \citet{1999ApJ...510..178T} &   &  24000 \\
32 & 3C 120 & 04 33 11.1 & 05 21 15.5 & Sy1 & S0, Int? & ASCA & \citet{1997ApJ...487..636G} &  1.861  &  50000 \\
33 & 2MASX J04440903+2813003 & 04 44 09.0 & 28 13 00.5 & Sy2 & S & XRT & 00035175005, 06, 07 &  0.114  &  76404 \\
34 & MCG -01-13-025 & 04 51 41.5 & -03 48 33.8 & Sy1.2 & SAB(s)0+ pec & XMM & \citet{2006MNRAS.365..688G} &   &  2363\\
35 & 1RXS J045205.0+493248 & 04 52 05.0 & 49 32 45.2 & Sy1 & ? & XRT & 00035281002 &  0.95  &  2200 \\
36 & XSS J05054-2348 & 05 05 45.7 & -23 51 14.0 & Sy2 & ? & XRT & 00035206004 &  0.12  &  3244 \\
37 &  4U 0517+17 & 05 10 45.5 & 16 29 55.7 & Sy1.5 & ? & XMM & not public &   &   \\ 
38 & Ark 120 & 05 16 11.4 & 00 08 59.3 & Sy1 & Sb/pec, Int & ASCA & 72000000 &  1.248  &  47850 \\
39 & ESO 362-G018 & 05 19 35.8 & -32 39 28.1 & Sy1.5 & S0/a, Int & XMM & \citet{2008ApJ...674..686W} &  0.95  &  8921\\
40 & Pictor A & 05 19 49.7 & -45 46 44.4 & Sy1 & (R')SA0:pec & ASCA & \citet{1998ApJ...505..577E} &  0.56  &  61877\\
41 & ESO 362-G021 & 05 22 58.0 & -36 27 31.0 & BL Lac & N & ASCA & 73055010 &  0.40  &  5133 \\
42 & PKS 0537-441 & 05 38 50.4 & -44 05 08.9 & BL Lac & ? & XRT & 00050150011 &  0.317  &  22570 \\
43 & $[$HB89$]$ 0537-286 & 05 39 54.3 & -28 39 55.8 & Blazar & ? & ASCA & \citet{1996AA...307....8S} &  0.04  &  29000 \\
44 & PKS 0548-322 & 05 50 40.8 & -32 16 17.8 & BL Lac & ?, Int & ASCA & \citet{1998ApJ...502..630S} &  1.70  &  31533 \\
45 & NGC 2110 & 05 52 11.4 & -07 27 22.3 & Sy2 & SAB0- & ASCA & \citet{1997ApJS..113...23T} &  0.319  &  36300 \\
46 & MCG +08-11-011 & 05 54 53.6 & 46 26 21.5 & Sy1.5 & SB0 & ASCA & \citet{1998ApJ...498..220G} &  0.554 &  10000 \\
47 & EXO 055620-3820.2 & 05 58 02.0 & -38 20 04.6 & Sy1 & ?, Int & ASCA & \citet{1996ApJ...463..134T} &   &  45000\\
48 & IRAS 05589+2828 & 06 02 10.7 & 28 28 22.1 & Sy1 & ? & XRT & 00035255001 &  0.35  &  6446\\
49 & ESO 005-G004 & 06 05 41.6 & -86 37 54.8 & Sy2 & Sb & Suzaku &  &   &  \\
50 & Mrk 3 & 06 15 36.4 & 71 02 15.0 & Sy2 & S0 & ASCA & \citet{1997ApJS..113...23T} &  0.044  &  27300 \\
51 & ESO 121-G028 & 06 23 45.6 & -60 58 44.4 & Sy2 & SB(l)0/a?, Int & XRT & 00036296007 &  0.02  &  8716 \\
52 & ESO 490-G026 & 06 40 11.7 & -25 53 43.4 & Sy1.2 & Pec, Int & XMM & \citet{2008ApJ...674..686W} &  5.68 &  9192 \\
53 & 2MASX J06403799-4321211 & 06 40 38.0 & -43 21 20.9 & ? & ?, Int & XRT & 00035601001, 02 &  0.01  &  18897 \\
54 & 2MASX J06411806+3249313 & 06 41 18.0 & 32 49 31.4 & Sy2 & ?, Int & XMM & \citet{2008ApJ...674..686W} &  0.25 &  10696\\
55 & Mrk 6 & 06 52 12.2 & 74 25 37.6 & Sy1.5 & SAB0+ & XRT & 00035461003 &  0.114  &  5367 \\
56 & Mrk 79 & 07 42 32.8 & 49 48 34.9 & Sy1.2 & SBb & XMM & \citet{2005MNRAS.363...64G} &  12.57  &  1680 \\
57 & SDSS J074625.87+254902.2 & 07 46 25.9 & 25 49 02.3 & Blazar & ? & XRT & 00035422002 &  0.143  &  13365 \\
58 & IGR J07597-3842 & 07 59 41.0 & -38 45 36.0 & Sy1.2 & ? & XRT & 00035223004 &  0.6055  &  7534 \\
59 & 4C +71.07 & 08 41 24.4 & 70 53 42.4 & Blazar & ? & ASCA & \citet{1997ApJ...478..492C} &  0.30  &  10540 \\ 
60 & Mrk 18 & 09 01 58.4 & 60 09 06.1 & Sy2 & S?, Int & XMM & \citet{2008ApJ...674..686W} &  0.50  &  9910 \\
61 & 2MASX J0904699+5536025 & 09 04 36.9 & 55 36 02.5 & Sy1 & ?, Int & XMM & \citet{2008ApJ...674..686W} &  1.52  &  7142 \\
62 & 2MASX J09112999+4528060 & 09 11 30.0 & 45 28 05.9 & Sy2 & ? & XMM & \citet{2008ApJ...674..686W} &  0.05 (mos)  &  11530 \\
63 & IRAS  09149-6206 & 09 16 08.9 & -62 19 29.6 & Sy1 & ? & XRT & 00035233001 &  0.21  &  7399 \\
64 & 2MASX J09180027+0425066 & 09 18 00.3 & 04 25 06.2 & QSO2 & ?, Int? & Suzaku & 702076010 &  0.029 (XIS0+3)  &  4989 \\
65 & MCG -01-24-012 & 09 20 46.2 & -08 03 22.0 & Sy2 & SAB(rs)c, Int & XRT & 00035262001 &  0.08  &  8522 \\
66 & MCG +04-22-042 & 09 23 43.0 & 22 54 32.4 & Sy1.2 & E? & XMM & \citet{2008ApJ...674..686W} &  13.99  &  9012 \\
67 & Mrk 110 & 09 25 12.9 & 52 17 10.7 & Sy1 & S?, Int & ASCA & 73091000 &  1.15  &  20747 \\
68 & NGC 2992 & 09 45 42.0 & -14 19 35.0 & Sy2 & Sa pec, Int & ASCA & \citet{1997ApJS..113...23T} &  0.039  &  27300 \\
69 & MCG -05-23-016 & 09 47 40.2 & -30 56 56.0 & Sy2 & (RL)SA(l) & ASCA & \citet{1997ApJS..113...23T} &  1.583  &  34000 \\
70 & NGC 3081 & 09 59 29.5 & -22 49 34.7 & Sy2 & SAB(r)0/a & ASCA & 74043000 &  0.03  &  34738 \\
71 & NGC 3227 & 10 23 30.6 & 19 51 54.0 & Sy1.5 & SAB(s) pec, Int & ASCA & 73068000 &  0.42  &  34312 \\
72 & NGC 3281 & 10 31 52.1 & -34 51 13.3 & Sy2 & SAB(rs+)a & ASCA & 74058000 &  0.015  &  16959 \\
73 & 2MASX J10384520-4946531 & 10 38 44.5 & -49 46 57.4 & Sy1 & ? & XRT & 00035225003 &  0.087  &  16907 \\
74 & LEDA 093974 & 10 40 22.5 & -46 25 25.7 & Sy2 & S0 & XRT & 00035345003 &  0.058  &  14377 \\
75 & Mrk 417 & 10 49 30.9 & 22 57 51.8 & Sy2 & Sa & XMM & \citet{2008ApJ...674..686W} &  0.145  &  7437 \\
76 & Mrk 421 & 11 04 27.3 & 38 12 31.7 & BL Lac & ? & ASCA & 70008000 &  2.20  &  32769 \\
77 & NGC 3516 & 11 06 47.5 & 72 34 07.0 & Sy1.5 & (R)SB(s) & ASCA & 71007000 &  2.38  &  29842 \\
78 & 1RXS J112716.6+190914 & 11 27 16.3 & 19 09 20.2 & Sy1.8 & ? & XRT & 00037088002 &  0.11  &  10062 \\
79 & NGC 3783 & 11 39 01.7 & -37 44 19.0 & Sy1 & (R')SB(r)a & ASCA & 74054010 &  1.71  &  18716 \\
80 & SBS 1136+594 & 11 39 09.0 & 59 11 54.6 & Sy1.5 & ? & XRT & 00035265001 &  0.34  &  8991 \\
81 & UGC 06728 & 11 45 16.0 & 79 40 53.4 & Sy1.2 & SB0/a, Int? & XMM & \citet{2008ApJ...674..686W} &  4.25  &  7220 \\
82 & 2MASX J11454045-1827149 & 11 45 40.5 & -18 27 15.5 & Sy1 & Irr, Int & XRT & 00035645001 &  0.72  &  9402 \\
83 & CGCG 041-020 & 12 00 57.9 & 06 48 23.0 & Sy2 & S? & XMM & \citet{2008ApJ...674..686W} &  0.465  &  9777 \\
84 &  IGR J12026-5349 & 12 02 47.6 & -53 50 07.8 & Sy2 & SB0 pec, Int & chandra not public &  &   &  \\
85 & NGC 4051 & 12 03 09.6 & 44 31 52.7 & Sy1.5 & SAB(rs)bc & ASCA & 70001000 &  0.917  &  25970 \\
86 & Ark 347 & 12 04 29.7 & 20 18 58.3 & Sy2 & S0: pec & XRT & 00035599001, 02 &  0.013  &  21706 \\
87 &  NGC 4102 & 12 06 23.1 & 52 42 39.2 & LINER & SAB(s)b? & Chandra & too short? &   &  \\
88 & NGC 4138 & 12 09 29.8 & 43 41 07.1 & Sy1.9 & SA(r)0+ & XMM & \citet{2006AA...446..459C} &   &  8856 \\
89 & NGC 4151 & 12 10 32.6 & 39 24 20.5 & Sy1.5 & (R')SAB(rs)ab & ASCA & 78001003 &  1.255  &  46006 \\
90 & Mrk 766 & 12 18 26.5 & 29 48 46.4 & Sy1.5 & (R')SB(s)a & ASCA & 71046000 &  0.819  &  32970 \\
91 & NGC 4388 & 12 25 46.8 & 12 39 43.6 & Sy2 & SA(s)b & ASCA & 73073000 &  0.053  &  26040 \\ 
92 & NGC 4395 & 12 25 48.9 & 33 32 48.5 & Sy1.9 & SA(s)m, Int & ASCA & 78009000 &  0.079  &  36084 \\
93 & 3C 273 & 12 29 06.7 & 02 03 08.6 & Blazar & ? & ASCA & 70023000 &  3.849  &  31964 \\
94 & NGC 4507 & 12 35 36.6 & -39 54 33.5 & Sy2 & SAB(s)ab & ASCA & \citet{1997ApJS..113...23T} &  0.1421  &  24800 \\
95 & ESO 506-G027 & 12 38 54.6 & -27 18 28.1 & Sy2 & S pec sp & XMM & \citet{2008ApJ...674..686W} &  0.27  &  8919 \\
96 & XSS J12389-1614 & 12 39 06.3 & -16 10 47.6 & Sy2 & ? & XRT & 00035208002 &  0.103  &  6622 \\
97 & NGC 4593 & 12 39 39.4 & -05 20 39.1 & Sy1 & (R)SB(rs)b & ASCA & 71024000 &  1.478  &  21185 \\
98 & WKK 1263 & 12 41 25.7 & -57 50 03.5 & Sy2? & ? & XMM & \citet{2008ApJ...674..686W} &  2.72  &  8902 \\
99 & 3C 279 & 12 56 11.2 & -05 47 21.5 & Blazar & ? & ASCA & 70026000 &  0.299  &  37195 \\
100 & SBS 1301+540 & 13 03 59.5 & 53 47 30.1 & Sy1 & ? & XMM & \citet{2008ApJ...674..686W} &  10.30  &  8408 \\
101 & NGC 4945 & 13 05 27.5 & -49 28 05.5 & Sy2 & SB(s)cd & ASCA & \citet{1997ApJS..113...23T} &  0.045  &  6100 \\
102 & NGC 4992 & 13 09 13.0 & 11 38 45.2 & Gal & Sa & XMM & \citet{2008ApJ...674..686W} &  0.137  &  12849 \\
103 & MCG -03-34-064 & 13 22 24.5 & -16 43 43.0 & Sy1.8 & SB?, Int & ASCA & 73029000 &  0.044  &  36174 \\
104 &  Cen A & 13 25 27.6 & -43 01 09.1 & Sy2 & S0 pec, Int &  &  &   &  \\
105 & MCG -06-30-015 & 13 35 53.8 & -34 17 44.2 & Sy1.2 & E-S0 & ASCA & 70016010 &  1.18  &  30026 \\
106 & NGC 5252 & 13 38 16.0 & 04 32 33.4 & Sy1.9 & S0 & ASCA & \citet{1997ApJS..113...23T} &  0.063  &  27300\\
107 & 4U 1344-60 & 13 47 24.0 & -60 38 24.0 & Sy1.5 & S & XRT & 00035284002 &  0.388  &  14065 \\
108 & IC 4329A & 13 49 19.3 & -30 18 34.6 & Sy1.2 & SA0+, Int & ASCA & 70005000 &  2.025  &  27212 \\
109 & Mrk 279 & 13 53 03.5 & 69 18 29.5 & Sy1.5 & S0, Int & ASCA & 72028000 &  1.342  &  21357 \\
110 & NGC 5506 & 14 13 14.9 & -03 12 27.0 & Sy1.9 & Sa pec, Int? & ASCA & 75033000 &  0.9316  &  39272 \\ 
111 & RBS 1366 & 14 17 56.7 & 25 43 26.4 & BL Lac & ? & XRT & 00035270001 &  0.623  &  7930 \\
112 & NGC 5548 & 14 17 59.5 & 25 08 12.5 & Sy1.5 & (R')SA(s)0/a & ASCA & 76029000 &  2.633  &  21200 \\
113 & ESO 511-G030 & 14 19 22.4 & -26 38 40.9 & Sy1 & SA(rs)c pec, Int & ASCA & 76067000 &  0.467  &  16358 \\
114 & RBS 1399 & 14 28 32.7 & 42 40 20.6 & BL Lac & ? & ASCA & 76053000 &  1.44  &  36000 \\
115 & NGC 5728 & 14 42 23.9 & -17 15 11.5 & Sy2 & (R)SAB(r)a & Chandra & \citet{2006AA...450..933Z} &   &  18700 \\
116 & Mrk 841 & 15 04 01.2 & 10 26 16.1 & Sy1 & E & XMM & \citet{2007AA...470..889P} &  17.9  &  5861 \\
117 & Mrk 290 & 15 35 52.4 & 57 54 09.4 & Sy1 & E1?, Int? & ASCA & 72027000 &  0.308  &  41018 \\
118 & Mrk 1498 & 16 28 04.1 & 51 46 31.4 & Sy1.9 & S? & XRT & 00035271002 &  0.0613  &  19948 \\
119 & 2MASX J16481523-3035037 & 16 48 15.2 & -30 35 04.2 & Sy1 & ? & XRT & 00035348002 &  0.244  &  9203 \\
120 &  NGC 6240 & 16 52 58.9 & 02 24 02.9 & Sy & I0: pec, Int &  &  &   &  \\
121 & Mrk 501 & 16 53 52.2 & 39 45 36.7 & BL Lac & E? & ASCA & 76052000 &  5.55  &  5904\\
122 & NGC 6300 & 17 16 59.5 & -62 49 14.2 & Sy2 & SB(rs)b & XMM & \citet{2004ApJ...617..930M} &   &  \\
123 &  GRS 1734-292 & 17 37 24.3 & -29 10 48.0 & Sy1 & ? &  &  &   &  \\
124 & 1RXS J174538.1+290823 & 17 45 38.3 & 29 08 22.2 & Sy1 & ? & XRT & 00035273001 &  0.237  &  9344 \\
125 & 3C 382 & 18 35 02.2 & 32 41 50.3 & Sy1 & ? & ASCA & \citet{1999ApJ...526...60S} &  1.793  &  40500 \\
126 & ESO 103-035 & 18 38 20.3 & -65 25 39.4 & Sy2 & SA0, Int & ASCA & \citet{1999ApJ...523..521F} &  17.31  &  15700 \\
127 & 3C 390.3 & 18 42 09.0 & 79 46 17.0 & Sy1 & ? & ASCA & \citet{1999ApJ...526...60S} &  0.528  &  19200 \\
128 & NVSS 193013+341047 & 19 30 13.3 & 34 10 46.9 & Sy1 & ? & XRT & 00035274001, 02 &  0.013  &  10247 \\
129 & NGC 6814 & 19 42 40.7 & -10 19 24.6 & Sy1.5 & SAB(rs)bc & ASCA & 70012000 &  0.032  &  39082 \\
130 & 3C 403 & 19 52 15.8 & 02 30 24.5 & Sy2 & S0 & Chandra & \citet{2005ApJ...622..149K} &   &  45900 \\
131 & Cyg A & 19 59 28.3 & 40 44 02.0 & Sy2 & Sc, Int & ASCA & \citet{1999ApJ...526...60S} &  0.911  &  21100 \\
132 & 2MASX J19595975+6508547 & 19 59 59.9 & 65 08 54.6 & BL Lac & E & XRT & 00035025006 &  3.50  &  479.5 \\
133 & NGC 6860 & 20 08 46.9 & -61 06 00.7 & Sy1.5 & (R')SB(r)ab & XMM & \citet{2008ApJ...674..686W} &  0.46  &  11815 (mos) \\
134 & NGC 6921 & 20 28 28.9 & 25 43 24.3 & Sy2 & SA(r)0/a, Int & XMM & \citet{2008ApJ...674..686W} &  0.255  &  8789 \\
135 & MCG +04-48-002 & 20 28 35.1 & 25 44 01.0 & Sy2 & Scd, Int & XMM & \citet{2008ApJ...674..686W} &  0.096  &  8789 \\ 
136 & 4C +74.26 & 20 42 37.3 & 75 08 02.4 & Sy1 & ? & ASCA & 74097000 &  0.615  &  21210 \\
137 & Mrk 509 & 20 44 09.7 & -10 43 24.6 & Sy1.2 & S? & ASCA & 71013000 &  1.96  &  40071 \\
138 & IC 5063 & 20 52 02.3 & -57 04 07.7 & Sy2 & SA(s)0+ & ASCA & \citet{1997ApJS..113...23T} &  0.096  &  23900 \\ 
139 & 2MASX J21140128+8204483 & 21 14 01.2 & 82 04 48.4 & Sy1 & ? & XRT & 00035624002 &  0.448  &  5179 \\
140 & IGR J21247+5058 & 21 24 38.1 & 50 58 58.1 & Sy1 & ? & XRT & 00035626002 &  0.708  &  6690 \\
141 & IGR 21277+5656 & 21 27 45.9 & 56 56 34.4 & Sy1 & ? & XRT & 00035215003 &  0.426  &  8372 \\
142 & RX J2135.9+4728 & 21 35 55.0 & 47 28 23.2 & Sy1 & ? & XRT & 00035628003 &  0.138  &  9115 \\
143 & PKS 2149-306 & 21 51 55.5 & -30 27 54.0 & Blazar & ? & ASCA & \citet{1996AA...307....8S} &  0.24  &  19000 \\
144 & UGC 11871 & 22 00 41.4 & 10 33 08.6 & Sy1.9 & Sb, Int & XRT & 00035278003 &  0.078  &  8928 \\
145 & NGC 7172 & 22 02 01.9 & -31 52 11.3 & Sy2 & Sa pec sp & ASCA & \citet{1997ApJS..113...23T} &  0.0066  &  14900 \\
146 & NGC 7213 & 22 09 16.2 & -47 10 00.1 & Sy1.5 & Sa & ASCA & 71026000 &  1.226  &  27091 \\
147 & NGC 7314 & 22 35 46.2 & -26 03 00.7 & Sy1.9 & SAB(rs)bc, Int & ASCA & \citet{1997ApJS..113...23T} &  0.9284  &  45300 \\
148 & NGC 7319 & 22 36 03.6 & 33 58 32.5 & Sy2 & SB(s)bc pec, Int & XRT & 00035083004 &  0.012  &  14731 \\
149 & 3C 452 & 22 45 48.8 & 39 41 15.7 & Sy2 & S? & Chandra &  &   &  \\ 
150 & 3C 454.3 & 22 53 57.7 & 16 08 53.5 & Blazar & ? & XRT & 00035030001 &  1.151  &  13668 \\
151 & MR 2251-178 & 22 54 05.8 & -17 34 54.8 & Sy1 & ? & ASCA & 74028000 &  0.649  &  17760 \\
152 & NGC 7469 & 23 03 15.6 & 08 52 26.4 & Sy1.2 & (R')SAB(rs)a, Int & ASCA & 71028010 &  1.593  &  18707 \\ 
153 & Mrk 926 & 23 04 43.5 & -08 41 08.5 & Sy1.5 & Sbc, Int & ASCA & 75049000 &  1.218  &  37221 \\
154 & NGC 7582 & 23 18 23.5 & -42 22 14.2 & Sy2 & (R')SB(s)ab, Int & ASCA & \citet{1997ApJS..113...23T} &  0.0906  &  17800 \\

\end{longtable}
\end{center}
\clearpage
\noindent$^1${AGN type is from \citet{2007arXiv0711.4130T}.  For AGN types, optical identifications
are listed, where available.  Where ``Gal'' is indicated, \\there are no optical emission lines indicative
of the presence of an AGN.  The optical spectrum looks like a galaxy spectrum.  Additional host \\
galaxy classifications were obtained from the NED and LEDA databases.  Where ``?'' is indicated, there is no available classification.\\}
$^2${Observations were collected from archival or published work from ASCA, Chandra, XMM-Newton, Suzaku, and the SWIFT XRT instrument.\\}

\end{landscape}
\normalsize

\clearpage
\begin{deluxetable}{llllllll}
\tabletypesize{\scriptsize}
\tablecaption{X-ray Fits of SWIFT XRT spectra\label{tbl-2}}
\tablewidth{0pt}
\tablehead{
\colhead{Source}  & \colhead{N$_{H(Gal)}$\tablenotemark{1}} &
\colhead{$N_{H(abs)}$} & \colhead{Cvr. Frac} & \colhead{$\Gamma$} & \colhead{$\chi^2$/dof} & \colhead{F$_X$\tablenotemark{2}} & \colhead{type}
}
\startdata
NGC 235A &  0.014 & $28.25^{+3.79}_{-3.37}$ & $0.9975^{+0.0021}_{-0.0029}$& $2.85^{+0.40}_{-0.66}$ & 20.6/21 & 0.06, 2.92 & C\\
NGC 454 & 0.026 & $15.90^{+4.91}_{-3.64}$ &  $0.997^{+0.003}_{-0.004}$ &  $2.81^{+1.33}_{-0.88}$ & 19.7/15 & 0.023, 1.964 & C\\
Mrk 1018 & 0.0258 & $0.00^{+0.01}_{-0.00}$ &  -- & $1.93^{+0.07}_{-0.05}$ &  99.5/99 & 8.73, 11.89 & S\\
LEDA 138501 & 0.169 & $0.030^{+0.040}_{-0.030}$ &  -- & $1.812^{+0.094}_{-0.089}$ &  87.6/100 & 5.08, 11.38 & S\\
$[$HB89$]$ 0212+735 & 0.259 & $0.350^{+0.220}_{-0.200}$ & -- & $1.376^{+0.259}_{-0.245}$ & 30.6/23 & 0.686, 5.54 & S\\
QSO B0241+622 & 0.742 & 0.100$^{+0.070}_{-0.060}$ & -- & $1.733^{+0.083}_{-0.080}$ &  197.8/185 & 3.89, 25.82 & S\\
NGC 1365 & 0.0142 & 104.8 & 0.74$^{+0.25}_{-0.15}$ & $1.71^{+0.29}_{-0.23}$ &  55.0/43 & 0.40, 0.88 & C\\
PGC 13946 & 0.0156 & $14.4^{+7.4}_{-5.9}$ & $0.94^{+0.05}_{-0.19}$& $1.35^{+1.14}_{-0.70}$ & 9/6 & 0.08, 2.96 & C\\
2MASX J03565655-4041453 & 0.026 & $3.27^{+0.92}_{-0.61}$ & -- & $1.94^{+0.42}_{-0.30}$ & 29.4/36 & 0.19, 5.58 & S\\
2MASX J04440903+2813003 & 0.197 & $3.39^{+0.31}_{-0.25}$ & 0.98996$^{+0.003}_{-0.003}$ & $1.37^{+0.11}_{-0.08}$ & 329.8/331 & 0.02, 12.34 & C\\
1RXS J045205.0+493248 & 0.531 & $0.001^{+0.22}_{-0.001}$ & -- & $1.95^{+0.16}_{-0.11}$ & 110.1/86 & 11.4, 37.8 & S\\
XSS J05054-2348 & 0.0228 &  $6.57^{+3.03}_{-2.39}$ & -- & $1.51^{+0.56}_{-0.51}$ & 21.9/16 & 0.15, 13.4 & S\\
PKS 0537-441 & 0.0394 &  -- & -- & $1.78^{+0.03}_{-0.03}$ & 274.8/237 & 4.41, 7.57 & S\\
IRAS 05589+2828 & 0.462 & $0.00^{+0.04}$ & -- & $1.60^{+0.06}_{-0.06}$ & 106.4/98 & 3.95, 16.42 & S\\
ESO 121-G028 & 0.0469 & $16.19^{+12.6}_{-9.4}$ & -- & $2.20^{+1.55}_{-1.32}$ & 4/6 & 0, 2.97 & S*\\
2MASX J06403799-4321211 & 0.0624 & $16.13^{+10.0}_{-12.3}$ & -- & $1.79^{+1.41}_{-1.51}$ & 6.6/7 & 0.0006, 1.58 & S*\\
Mrk 6 & 0.0623 & $3.26^{+1.33}_{-1.19}$ & 0.909$^{+0.041}_{-0.075}$ & $1.31^{+0.40}_{-0.37}$ & 31.6/26 & 0.44, 11.36 & C\\
SDSS J074625.87+254902.2  & 0.04 & $0.062^{+0.04}_{-0.03}$ & -- & $1.29^{+0.09}_{-0.08}$ & 95.8/84 & 1.63, 6.63 & S\\
IGR J07597-3842 & 0.62 & $0.0^{+0.03}$ & -- & $1.88^{+0.06}_{-0.05}$ & 185/181 & 6.47, 27.81 & S\\
IRAS  09149-6206 & 0.189 & $0.85^{+0.26}_{-0.17}$ & -- & $1.74^{+0.20}_{-0.14}$ & 75.7/68 & 1.81, 11.59 & S\\
MCG -01-24-012 & 0.0355 & $11.44^{+2.82}_{-2.27}$ & -- & $2.17^{+0.50}_{-0.44}$ & 26.8/29 & 0.04, 8.28 & S*\\
2MASX J10384520-4946531 & 0.262 & $1.55^{+0.56}_{-0.58}$ & 0.880$^{+0.064}_{-0.063}$ & $1.44^{+0.18}_{-0.20}$ & 77.8/66 & 0.60, 6.17 & C\\
LEDA 093974& 0.141 & $4.24^{+1.06}_{-1.00}$ & 0.986$^{+0.007}_{-0.014}$ & $2.12^{+0.43}_{-0.40}$ & 57.7/35 & 0.16, 4.95 & C\\
1RXS J112716.6+190914 & 0.017 & $0.25^{+0.10}_{-0.08}$ & -- & $1.79^{+0.15}_{-0.15}$ & 58.4/48 & 1.38, 3.77 & S\\
SBS 1136+594 & 0.0112 & -- & -- & $1.94^{+0.05}_{-0.04}$ & 132.6/115 & 4.76, 6.15 & S\\
2MASX J11454045-1827149 &0.0349 & -- & -- & $1.92^{+0.03}_{-0.03}$ & 212.7/214 & 10.37, 14.33 & S\\
Ark 347 & 0.0236 & $30.0^{+8.76}_{-8.88}$ & 0.995$^{+0.005}_{-0.017}$ & $2.17^{+1.03}_{-0.80}$ & 6.3/13 & 0.03, 1.80 & C\\
XSS J12389-1614 & 0.0369 & $3.21^{+0.83}_{-0.71}$ & -- & $1.84^{+0.37}_{-0.35}$ & 19.2/30 & 0.30, 9.57 & S\\
4U 1344-60 & 1.05 & $1.45^{+0.20}_{-0.19}$ & -- & $1.69^{+0.09}_{-0.09}$ & 231.7/216 & 2.09, 29.20 & S\\
RBS 1366 & 0.0167 & $0.05^{+0.01}_{-0.01}$ & -- & $2.08^{+0.06}_{-0.06}$ & 195.1/176 & 9.26, 11.03 & S\\
Mrk 1498 & 0.021 & $17.84^{+2.37}_{-1.82}$ & 0.990$^{+0.005}_{-0.006}$ & $1.92^{+0.29}_{-0.22}$ & 55.8/55 & 0.11, 7.20 & C\\
2MASX J16481523-3035037 & 0.173 & $0.24^{+0.06}_{-0.06}$ & -- & $1.72^{+0.10}_{-0.10}$ & 106.3/96 & 2.93, 9.95 & S\\
1RXS J174538.1+290823 & 0.137 & $0.02^{+0.07}_{-0.02}$ & -- & $1.85^{+0.07}_{-0.07}$ & 84.8/89 & 3.25, 6.20 & S\\
NVSS 193013+341047 & 0.173 & $27.47^{+20.37}_{-13.74}$ & 0.926$^{+0.065}_{-0.426}$ &$0.958^{+1.56}_{-1.54}$ &   0.74/3 & 0.03, 2.22 & C\\
2MASX J19595975+6508547 & 0.103 & $0.07^{+0.04}_{-0.04}$ & -- & $1.93^{+0.11}_{-0.10}$ & 77.9/71 & 50.19, 91.75 & S\\
2MASX J21140128+8204483 & 0.0736 & $0.047^{+0.023}_{-0.021}$ & -- & $1.89^{+0.09}_{-0.09}$ & 99.9/96 & 6.27, 11.92 & S\\
IGR J21247+5058 & 1.17 & $1.30^{+0.23}_{-0.21}$ & -- & $1.33^{+0.09}_{-0.09}$ & 202.5/194 & 3.01, 63.18 & S\\
IGR 21277+5656 & 0.806 & $0.424^{+0.11}_{-0.11}$ & -- & $2.02^{+0.10}_{-0.10}$ & 141.9/144 & 4.29, 20.25 & S\\
RX J2135.9+4728 & 0.385 & $0.44^{+0.17}_{-0.15}$ & -- & $1.66^{+0.16}_{-0.15}$ & 52.1/56 & 1.37, 7.37 & S\\
UGC 11871 & 0.0544 & $2.39^{+0.71}_{-0.56}$ & -- & $1.53^{+0.27}_{-0.25}$ & 39.1/31 & 0.39, 6.35 & S*\\
NGC 7319 & 0.083 & $86.48^{+30.87}_{-20.55}$ & 0.9965$^{+0.0027}_{-0.0055}$ &$2.87^{+0.53}_{-0.28}$ &   1.6/4 & 0.13, 1.49 & C\\
3C 454.3 & 0.0645 & $0.08^{+0.01}_{-0.01}$ & -- & $1.49^{+0.03}_{-0.03}$ & 446.7/407 & 14.31, 46.33 & S\\
\enddata

\tablenotetext{1}{Milky Way Galactic absorption obtained from the $N_H$ FTOOL on HEASARC in units of $10^{22}$\,cm$^{-2}$.\\}
\tablenotetext{2}{Observed flux in the 0.5--2\,keV and 2--10\,keV bands in units of $10^{-12}$\,\flux.  The statistical errors on fluxes are very small and the systematic errors are dominated by model uncertainties.\\}

\end{deluxetable}

\clearpage
\begin{landscape}
\scriptsize
\begin{center}
\begin{longtable}{llllllllll}

\caption{X-ray Spectral Fits from the Literature/Suzaku XIS/{\it TARTARUS} Fits}\label{tbl-3}\\

\hline \hline \\[-2ex]
\multicolumn{1}{c}{Source} &
\multicolumn{1}{c}{n$_{H(abs)}$} & \multicolumn{1}{c}{$\Gamma$} & \multicolumn{1}{c}{par$_{model}$\tablenotemark{3}} & \multicolumn{1}{c}{Fe K E} & \multicolumn{1}{c}{Fe K EW} & \multicolumn{1}{c}{$\chi^2$/dof} & \multicolumn{1}{c}{F$_X$\tablenotemark{2}}& \multicolumn{1}{c}{model} & \multicolumn{1}{c}{type}
\\[0.5ex] \hline \\[-1.8ex]
\endfirsthead

\multicolumn{10}{c}{{\tablename} \thetable{} -- Continued} \\[0.5ex]
\hline \hline \\[-2ex]
\multicolumn{1}{c}{Source} &
\multicolumn{1}{c}{n$_{H(abs)}$} & \multicolumn{1}{c}{$\Gamma$} & \multicolumn{1}{c}{par$_{model}$\tablenotemark{3}} & \multicolumn{1}{c}{Fe K E} & \multicolumn{1}{c}{Fe K EW} & \multicolumn{1}{c}{$\chi^2$/dof} & \multicolumn{1}{c}{F$_X$\tablenotemark{2}}& \multicolumn{1}{c}{model} & \multicolumn{1}{c}{type}
\\[0.5ex] \hline \\[-1.8ex]
\endhead

\multicolumn{10}{l}{{Continued on Next Page\ldots}} \\
\endfoot

\\[-1.8ex] \hline \hline
\endlastfoot

Mrk 348 & $16^{+4}_{-3}$ & $1.69^{+0.36}_{-0.39}$ & -- & $6.27^{+0.08}_{-0.07}$	 &$215^{+75}_{-75}$ & 117.4/116	& 0.07, 4.84 & double pow & C\\
Fairall 9 & -- & $1.91^{+0.01}_{-0.01}$ & -- & $6.37^{+0.11}_{-0.13}$ & $350^{+150}_{-100}$ & 1283/1149 & 18.2\tablenotemark{4}, 19.0 & pow & S\\
NGC 526A & $1.50^{+0.14}_{-0.14}$ & $1.82^{+0.16}_{-0.14}$ & 32.5 & 6.4 & $111^{+33}_{-56}$ & 1676/1586 & 3.5\tablenotemark{4}, 35.4 & double pow & C\\
ESO 297-018 & $41.71^{+4.70}_{-2.90}$ & $1.28^{+0.14}_{-0.16}$ & $0.993^{+0.004}_{-0.006}$ & $6.38^{+0.02}_{-0.02}$ & $276^{+57}_{-51}$ & 180.7/176 & 0.03, 3.26 & pcfabs & C\\
NGC 788 & $46.89^{+4.68}_{-4.47}$ & $1.89^{+0.24}_{-0.16}$ & $0.993^{+0.003}_{-0.005}$ & $6.41^{+0.09}_{-0.12}$ & $148^{+114}_{-45}$ & 1676/1586 & 0.11, 4.95 & pcfabs & C\\
Mrk 590 & -- & 1.62$^{+0.02}_{-0.02}$ & -- & 6.39$^{+0.02}_{-0.02}$ & 121$^{+11}_{-16}$ & 948/964 & 4.4, 6.4 & pow & S\\
NGC 931 & 0.36$^{+0.08}_{-0.08}$ & 1.75$^{+0.11}_{-0.11}$ & 0.066$^{+0.011}_{-0.036}$ & 6.41$^{+0.15}_{-0.15}$ & 550$^{+250}_{-250}$ & 342/365 & 1.14\tablenotemark{4}, 5.1 & pow + bb & S\\
NGC 985 & -- & 1.60$^{+0.03}_{-0.03}$ & 0.010$^{+0.01}_{-0.01}$ &  -- & -- & 179/174 & 63, 54 & pow + bb & S\\
ESO 416-G002 & $< 0.027$ & 1.63$^{+0.06}_{-0.04}$ & 0.105$^{+0.006}_{-0.004}$ & -- & -- & 647/634 & 7.53 & pow + bb & S\\
ESO 198-024 & -- & 1.84$^{+0.02}_{-0.02}$ & 0.172$^{+0.007}_{-0.007}$ & 6.41 & 77$^{+23}_{-23}$ & 1667.9/1535 & 15.8, 11.0 & pow + bb & S\\
RBS 476 & -- & 1.87$^{+0.2}_{-0.2}$ & -- & -- & -- & 47.2/50 & 10.30, 18.02 & pow & S\\

3C 105 & $28.16^{+1.41}_{-3.23}$ & $1.36^{+0.18}_{-0.36}$ & $0.973^{+0.005}_{-0.004}$ & $6.42^{+0.04}_{-0.06}$ & $216^{+98}_{-61}$ & 315.4/317 & 0.04, 2.08 & pcfabs & C\\
3C 111 & 0.62$^{+0.03}_{-0.02}$ & 1.72$^{+0.03}_{-0.02}$ & -- & -- & -- & 1653/1724 & 5.0, 35 & pow & S\\
1H 0419-577 & -- & 1.48$^{+0.07}_{-0.07}$ & -- & -- & -- & 280/319 & 4.8, 9.4 & pow & S\\
3C 120 & 0.16$^{+0.01}_{-0.01}$ & 2.00$^{+0.02}_{-0.02}$ & -- & 6.06$^{+0.26}_{-0.21}$ & $> 400$ & 139.9/126 & 23.5\tablenotemark{4}, 45.9 & pow & S\\
MCG -01-13-025 & $< 0.0039$ & $1.70^{+0.03}_{-0.03}$ & $0.139^{+0.027}_{-0.032}$ & -- & -- & 819.3/874 & 21.3 & pow + bb & S\\
Ark 120 & -- & 1.90$^{+0.02}_{-0.03}$ & $0.20^{+0.01}_{-0.01}$ & $6.46^{+0.06}_{-0.08}$ & $123^{+38}_{-36}$ & 553/449 & 17.1, 27.8 & pow + bb & S\\
Pictor A & $0.06^{+0.02}_{-0.02}$ & 1.77$^{+0.04}_{-0.05}$ & -- & -- & -- & 113/106 & 6.97,11.33 & pow & S\\
ESO 362-G021 & -- & 1.58$^{+0.03}_{-0.03}$  & -- & -- & -- & 365.6/336 & 4.65,11.02 & pow & S\\
$[$HB89$]$ 0537-286  & -- & 1.60$^{+0.07}_{-0.07}$  & -- & -- & -- & 142.8/140 & 0.57,1.75 & pow & S\\
PKS 0548-322 & $0.103^{+0.004}_{-0.004}$ & 2.05$^{+0.02}_{-0.02}$ & -- & -- & -- & 862.9/846 &25, 28 & pow & S \\
NGC 2110 & $2.84^{+0.19}_{-0.16}$ & 1.54$^{+0.08}_{-0.07}$ & $0.952^{+0.006}_{-0.002}$ & 6.4 & $103^{+29}_{-29}$ & 628/682 &0.86\tablenotemark{4}, 25.5 & pcfabs & C \\
MCG +08-11-011 & $0.25^{+0.016}_{-0.015}$ & 1.64$^{+0.03}_{-0.03}$ & -- & 6.43$^{+0.08}_{-0.08}$ & $380^{+102}_{-83}$ & 896/860 & 5.5, 20.3 & pow & S \\
EXO 055620-3820.2 & $2.57^{+0.14}_{-0.14}$ & 1.82$^{+0.06}_{-0.07}$ & $0.966^{+0.005}_{-0.005}$ & 6.26 & 200$^{+100}_{-100}$ & 1177/1215& 1.44, 22.47 & pcfabs & C \\
ESO 005-G004 & $5.58^{+0.16}_{-0.16}$ & 1.48$^{+0.04}_{-0.04}$ & $0.984^{+0.002}_{-0.003}$ & 6.39$^{+0.05}_{-0.05}$ & 79$^{+9}_{-15}$ & 1316.1/1161 & 0.15, 10.41 & pcfabs & C \\
Mrk 3 & $1.24^{+5.67}_{-1.24}$ & 2.94$^{+0.55}_{-0.39}$ & 0.076 & 6.4 & 977$^{+193}_{-137}$ & 134.1/253& 0.58\tablenotemark{4}, 2.15 & double pow & C \\
Mrk 79 & $< 0.0063$ &  $1.85^{+0.04}_{-0.04}$ & $0.122^{+0.007}_{-0.007}$ & $6.33^{+0.08}_{-0.09}$ & $117^{+0.15}_{-0.15}$ & 436.8/480  &  42.75 & pow + bb & S \\
4C +71.07 &  $0.11^{+0.03}_{-0.03}$ & 1.45$^{+0.05}_{-0.05}$ & -- & 6.4 & $<110$ & 437.6/503 & 4.25, 14.0 & pow & S \\
2MASX J09180027+0425066 & $11.13^{+1.55}_{-1.18}$ & 1.77$^{+0.22}_{-0.15}$ & $0.987^{+0.006}_{-0.005}$ & 6.39$^{+0.05}_{-0.05}$ & 65.6$^{+39.4}_{-30.2}$ & 310.6/333& 0.02, 1.38 & pcfabs & C \\
Mrk 110 & $0.02^{+0.01}_{-0.01}$ & 1.78$^{+0.02}_{-0.02}$  & -- & -- & -- & 484.4/451 & 16.55, 27.93 & pow & S\\
NGC 2992 & $1.19^{+2.21}_{-0.96}$ & 1.54$^{+0.41}_{-0.25}$ & $0.476^{+0.209}_{-0.368}$ & 6.4 & $656^{+204}_{-191}$ & 252/257 &0.86\tablenotemark{4}, 3.51 & pcfabs & C \\
MCG -05-23-016 & $1.60^{+0.005}_{-0.006}$ & 1.90$^{+0.04}_{-0.04}$ & 139 & 6.4 & $79^{+16}_{-14}$ & 1995/1744 & 7.84\tablenotemark{4}, 85.1 & double pow & C \\
NGC 3081 & $94.2^{+6.2}_{-7.2}$ & 1.60$^{+0.12}_{-0.17}$ & $0.990^{+0.011}_{-0.002}$ & 6.38$^{+0.03}_{-0.05}$ & $191^{+66}_{-58}$ & 303/245 &0.18, 5.53 & pcfabs & C \\
NGC 3227 & $1.74^{+0.12}_{-0.09}$ & 1.35$^{+0.04}_{-0.04}$ & $0.852^{+0.006}_{-0.010}$ & 6.36$^{+0.03}_{-0.06}$ & $142^{+22}_{-40}$ & 488.1/445 &2.44, 28.29 & pcfabs & C \\
NGC 3281 & $86.3^{+16.32}_{-16.12}$ & 1.67$^{+0.44}_{-0.50}$ & $0.981^{+0.012}_{-0.025}$ & 6.36$^{+0.06}_{-0.06}$ & $617^{+225}_{-181}$ & 153.6/135 & 0.01, 2.58 & pcfabs & C \\
Mrk 421 & $0.077^{+0.007}_{-0.007}$ & 2.95$^{+0.01}_{-0.01}$ & -- & -- & -- & 621.1/439 & 61.22, 21.19 & pow & S \\
NGC 3516 & $0.353^{+0.32}_{-0.12}$ & 1.73$^{+0.01}_{-0.01}$ & $0.99998^{+2e-5}_{-0.0462}$ & 6.32$^{+0.09}_{-0.02}$ & $104^{+20}_{-20}$ & 584.2/413 &2.44, 28.29 & pcfabs, bb & C \\
NGC 3783 & $0.57^{+0.21}_{-0.14}$ & 1.60$^{+0.04}_{-0.03}$ & $0.722^{+0.107}_{-0.082}$ & 6.34$^{+0.07}_{-0.04}$ & $80^{+26}_{-29}$ & 559.9/421 &23.96, 70.49 & pcfabs, bb & C \\
NGC 4051 & -- &  $1.95^{+0.02}_{-0.02}$ & $0.087^{+0.005}_{-0.003}$ & $6.43^{+0.02}_{-0.02}$ & $147^{+41}_{-40}$ & 509.7/434  &  24.11, 21.86 & pow + bb & S \\
NGC 4138 & $8.00^{+1.00}_{-1.00}$ & 1.5$^{+0.1}_{-0.1}$ & $0.3^{+0.1}_{-0.1}$ & 6.4 & 83$^{+30}_{-30}$ & 408/408 & 0.06, 5.5 & pow + bb & S* \\
NGC 4151 & $5.32^{+0.07}_{-0.08}$ & 1.46$^{+0.02}_{-0.02}$ & $0.959^{+0.002}_{-0.001}$ & 6.36$^{+0.01}_{-0.04}$ & $137^{+10}_{-14}$ & 692/413 & 3.31, 129.36 & pcfabs & C \\
Mrk 766 & -- &  $1.78^{+0.01}_{-0.01}$ & $0.043^{+0.007}_{-0.005}$ & $6.38^{+0.07}_{-0.06}$ & $75^{+40}_{-32}$ & 498.2/441  &  15.65, 20.09 & pow + bb & S \\
NGC 4388 & $36.17^{+3.81}_{-3.82}$ & 1.79$^{+0.17}_{-0.17}$ & $0.954^{+0.013}_{-0.016}$ & 6.44$^{+0.04}_{-0.04}$ & 466$^{+88}_{-92}$ & 396.5/344 & 0.52, 5.90 & pcfabs & C \\
NGC 4395 & $3.30^{+0.68}_{-0.67}$ & 1.07$^{+0.12}_{-0.06}$ & $0.678^{+0.056}_{-0.062}$ & 6.22$^{+0.11}_{-0.16}$ & 101$^{+69}_{-64}$ & 318.6/359 & 0.51, 5.95 & pcfabs & C \\
3C 273 & -- &  $1.61^{+0.01}_{-0.01}$ & $0.154^{+0.021}_{-0.013}$ & -- & -- & 574.06/452  &  57.44, 116.13 & pow + bb & S \\
NGC 4507 &  $34.28^{+4.50}_{-4.57}$ & $1.65^{+0.35}_{-0.38}$ & 66 & $6.4$ &$167^{+34}_{-33}$ & 564/490	& 0.51\tablenotemark{4}, 22.8 & double pow & C\\
NGC 4593 & $0.031^{+0.011}_{-0.012}$ &  $1.86^{+0.01}_{-0.02}$ & -- & $6.41^{+0.06}_{-0.08}$ & $105^{+49}_{-23}$ & 583.8/444  &  21.11, 34.19 & pow & S \\
3C 279 & $0.020^{+0.017}_{-0.016}$ &  $1.78^{+0.04}_{-0.03}$ & -- & -- & -- & 449.2/430  &  4.59, 8.06 & pow & S \\
NGC 4945 & $0.793^{+0.957}_{-0.495}$ & 1.75$^{+0.31}_{-0.14}$ & $0.691^{+0.219}_{-0.590}$ & 6.4 & 1020$^{+210}_{-298}$ & 255/233 & 0.76\tablenotemark{4}, 2.91 & pcfabs & C \\
MCG -03-34-064 & $40.73^{+4.79}_{-4.30}$ & 1.83$^{+0.28}_{-0.23}$ & $0.961^{+0.014}_{-0.008}$ & $6.44^{+0.09}_{-0.10}$ & 191$^{+123}_{-74}$ & 306.9/261 & 0.41, 4.78 & pcfabs, apec & C \\
MCG -06-30-015 & $0.19^{+0.03}_{-0.01}$ & 1.78$^{+0.03}_{-0.02}$ & 0.057$^{+0.003}_{-0.003}$ & 6.41 & 71.4$^{+29}_{-40}$ & 488.2/450 & 20.41, 37.65 & pow + bb & S\\
NGC 5252 & $4.34^{+0.52}_{-0.42}$ & 1.55$^{+0.16}_{-0.13}$ & $0.962^{+0.009}_{-0.009}$ & 6.4 & 76$^{+54}_{-53}$ & 483/478 & 0.173\tablenotemark{4}, 5.81 & pcfabs & C \\
IC 4329A & $0.61^{+0.03}_{-0.03}$ & 1.80$^{+0.01}_{-0.01}$ & $0.941^{+0.016}_{-0.015}$ & $6.40$ & 99$^{+123}_{-18}$ & 862.8/452 & 21.91, 81.48 & pcfabs & C \\
Mrk 279 & $< 0.013$ & $2.04^{+0.02}_{-0.02}$ & -- & 6.4 & 223$^{+28}_{-66}$ & 644.9/447  &  22.01, 25.98 & pow & S \\
NGC 5506 & $2.78^{+0.05}_{-0.05}$ & 1.76$^{+0.03}_{-0.02}$ & $0.9893^{+0.001}_{-0.001}$ & $6.40$ & 137$^{+14}_{-24}$ & 426.4/413 & 2.81, 67.59 & pcfabs & C \\
NGC 5548 & $0.07^{+0.04}_{-0.05}$ & 1.88$^{+0.01}_{-0.01}$ & 0.031$^{+0.016}_{-0.003}$ & 6.41 & 73$^{+14}_{-44}$ & 502.3/423 & 40.34, 59.60 & pow + bb & S\\
ESO 511-G030 & $0.098^{+0.021}_{-0.021}$ & 1.82$^{+0.04}_{-0.04}$ & -- & 6.41 & 125$^{+29}_{-88}$ & 384.5/388 & 6.26, 13.04 & pow & S\\
RBS 1399 & $0.102^{+0.007}_{-0.007}$ & 1.95$^{+0.01}_{-0.01}$ & -- & -- & -- & 499.0/417 & 19.49, 33.79 & pow & S\\
NGC 5728 & $82.0^{+5.3}_{-5.0}$ & 0.94$^{+0.08}_{-0.07}$ & 0.011 & $6.32^{+0.04}_{-0.04}$ & 890 & 16/16 & 0.30, 1.44 & pcfabs & C \\
Mrk 841 & -- & 1.81$^{+0.07}_{-0.09}$ & 0.20$^{+0.01}_{-0.01}$ & 6.25$^{+15}_{-14}$ & 90$^{+210}_{-35}$ & 548/244 & 16, 10 & pow + bb & S\\
Mrk 290 & $0.15^{+0.03}_{-0.05}$ & 1.61$^{+0.04}_{-0.04}$ & 0.05$^{+0.01}_{-0.01}$ & 6.4 & 176$^{+79}_{-76}$ & 523.3/440 & 4.84, 9.31 & pow + bb & S\\
Mrk 501 & $0.18^{+0.01}_{-0.01}$ & 2.01$^{+0.01}_{-0.01}$ & -- & -- & -- & 523.8/416 & 74.96, 126.03 & pow & S\\
NGC 6300 & $21.5^{+0.8}_{-0.9}$ & 1.83$^{+0.08}_{-0.08}$ & 0.03 & $6.43^{+0.01}_{-0.02}$ & 148$^{+18}_{-18}$ & 423.5/420 & 0.037, 8.6 & ** & C\\
3C 382 & -- & 1.12$^{+0.10}_{-0.29}$ & 1.94$^{+0.24}_{-0.06}$ & $6.30^{+0.21}_{-0.32}$ & 489$^{+209}_{-210}$ & 1011/982 & 18.00, 31.00 & pow + brems & S\\
ESO 103-035 &  $21.6^{+2.6}_{-2.5}$ & $2.08^{+0.29}_{-0.28}$ & 0.076 & $6.363^{+0.270}_{-0.144}$ &$200^{+230}_{-110}$ & 534/527 & 0.10\tablenotemark{4}, 24.10 & double pow & C\\
3C 390.3 & $0.12^{+0.03}_{-0.03}$ & 1.50$^{+0.03}_{-0.03}$ & -- & $6.52^{+0.10}_{-0.10}$ &$108^{+60}_{-57}$ & 807.8/792 & 7.48\tablenotemark{4}, 19.4 & pow & S\\
NGC 6814 & -- & 1.74$^{+0.07}_{-0.06}$ & -- & $6.4$ &$545^{+217}_{-216}$ & 221.8/250 & 0.53, 1.25 & pow & S\\
3C 403 & $45.0^{+7.0}_{-6.0}$ & $1.76^{+0.2}_{-0.2}$ & 0.0044 & $6.32^{+0.2}_{-0.2}$ &$244^{+20}_{-20}$ & 31.1/50 & 0.05, 0.75\tablenotemark{5} & double pow & C\\
Cyg A & $11.0^{+21.0}_{-6.0}$ & $1.80^{+0.28}_{-0.43}$ & -- & -- & -- & 969.3/932 & 16.0, 24.0 & pow + R-S& C\\
4C +74.26 & $0.177^{+0.035}_{-0.016}$ & 1.86$^{+0.03}_{-0.03}$ & $1.00_{-0.091}$ & 6.4 & $110^{+42}_{-41}$ & 413/405 & 6.67, 19.61 & pcfabs & C \\
Mrk 509 & $0.015^{+0.008}_{-0.008}$ & 1.87$^{+0.01}_{-0.01}$ & -- & $6.4$ &$86.1^{+50}_{-15}$ & 573/445 & 28.32, 45.83 & pow & S\\
IC 5063 & $21.78^{+2.24}_{-2.06}$ & $1.63^{+0.22}_{-0.22}$ & 63.1 & $6.4$ &$108^{+56}_{-54}$ & 526/461 & 0.26\tablenotemark{4}, 13.1 & double pow & C\\
PKS 2149-306 & $0.07^{+0.02}_{-0.02}$ & 1.57$^{+0.05}_{-0.05}$ & -- & -- &-- & 364.6/344 & 3.39\tablenotemark{4}, 10.6 & pow & S\\
NGC 7172 & $8.19^{+3.42}_{-3.30}$ & 1.69$^{+0.08}_{-0.09}$ & -- & 6.4 & $68^{+36}_{-35}$ & 690/729 & 0.128\tablenotemark{4}, 37.0 & pow & S \\
NGC 7213 & $0.025^{+0.011}_{-0.012}$ & 1.75$^{+0.02}_{-0.02}$ & -- & 6.4 & $133^{+49}_{-24}$ & 519.9/445 & 16.65, 31.18 & pow & S \\
NGC 7314 & $1.16^{+0.01}_{-0.14}$ & 2.41$^{+0.05}_{-0.02}$ & 18.5 & 6.4 & $182^{+46}_{-47}$ & 1702/1561 & 7.35\tablenotemark{4}, 35.6 & double pow & C \\
3C 452 & $22.98^{+2.78}_{-2.83}$ & -0.08$^{+0.08}_{-0.25}$ & 0.936$^{+0.012}_{-0.034}$ & 6.4 & 100 & 135.5/110 & 0.01, 2.17 & pcfabs & C \\
MR 2251-178 & $0.28^{+0.11}_{-0.08}$ & 1.41$^{+0.04}_{-0.05}$ & 0.09$^{+0.03}_{-0.06}$ & 6.4 & 67 & 406.8/437 & 6.39, 27.24 & pow + bb & S\\
NGC 7469 & -- & 1.98$^{+0.01}_{-0.01}$ & -- & 6.4 & $131^{+38}_{-39}$ & 562.9/444 & 24.10, 32.55 & pow & S\\
Mrk 926 & $0.035^{+0.01}_{-0.01}$ & 1.79$^{+0.02}_{-0.02}$ & -- & 6.4 & $57^{+28}_{-28}$ & 574.3/416 & 15.70, 33.13 & pow & S \\
NGC 7582 & $7.39^{+1.46}_{-1.00}$ & 1.38$^{+0.23}_{-0.20}$ & 23.5 & 6.4 & $106^{+66}_{-64}$ & 497/392 & 0.45\tablenotemark{4}, 13.2 & double pow & C \\
\end{longtable}
\end{center}
\vspace{-0.2cm}
\footnotesize
\noindent$^1${Milky Way Galactic absorption obtained from the n$_H$ FTOOL on HEASARC
in units of $10^{22}$\,cm$^{-2}$.\\}
$^2${Observed flux in the 0.5--2\,keV and 2--10\,keV bands in units of $10^{-12}$\,\flux.  Where only one number is shown, the value is the 0.3--10\,keV flux. The statistical errors on fluxes are very small and the systematic errors are dominated by model uncertainties.\\}
$^3${This parameter accounts from extra components referred to in the model column.  If the model includes: {\tt bbody} then this is the temperature in keV (kT), {\tt double pow} then this is the ratio of normalizations between the first and second power law components, and {\tt pcfabs} then this is the covering fraction.}\\
$^4${Estimate of 0.5-2\,keV flux from Tartarus, only 2-10\,keV flux was given in original paper.\\}

\end{landscape}
\normalsize

\clearpage
\begin{deluxetable}{llllllll}
\tabletypesize{\small}
\tablecaption{K-band Derived Mass and L$^{corr}_{2-10 keV}$/L$_{Edd}$\label{tbl-4}}
\tablewidth{0pt}
\tablehead{
\colhead{No.} & \colhead{Source} & 
\colhead{M/M$_{\odot}$\tablenotemark{1}} & \colhead{L$_{X}$/L$_{Edd}$\tablenotemark{2}}  & \colhead{No.} & \colhead{Source} &
\colhead{M/M$_{\odot}$\tablenotemark{1}} & \colhead{L$_{X}$/L$_{Edd}$
\tablenotemark{2}}
}
\startdata
1 & NGC 235A & 8.76 & -3.75 & 78 & 1RXS J112716.6+190914 & 9.00 & -3.21 \\ 
2 & Mrk 348 & 7.97 & -3.53 & 79 & NGC 3783 & 8.21 & -3.25 \\ 
3 & Mrk 352 & 7.26 & -2.69 & 80 & SBS 1136+594 & 7.53 & -2.03 \\ 
4 & NGC 454 & 6.23 & -2.08 & 81 & UGC 06728 & 6.81 & -3.22 \\ 
5 & Fairall 9 & 8.91 & -3.13 & 82 & 2MASX J11454045-1827149 & 6.70 & -1.36 \\ 
6 & NGC 526A & 8.02 & -2.75 & 83 & CGCG 041-020 & 8.46 & -3.32 \\ 
7 & NGC 612 & 8.47 & -3.17 & 85 & NGC 4051 & 7.27 & -4.05 \\ 
8 & ESO 297-018 & 9.68 & -4.66 & 86 & Ark 347 & 8.12 & -3.99 \\ 
9 & NGC 788 & 8.51 & -3.71 & 88 & NGC 4138 & 6.82 & -3.79 \\ 
10 & Mrk 1018 & 8.94 & -3.46 & 89 & NGC 4151 & 7.69 & -3.26 \\ 
12 & Mrk 590 & 8.87 & -4.08 & 90 & Mrk 766 & 7.85 & -3.18 \\ 
15 & NGC 931 & 8.55 & -4.25 & 91 & NGC 4388 & 8.53 & -4.24 \\ 
16 & NGC 985 & 8.94 & -2.78 & 92 & NGC 4395 & 5.30 & -3.30 \\ 
17 & ESO 416-G002 & 9.02 & -3.45 & 94 & NGC 4507 & 8.39 & -3.54 \\ 
18 & ESO 198-024 & 8.36 & -2.85 & 95 & ESO 506-G027 & 8.59 & -3.51 \\ 
20 & NGC 1142 & 9.36 & -4.01 & 96 & XSS J12389-1614 & 8.88 & -3.50 \\ 
24 & NGC 1365 & 8.88 & -4.75 & 97 & NGC 4593 & 8.61 & -4.02 \\ 
25 & ESO 548-G081 & 8.94 & -4.27 & 100 & SBS 1301+540 & 7.54 & -2.24 \\ 
27 & PGC 13946 & 8.75 & -3.74 & 102 & NGC 4992 & 8.56 & -3.69 \\ 
28 & 2MASX J03565655-4041453 & 8.64 & -2.88 & 103 & MCG -03-34-064 & 8.28 & -3.56 \\ 
29 & 3C 105 & 7.79 & -2.00 & 105 & MCG -06-30-015 & 7.36 & -2.87 \\ 
31 & 1H 0419-577 & 9.00 & -2.83 & 106 & NGC 5252 & 8.64 & -3.91 \\ 
32 & 3C 120 & 8.56 & -2.71 & 108 & IC 4329A & 8.52 & -3.05 \\ 
34 & MCG -01-13-025 & 8.06 & -3.18 & 109 & Mrk 279 & 8.62 & -3.09 \\ 
36 & XSS J05054-2348 & 7.53 & -2.04 & 110 & NGC 5506 & 7.77 & -3.12 \\ 
38 & Ark 120 & 8.74 & -3.13 & 112 & NGC 5548 & 8.42 & -3.03 \\ 
39 & ESO 362-G018 & 9.00 & -4.78 & 113 & ESO 511-G030 & 8.66 & -3.70 \\ 
40 & Pictor A & 7.60 & -2.30 & 115 & NGC 5728 & 8.53 & -4.90 \\ 
45 & NGC 2110 & 8.28 & -3.94 & 116 & Mrk 841 & 8.15 & -2.88 \\ 
47 & EXO 055620-3820.2 & 8.44 & -2.82 & 117 & Mrk 290 & 7.68 & -2.62 \\ 
49 & ESO 005-G004 & 7.89 & -3.99 & 118 & Mrk 1498 & 8.59 & -3.10 \\ 
50 & Mrk 3 & 8.48 & -3.66 & 124 & 1RXS J174538.1+290823 & 8.75 & -2.70 \\ 
51 & ESO 121-G028 & 9.00 & -3.81 & 125 & 3C 382 & 9.22 & -3.05 \\ 
53 & 2MASX J06403799-4321211 & -- & -- & 126 & ESO 103-035 & 7.73 & -3.12 \\ 
55 & Mrk 6 & 8.24 & -3.44 & 127 & 3C 390.3 & 8.52 & -2.58 \\ 
56 & Mrk 79 & 8.42 & -2.95 & 129 & NGC 6814 & 8.15 & -5.47 \\ 
60 & Mrk 18 & 7.45 & -3.76 & 133 & NGC 6860 & 8.24 & -3.75 \\ 
61 & 2MASX J0904699+5536025 & 7.70 & -2.64 & 136 & 4C +74.26 & 9.00 & -2.51 \\ 
62 & 2MASX J09112999+4528060 & 7.53 & -2.83 & 137 & Mrk 509 & 8.59 & -2.70 \\ 
64 & 2MASX J09180027+0425066 & 8.57 & -2.88 & 138 & IC 5063 & 7.68 & -3.08 \\ 
65 & MCG -01-24-012 & 7.16 & -2.26 & 139 & 2MASX J21140128+8204483 & 8.81 & -2.72 \\ 
66 & MCG +04-22-042 & 8.49 & -3.13 & 144 & UGC 11871 & 8.34 & -3.54 \\ 
67 & Mrk 110 & 7.80 & -2.11 & 145 & NGC 7172 & 8.31 & -3.58 \\ 
68 & NGC 2992 & 8.04 & -4.52 & 146 & NGC 7213 & 8.63 & -4.46 \\ 
69 & MCG -05-23-016 & 7.66 & -2.70 & 147 & NGC 7314 & 7.84 & -3.73 \\ 
70 & NGC 3081 & 7.96 & -3.66 & 148 & NGC 7319 & 8.54 & -3.55 \\ 
71 & NGC 3227 & 7.83 & -4.02 & 151 & MR 2251-178 & 8.76 & -2.56 \\ 
72 & NGC 3281 & 8.62 & -4.32 & 152 & NGC 7469 & 8.64 & -3.55 \\ 
75 & Mrk 417 & 8.04 & -2.87 & 153 & Mrk 926 & 8.95 & -2.93 \\ 
77 & NGC 3516 & 8.13 & -3.64 & 154 & NGC 7582 & 8.31 & -4.51 \\ 
\enddata

\tablenotetext{1}{The logarithm of $M/M_{\odot}$ where mass is derived from the 2MASS K-band stellar magnitudes (see Mushotzky {\it et al.} 2008).\\}
\tablenotetext{2}{The logarithm of $L^{corr}_{2-10 keV}/L_{Edd}$.  Here, L$^{corr}_{2- 10 keV}$ is the value supplied in the X-ray spectral fitting tables for sources with $N_H < 10^{22}$\,cm$^{-2}$ and the calculated unabsorbed flux for the higher column density sources.  The Eddington luminosity is calculated as $1.3 \times 10^{38} \times$\,M/M$_{\odot}$.\\}
\end{deluxetable}

\clearpage

\begin{figure}
\plotone{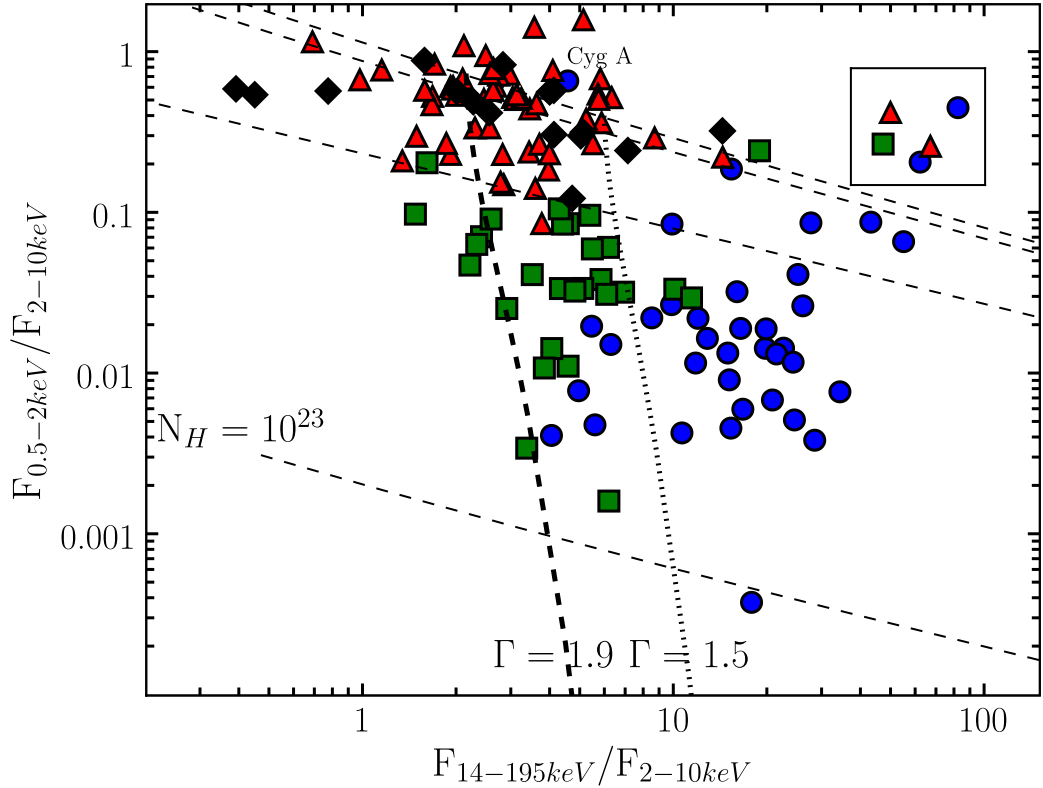}
\caption[The Color-Color Plot for the Entire Sample]{We plot all of the 9-month BAT AGN with soft and hard fluxes on the color-color plot initially presented in \citet{2008ApJ...674..686W}.  The triangles represent sources with $N_H < 10^{22}$\,cm$^{-2}$, while the circles represent sources with $N_H >10^{23}$\,cm$^{-2}$.  Squares indicate the intermediary hydrogen column sources.  Finally, diamonds are used to represent the 17 blazar/BL Lac sources, which all have low measured column densities.  In the text, we describe Cyg A (the circle labeled in the plot) and 5 other sources (NGC 1365 and NGC 5728 (circles), Mrk 3 (square), and NGC 4945 and NGC 6814 (triangles), within the box) to have unusual positions. \label{fig-2}}
\end{figure}

\begin{figure}
\includegraphics[width=9cm]{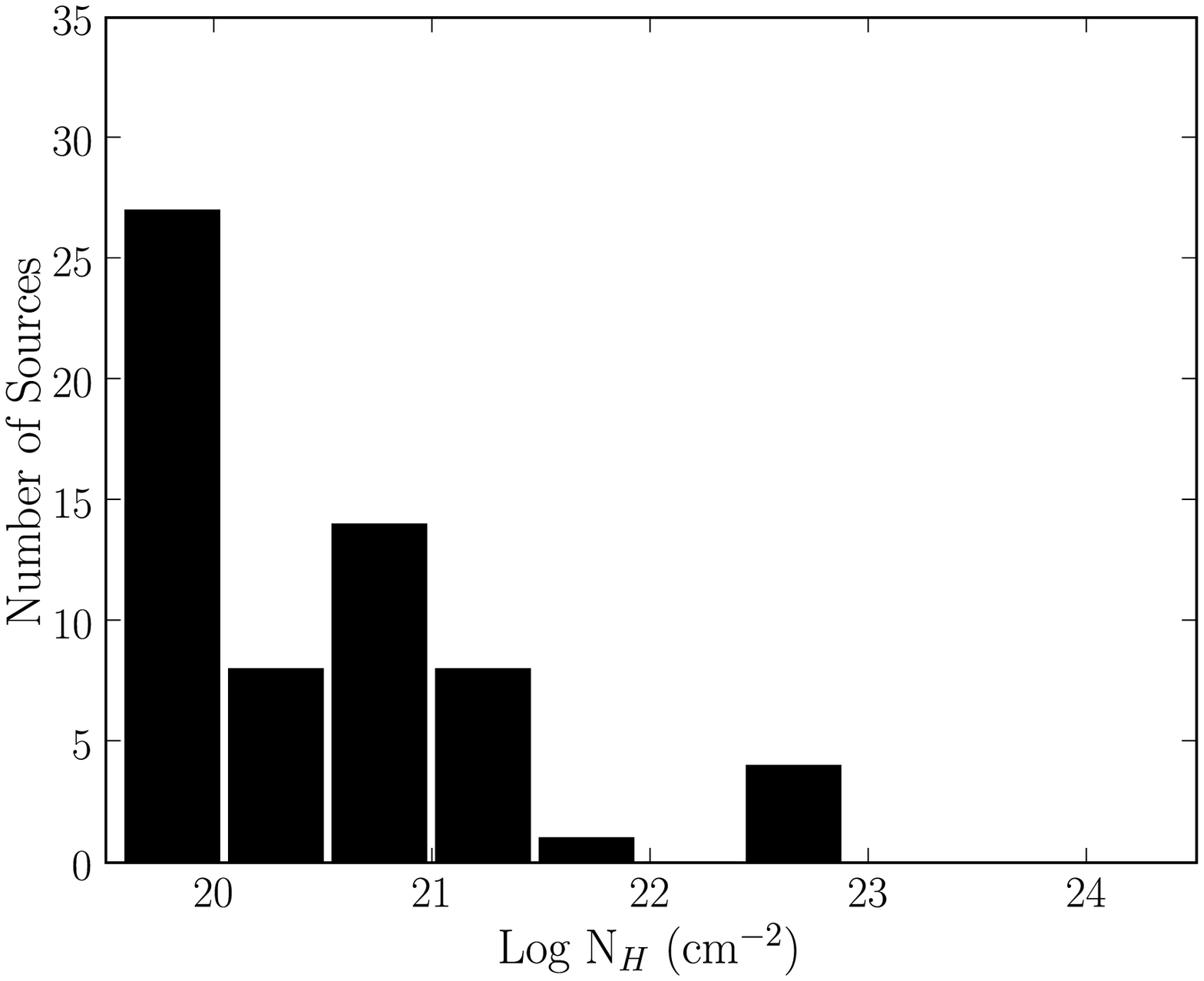}
\hspace{0.0cm}
\includegraphics[width=9cm]{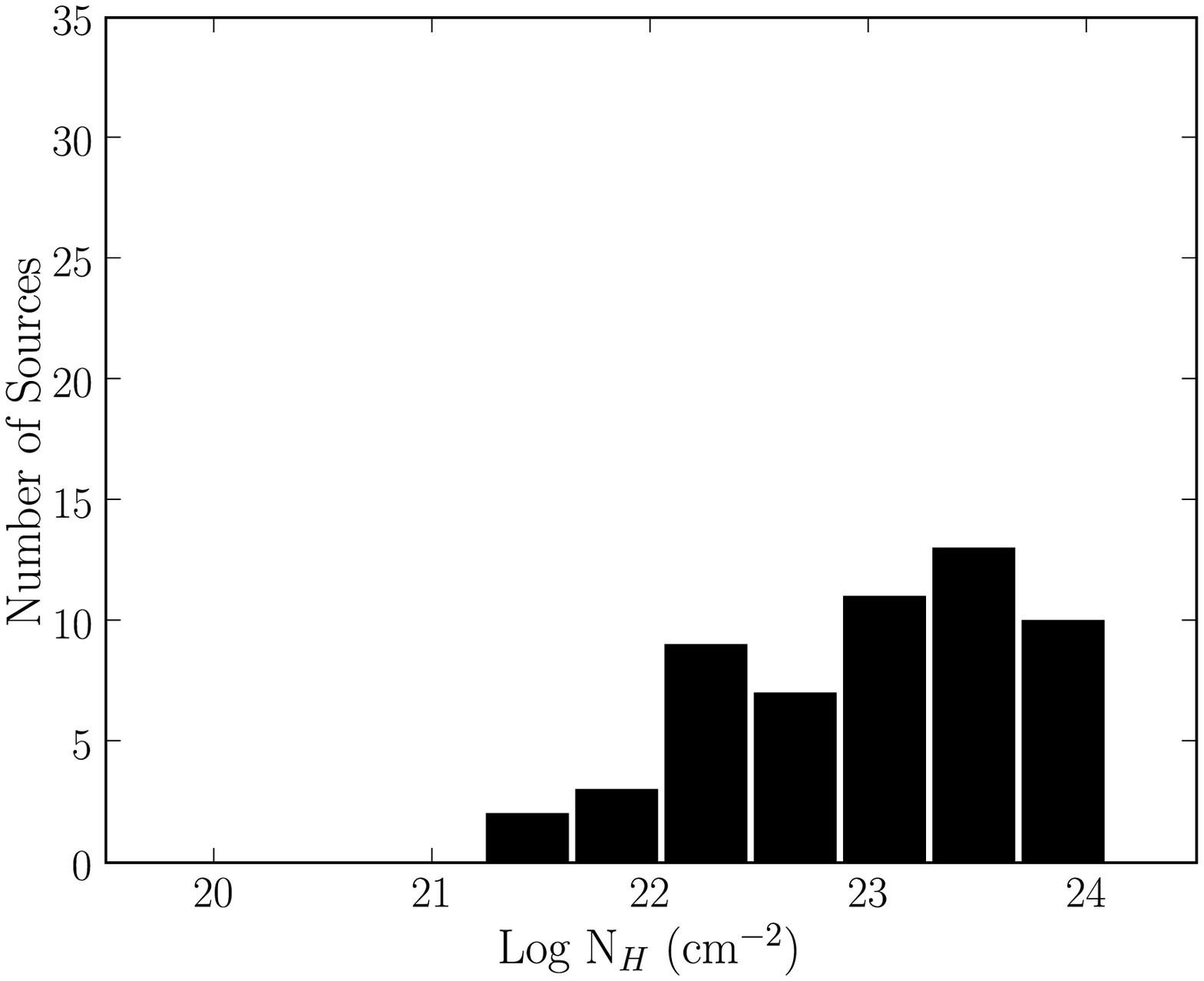}
\caption[Column Density Distributions]{We plot the distribution of the log of column densities for our uniform sample of 102 sources.  The left hand plot shows the simple model/power law sources, where the mean and standard deviation are $\mu = 20.58$ and $\sigma = 0.74$.  The right hand plot shows the distribution of complex/partial covering/double power law sources, where the mean and standard deviation are $\mu = 23.03$ and $\sigma = 0.71$.  Clearly, the simple model sources are dominated by low column density sources while the complex model sources mostly have higher column densities.   
\label{fig-nh}}
\end{figure}

\begin{figure}
\includegraphics[width=9cm]{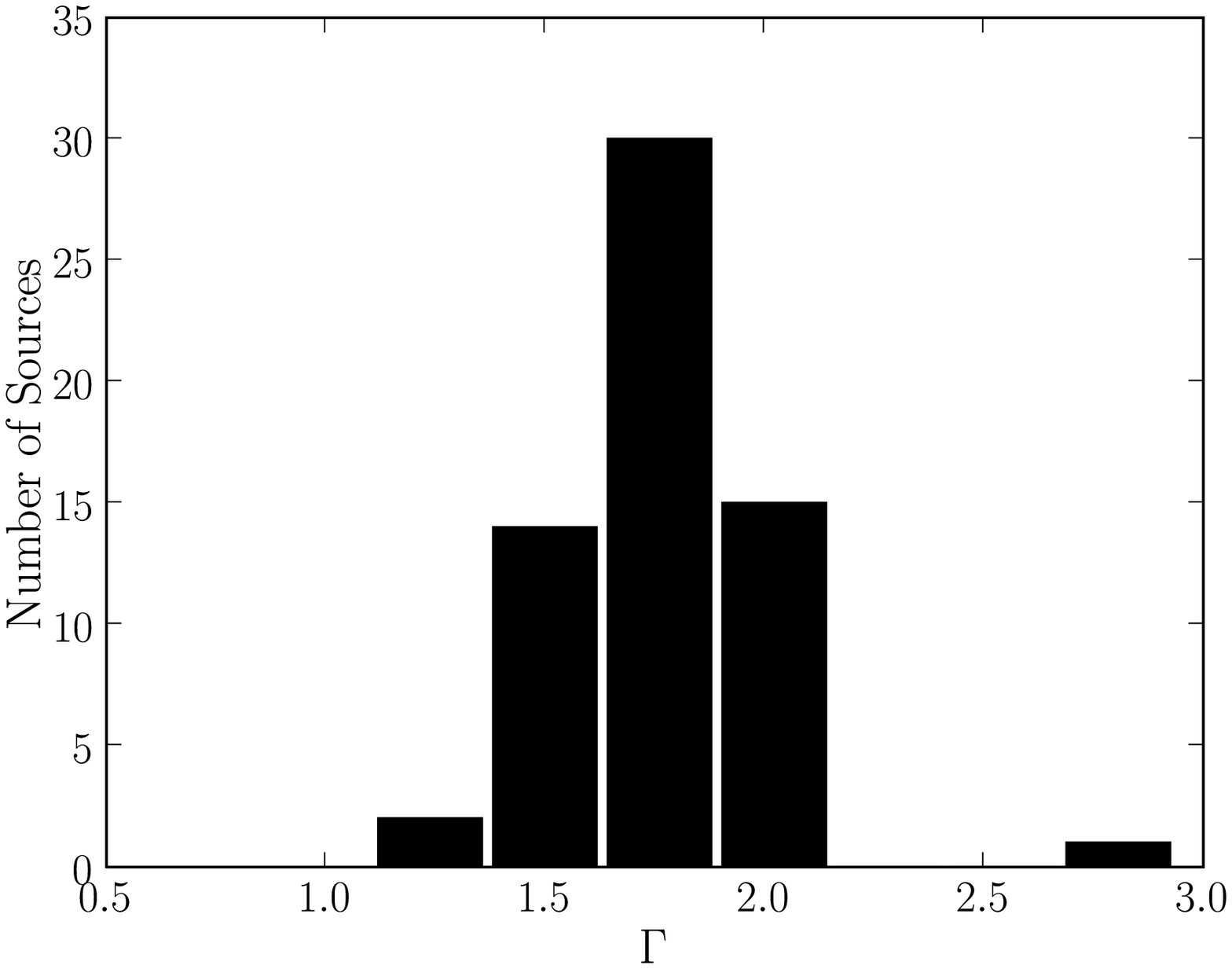}
\hspace{0.0cm}
\includegraphics[width=9cm]{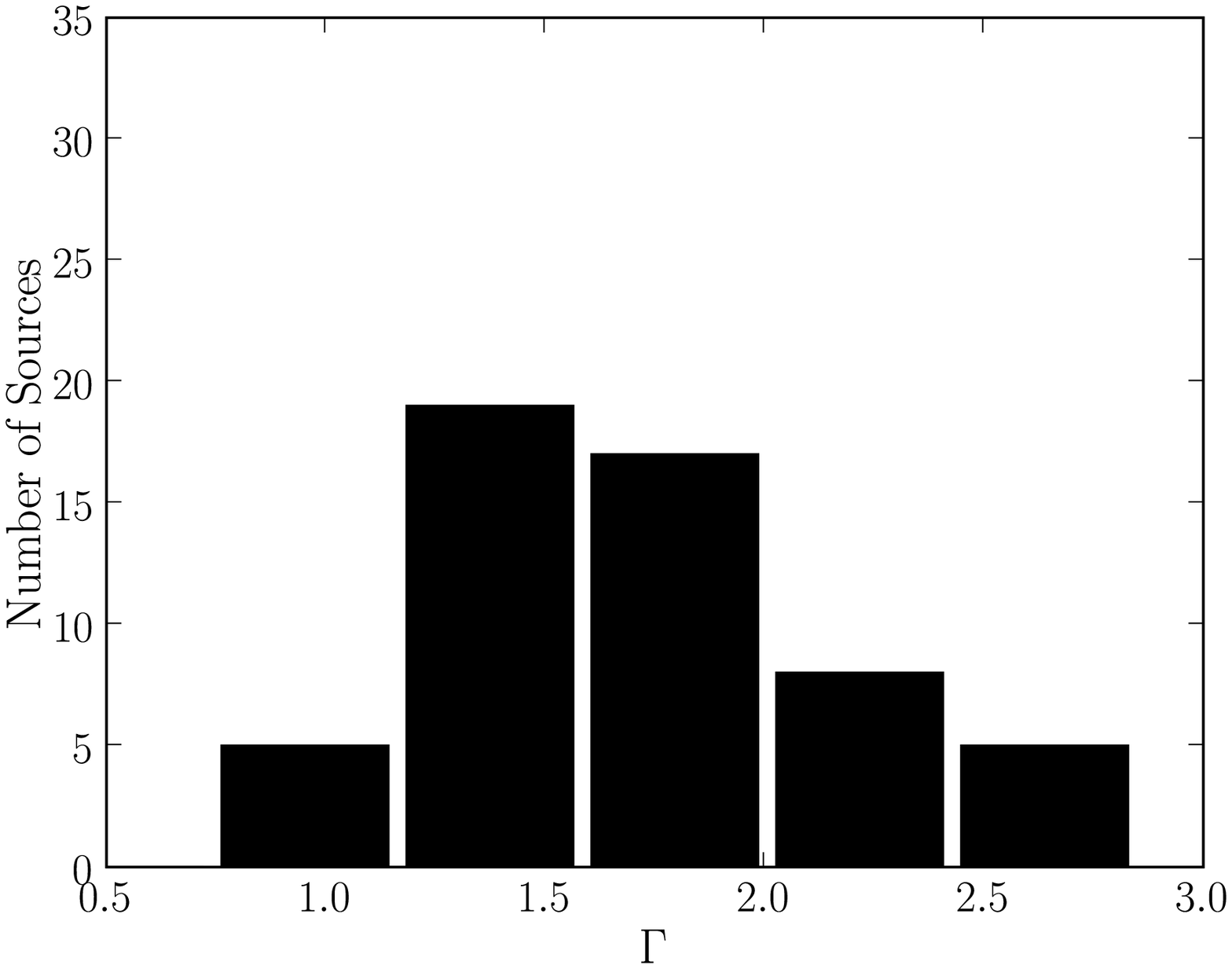}
\caption[$\Gamma$ Distributions]{We plot the distribution of the spectral index, $\Gamma$, for our uniform sample of 102 sources.  The left hand plot shows the simple model/power law sources, where the mean and standard deviation are $\mu = 1.78$ and $\sigma = 0.24$.  The right hand plot shows the distribution of complex/partial covering/double power law sources, where the mean and standard deviation are $\mu = 1.73$ and $\sigma = 0.45$.  The mean of the simple and complex models are very close, with only a 0.05 difference.  If the partial covering model is an accurate physical model, all of the AGN have roughly the same power law description for their X-ray spectra.     
\label{fig-pow}}
\end{figure}

\begin{figure}
\centering
\includegraphics[width=8.2cm]{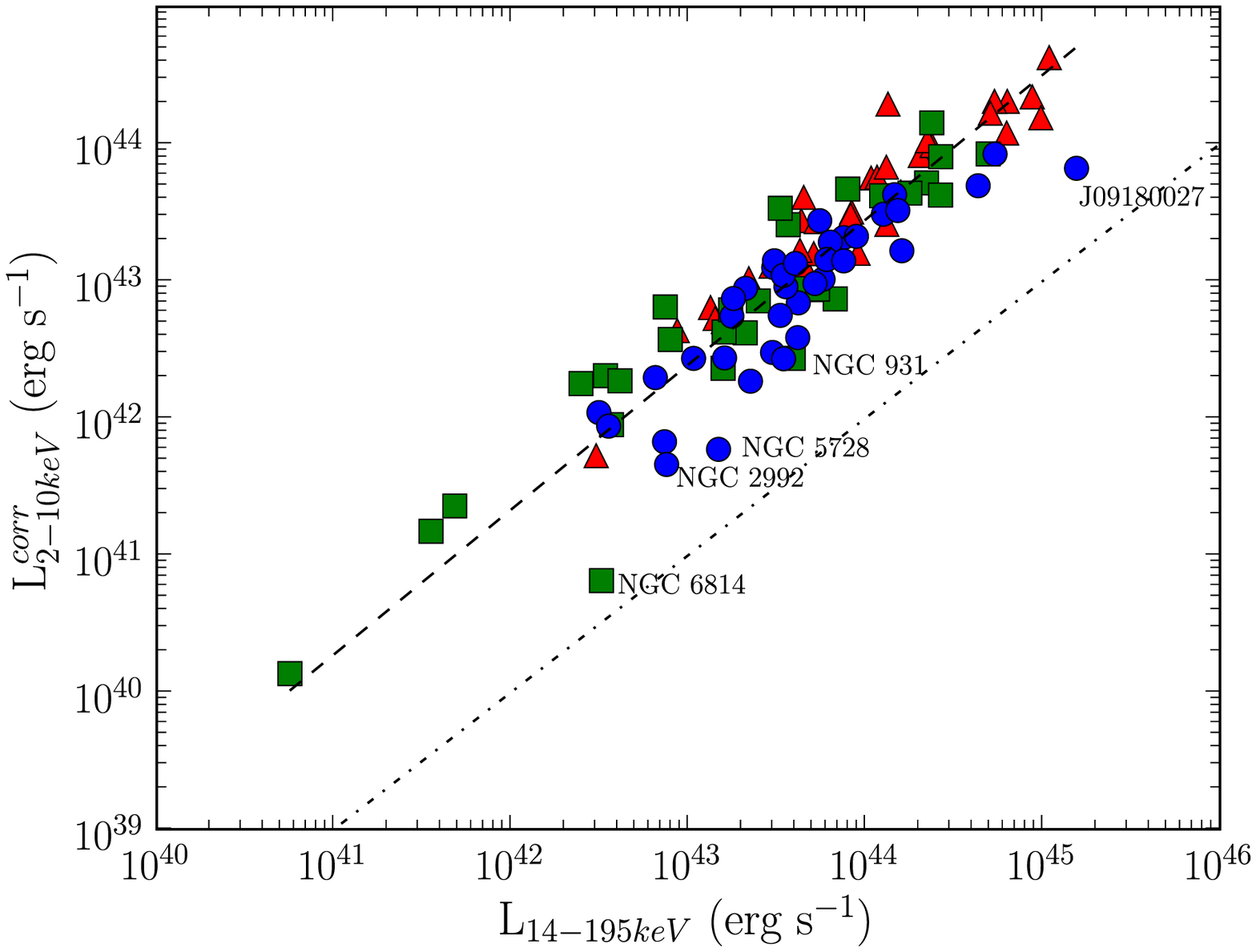}
\hspace{-0.3cm}
\includegraphics[width=8.0cm]{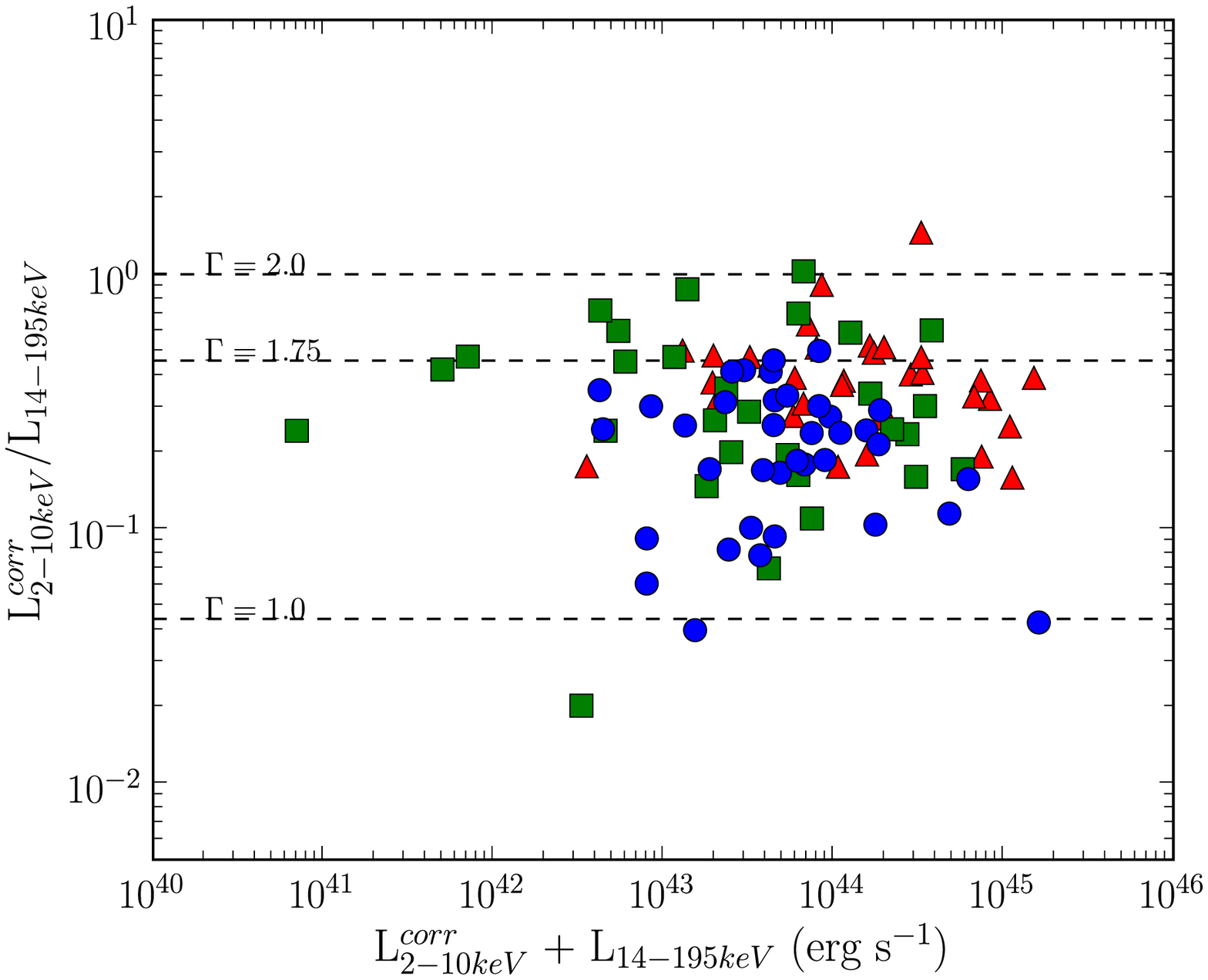}
\caption[Absorption Corrected L$^{corr}_{2-10 keV}$ versus L$_{14-195 keV}$]{(left)We plot the absorption corrected 2--10\,keV luminosity versus the 14--195\,keV luminosity from BAT.  The dashed line shows the best fit line to the data.  The fit shows a linear relationship ($L^{corr}_{2-10} \propto L_{14-195}$) with high significance ($R^2 = 0.85$).  The second line shows a slope of 2 in logarithmic space ($L^{corr}_{2-10} \propto 100 \times L_{14-195}$).  The 5 sources which deviate the most from this fit are labeled in the plot and discussed in the text. (right) We plot the ratio of L$^{corr}_{2-10 keV}$/L$_{14-195 keV}$ versus the total luminosity in the 2--10\,keV and 14--195\,keV bands.  The lines show values of constant $\Gamma$ between both bands at constant ratios of L$^{corr}_{2-10 keV}$/L$_{14-195 keV}$.
\label{fig-l210lbat}}
\end{figure}

\begin{figure}
\includegraphics[width=9cm]{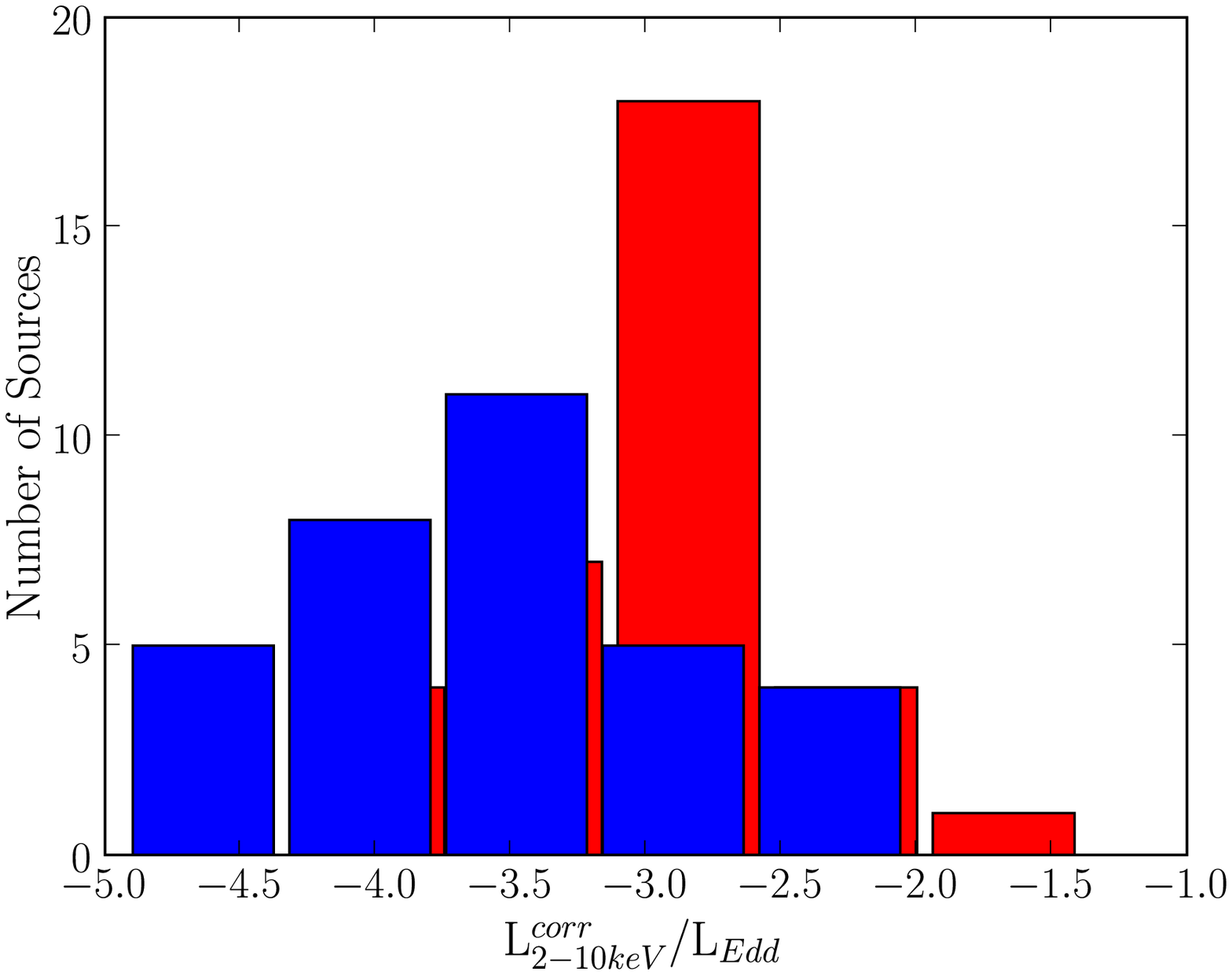}
\includegraphics[width=8.5cm]{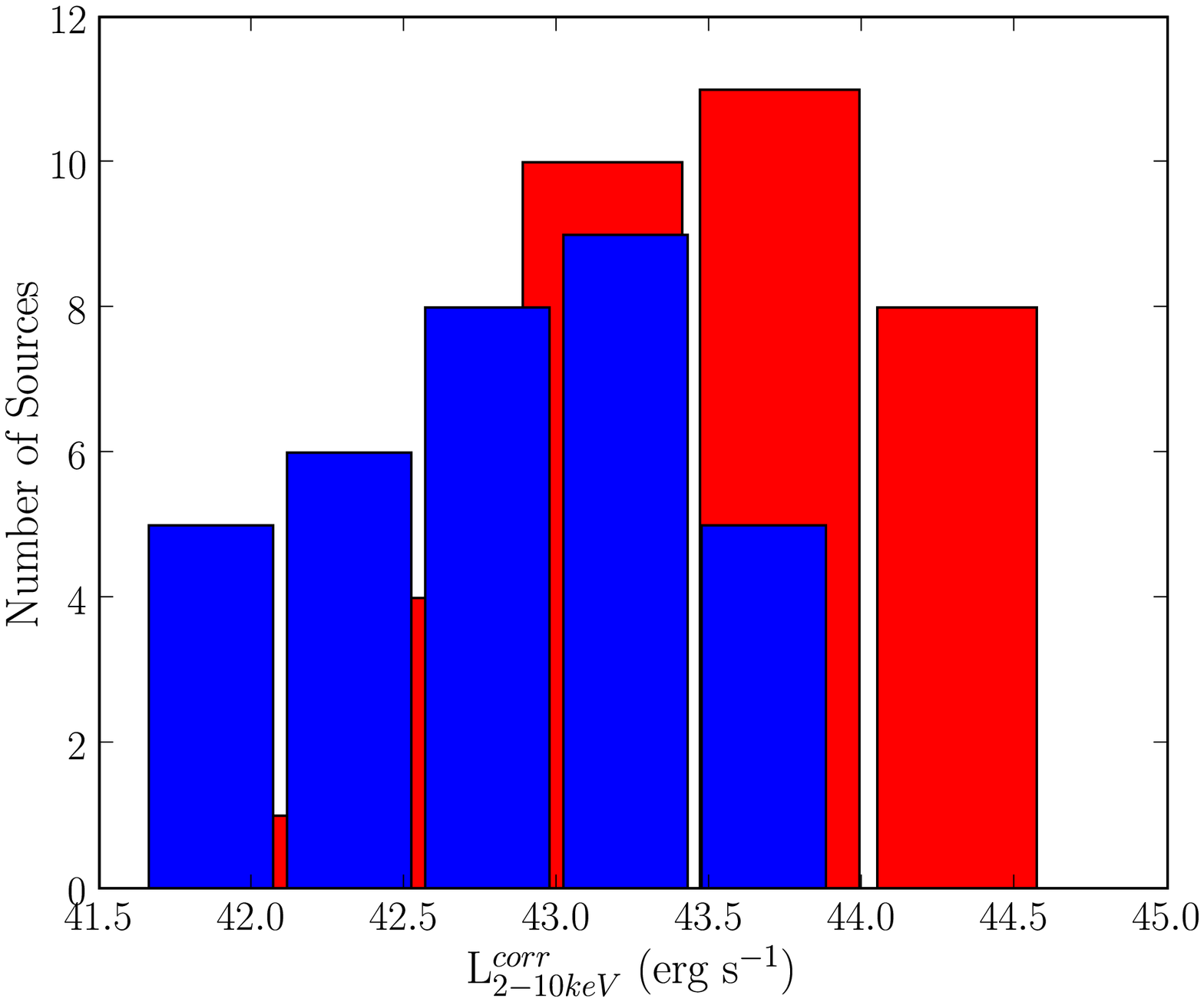}
\caption[$L^{corr}_{2-10keV}/L_{Edd}$ Distributions]{We plot the distribution of  $L^{corr}_{2-10keV}/L_{Edd}$  for our Seyfert 1 -- 1.2 (red) and Seyfert 2 (blue) sources.  From the plot, it is quite clear that the distribution of Seyfert 2s corresponds to lower $L^{corr}_{2-10keV}/L_{Edd}$.  The X-ray luminosities are unabsorbed values and so the result is not merely a result of absorption.  For both sets of distributions, we used a Kolmogorov-Smirnov test, finding a high probability of the distributions being different (the P values were $< 0.001$).
\label{fig-lledd}}
\end{figure}

\begin{figure}
\includegraphics[width=9cm]{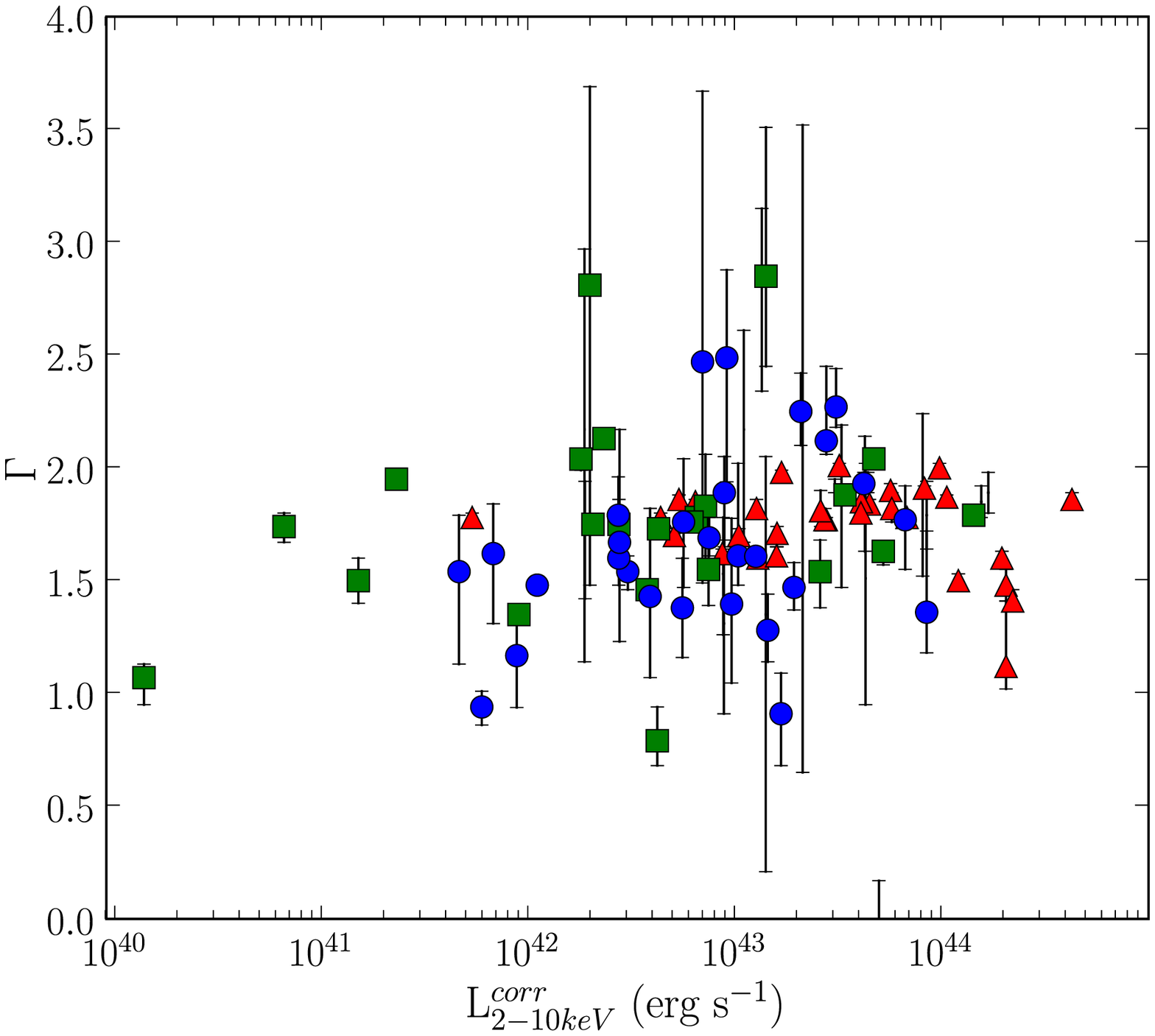}
\includegraphics[width=8.6cm]{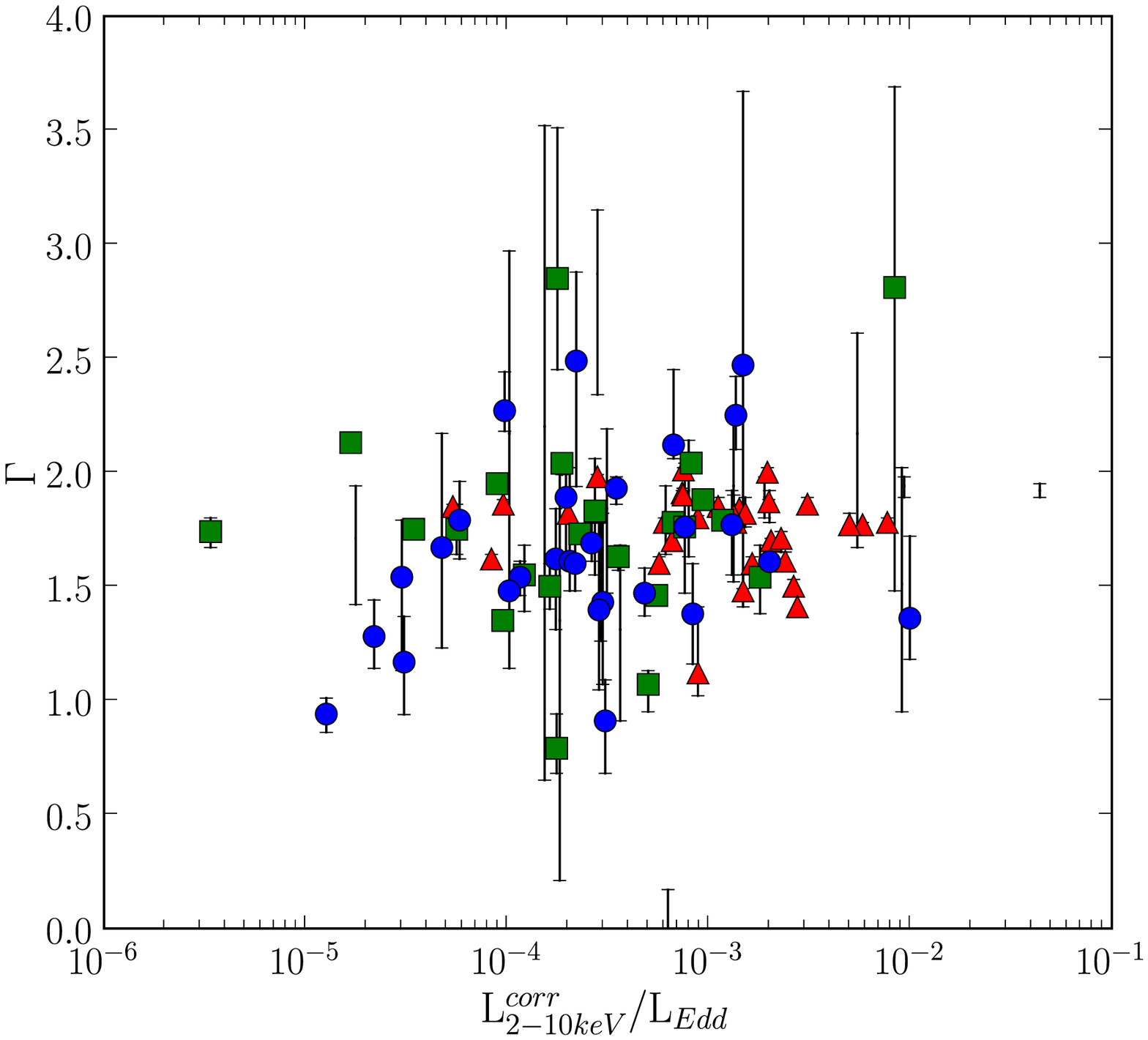}
\caption[$\Gamma$ versus L$^{corr}_{2- 10 keV}$ and L$^{corr}_{2- 10 keV}$/L$_{Edd}$]{We plot the unabsorbed 2--10\,keV luminosity versus power law index (left) and the ratio of  unabsorbed 2--10\,keV luminosity to Eddington luminosity versus power law index (right).  Eddington luminosity is calculated from an estimate of the black hole mass using 2MASS K band photometry.  We see no correlation between $\Gamma$ and L$^{corr}_{2- 10 keV}$ or L$^{corr}_{2- 10 keV}$/L$_{Edd}$.  Assuming that L$^{corr}_{2- 10 keV}$ is proportional to bolometric luminosity, our result shows no correlation between $\Gamma$ and accretion rate, contrary to those of \citet{2006ApJ...646L..29S}.  The triangles indicate optical Sy1--1.2 sources, the squares indicate Sy1.5--1.9 sources, and the circles indicate Sy2 sources.
\label{fig-lleddgamma}}
\end{figure}

\begin{figure}
\hspace{-0.7cm}
\includegraphics[width=6.0cm, height=5.0cm]{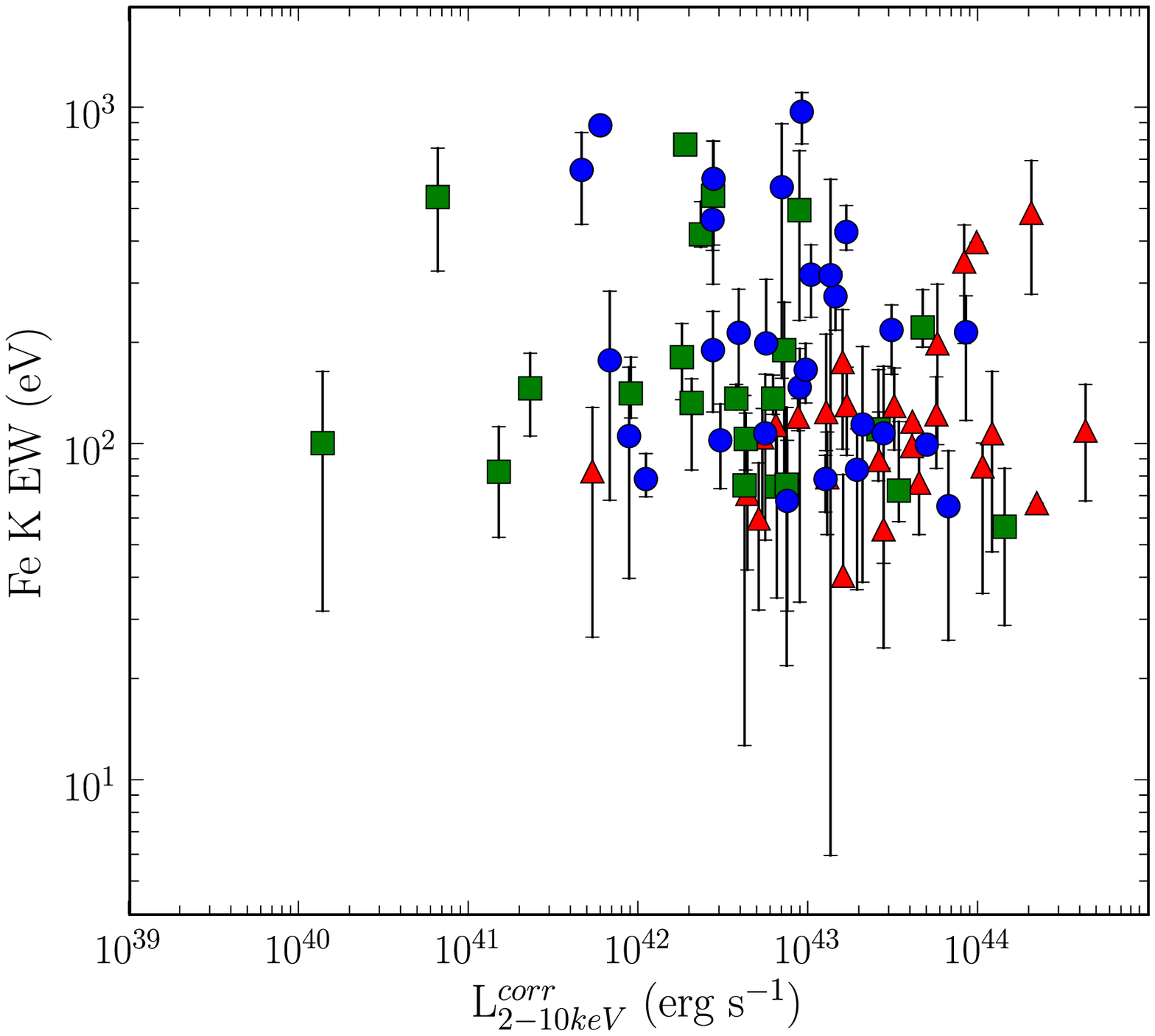}
\hspace{-0.75cm}
\includegraphics[width=6.0cm, height=5.0cm]{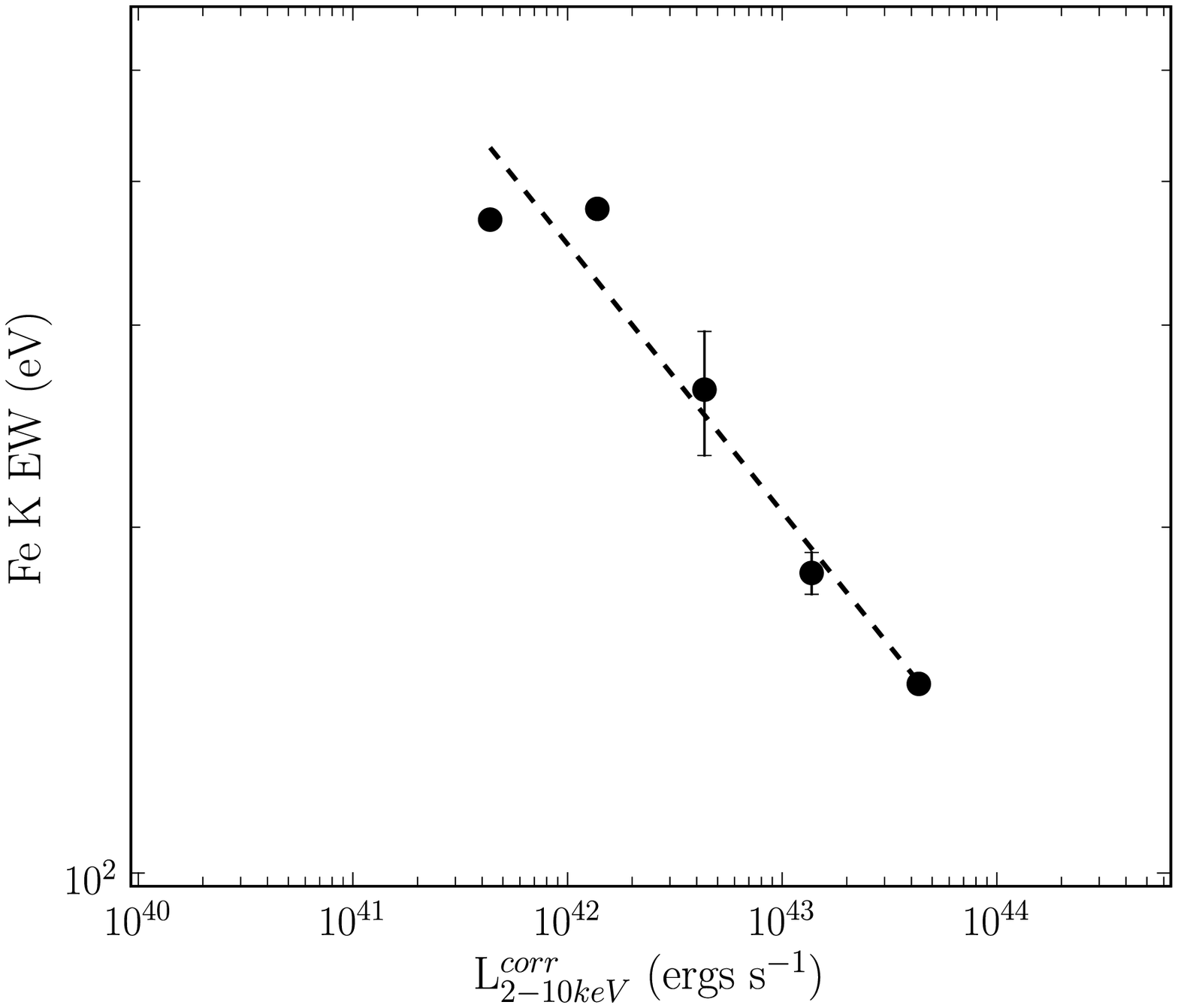}
\hspace{-0.75cm}
\includegraphics[width=6.0cm, height=5.0cm]{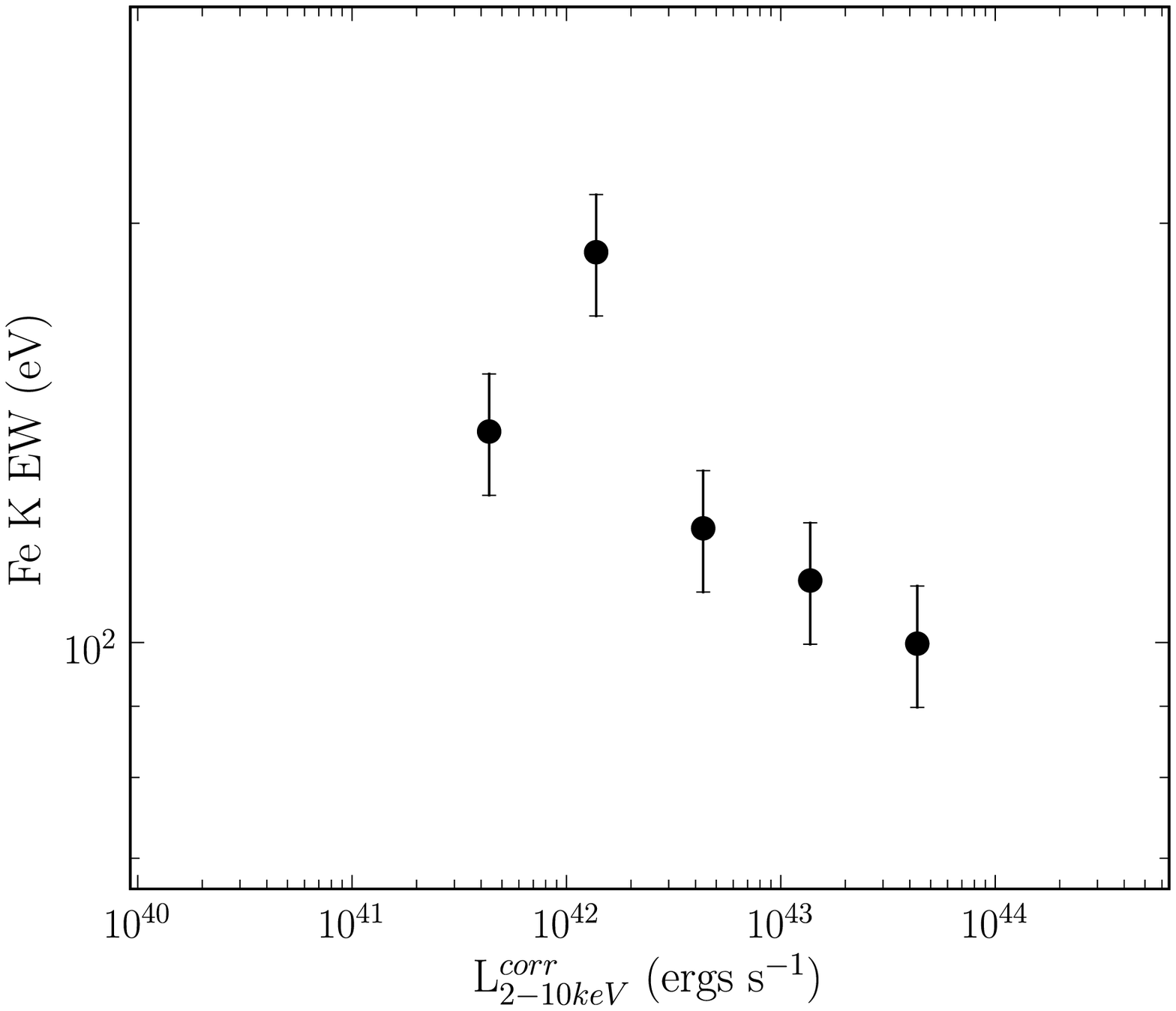}
\vspace{0.1cm}
\hspace{-0.7cm}
\includegraphics[width=6.0cm, height=5.0cm]{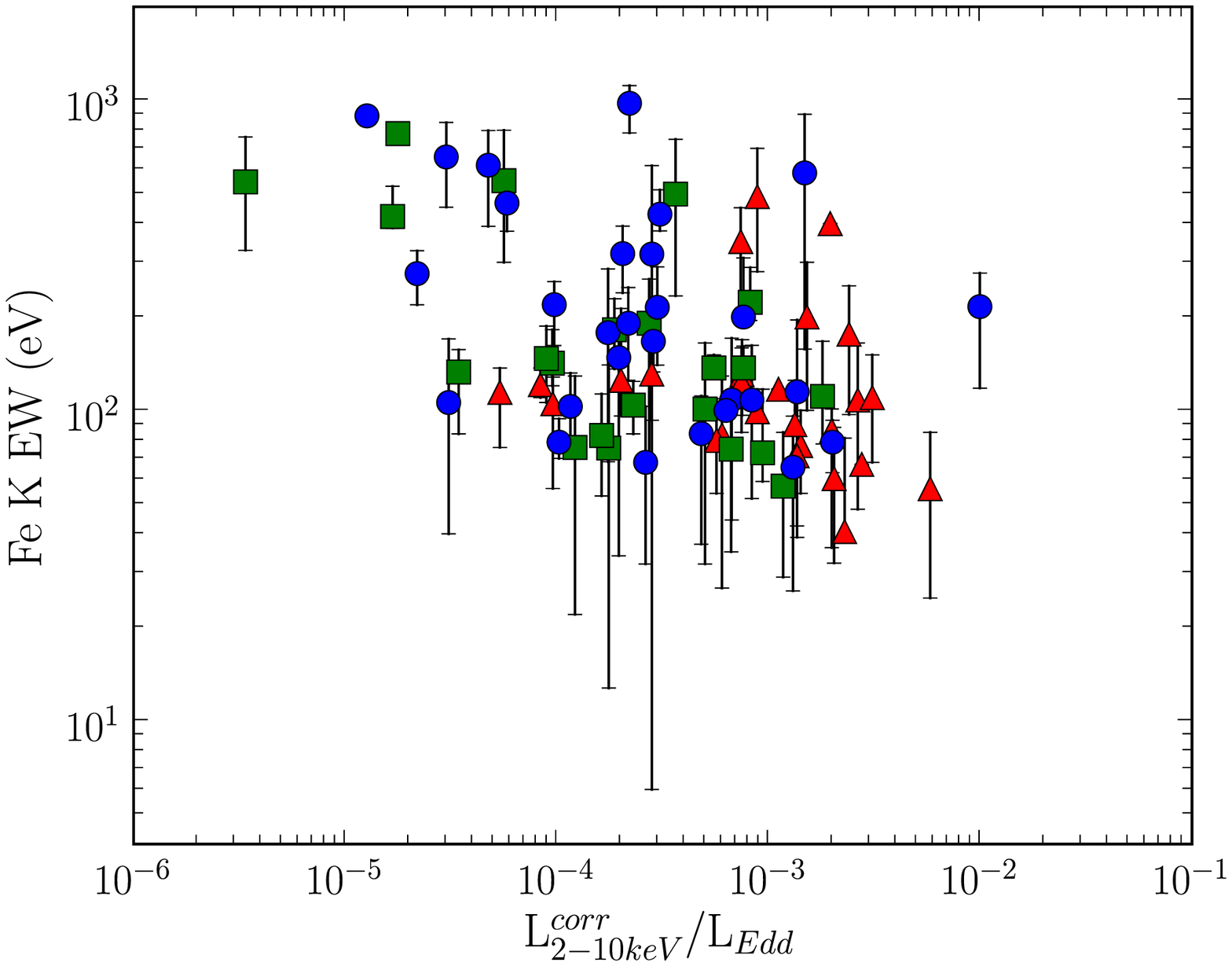} 
\hspace{-0.75cm}
\includegraphics[width=6.0cm, height=5.0cm]{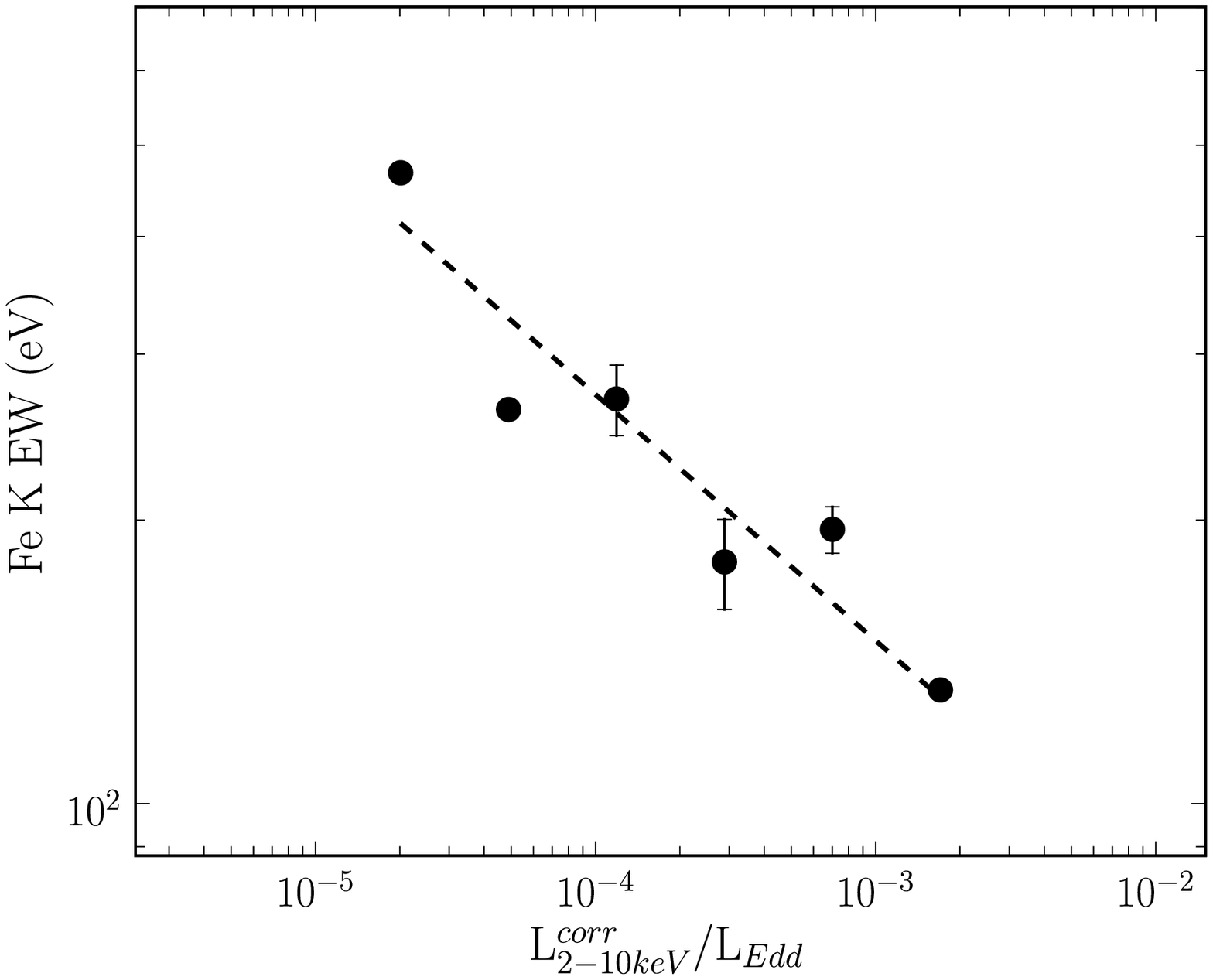}
\hspace{-0.75cm}
\includegraphics[width=6.0cm, height=5.0cm]{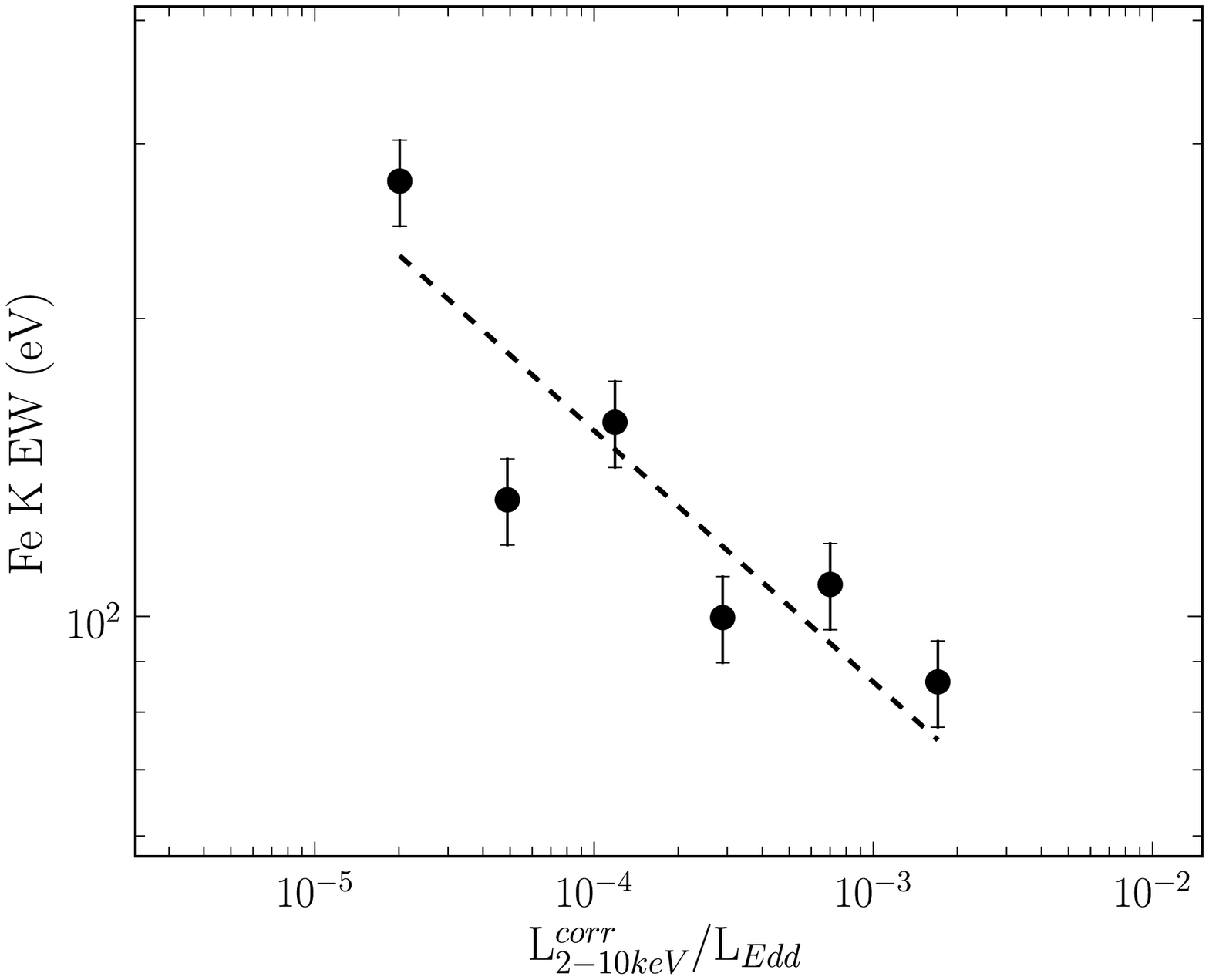}
\caption[Fe K-$\alpha$ EW versus L$^{corr}_{2- 10 keV}$ and L$^{corr}_{2- 10 keV}$/L$_{Edd}$]{We plot the narrow Fe K$\alpha$ line equivalent width (EW) versus unabsorbed 2 -- 10\,keV luminosity (top) and our proxy for Eddington rate (bottom).  In the left hand panels, we plot the distribution for all of our sources.  The triangles indicate optical Sy1--1.2 sources, the squares indicate Sy1.5--1.9 sources, and the circles indicate Sy2 sources.  The mean EW for the 76 sources plotted is $\mu = 200$\,eV.  In the middle and right panels, we plot the distribution after binning the values by luminosity or L$^{corr}_{2- 10 keV}$/L$_{Edd}$ using the mean Fe K EW (middle) or median EW (right).  While anti-correlations are seen in both sets of plots using the mean EW (middle), the relationship weakens substantially for the luminosity plot when the median EW is used (top right).  Thus, our data does not show evidence of the X-ray Baldwin effect but does show an anti-correlation in the binned EW vs. L$^{corr}_{2- 10 keV}$/L$_{Edd}$ relation.
\label{fig-fek}}
\end{figure}

\begin{figure}
\includegraphics[width=9cm]{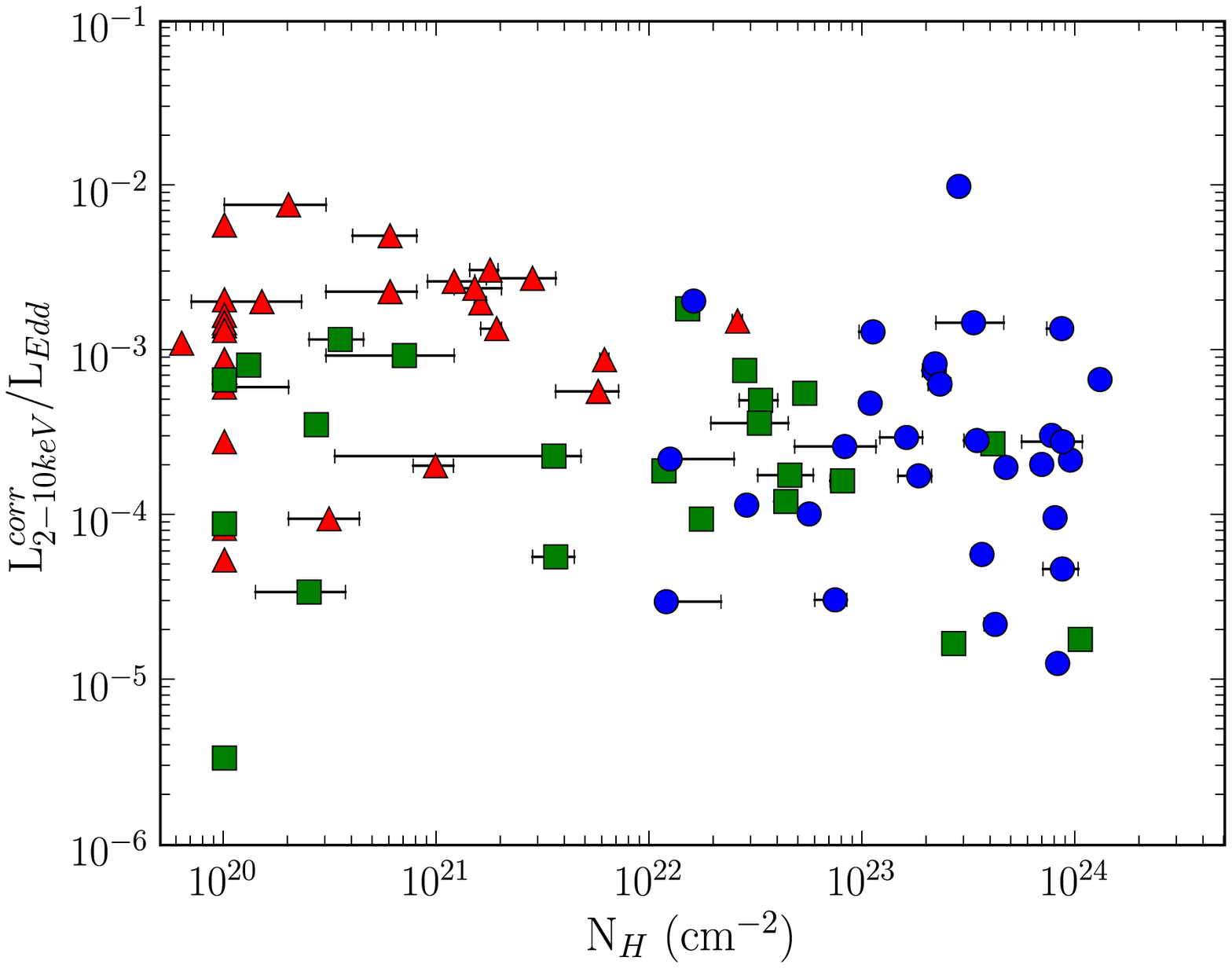} 
\includegraphics[width=9cm]{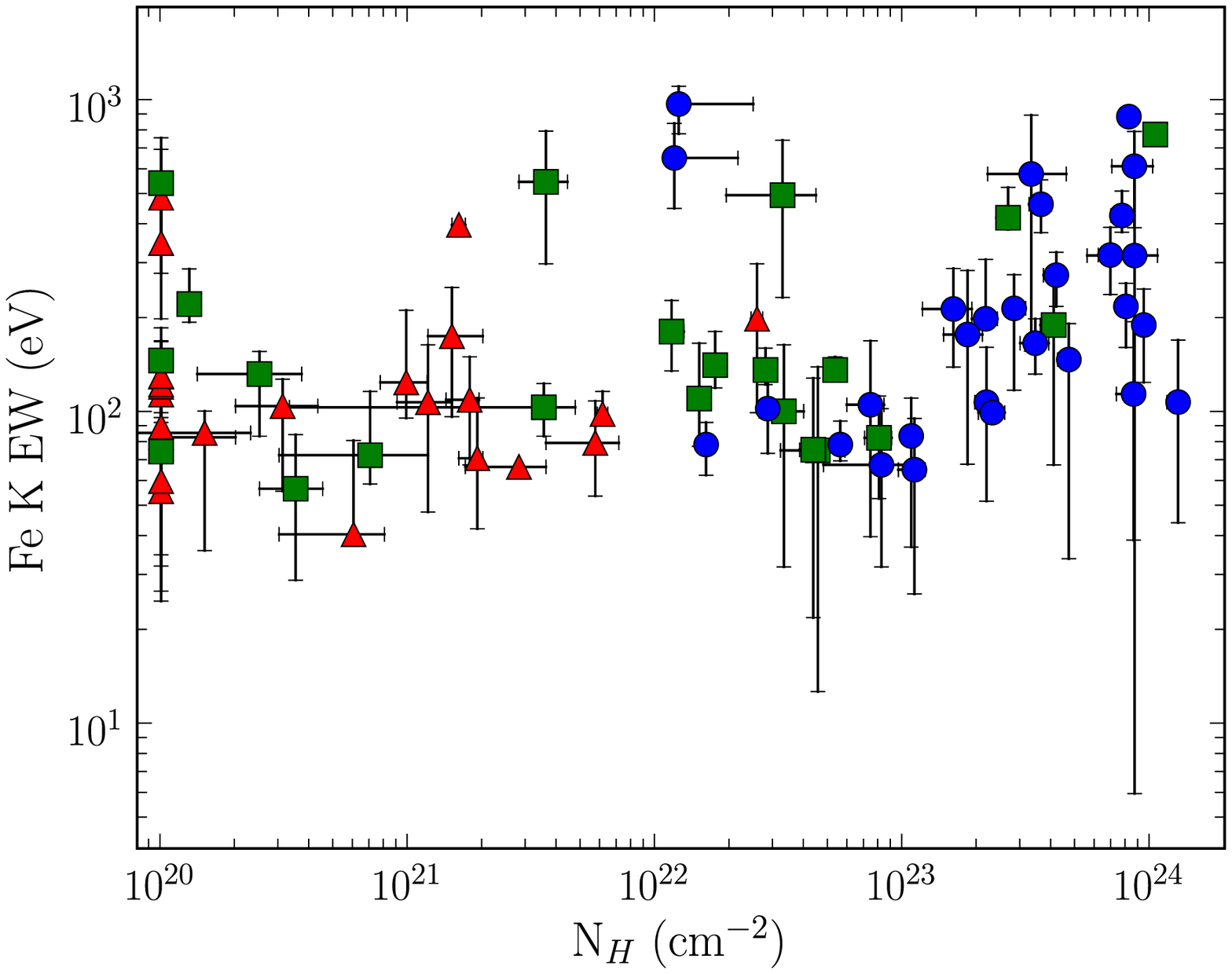}
\caption{We plot L$^{corr}_{2- 10 keV}$/L$_{Edd}$ (left) and the narrow Fe K$\alpha$ line equivalent width (EW) (right) versus the X-ray measured hydrogen column density.  We find no correlation in either relation.  Here, the triangles indicate optical Sy1--1.2 sources, the squares indicate Sy1.5--1.9 sources, and the circles indicate Sy2 sources.  
\label{fig-LLEddnh}}
\end{figure}

\begin{figure}
\includegraphics[width=9cm]{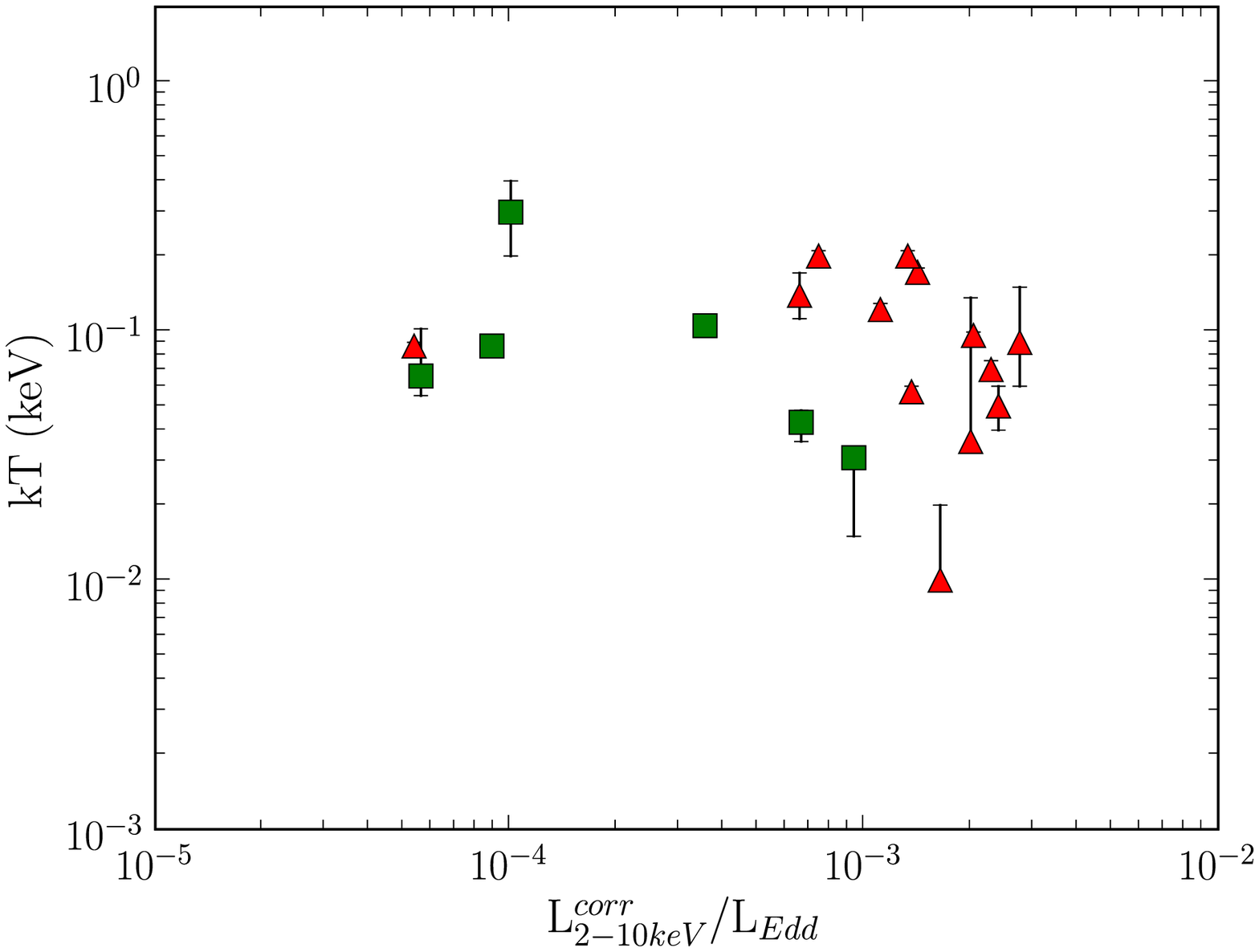}
\includegraphics[width=9cm]{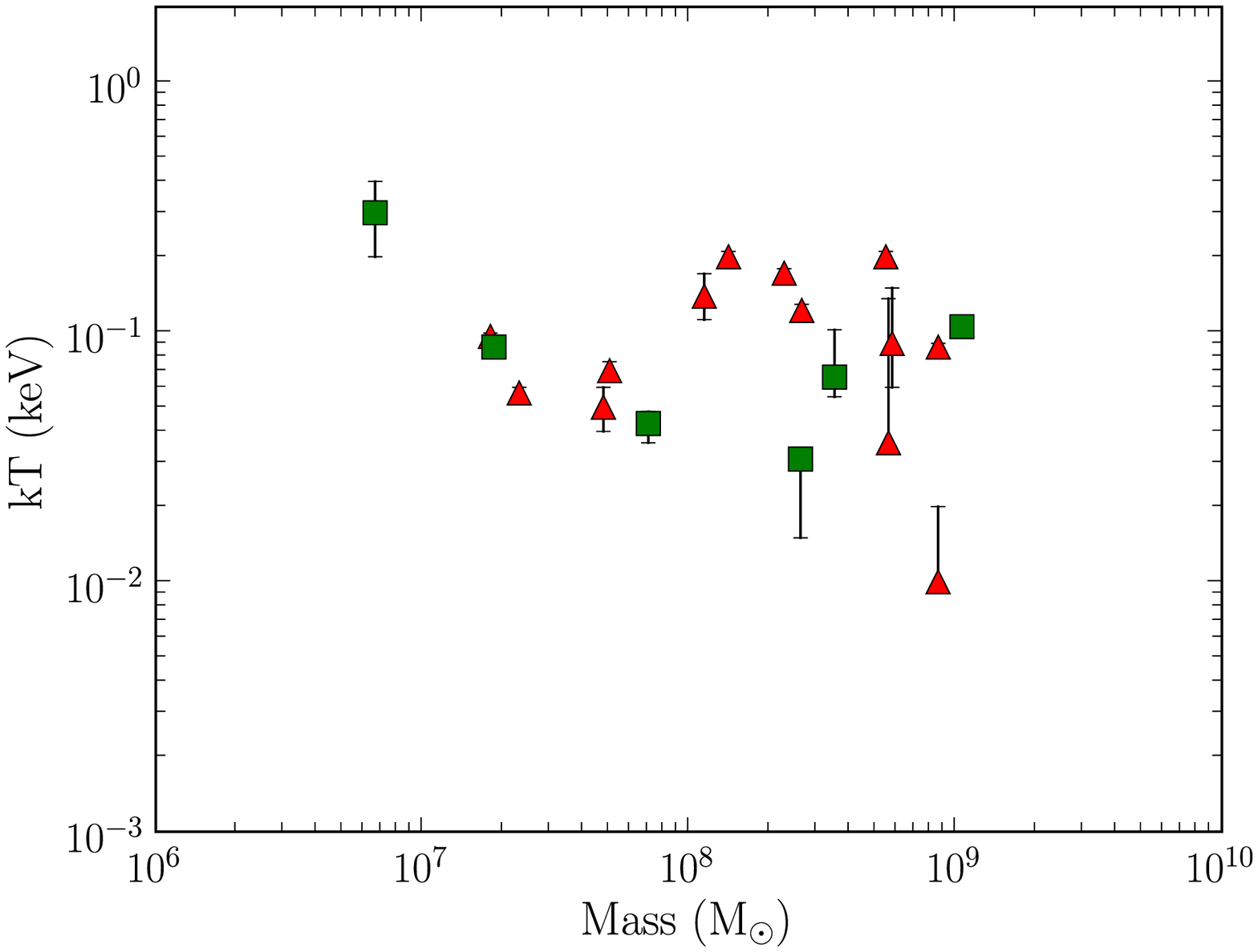} 
\includegraphics[width=9cm]{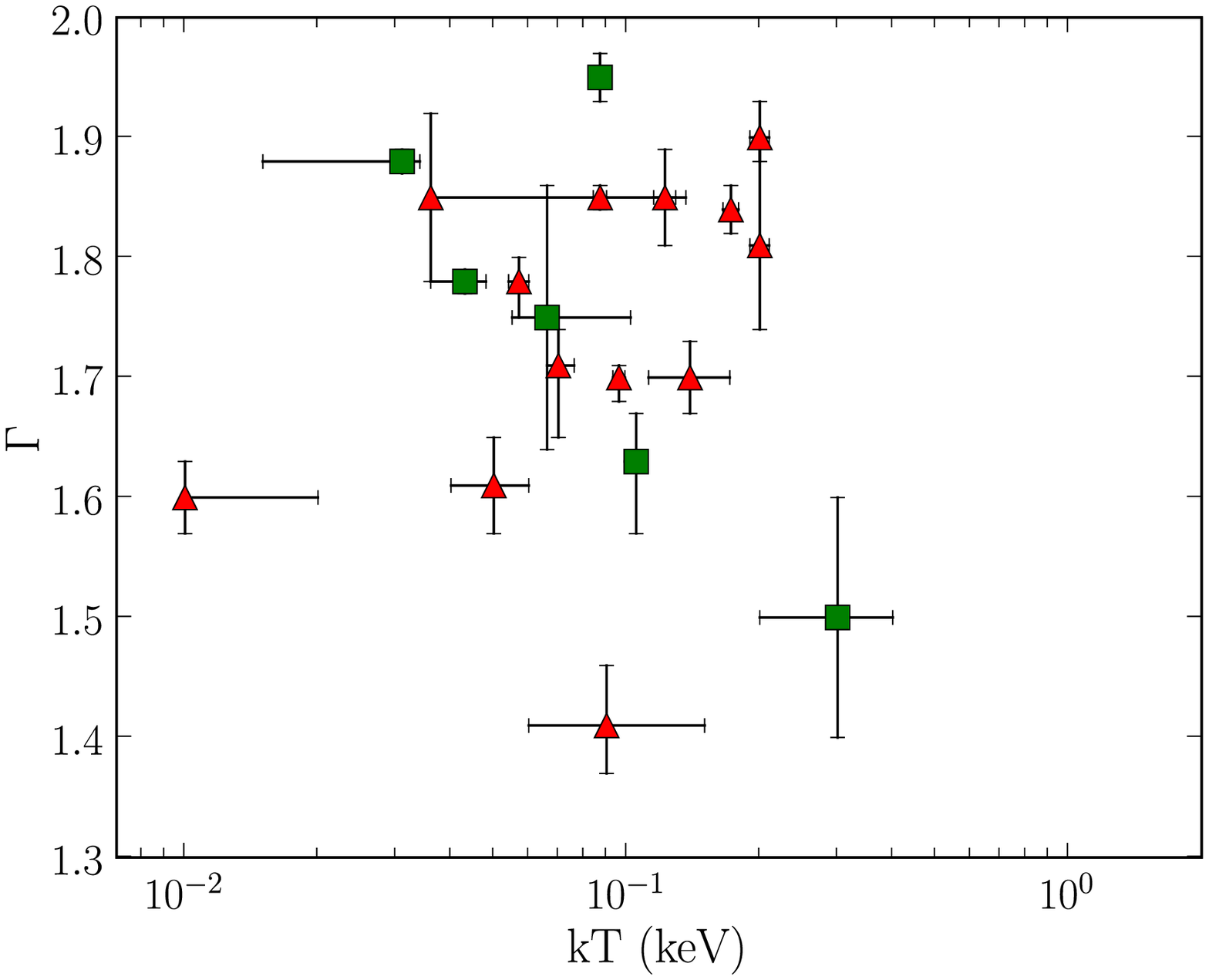}
\includegraphics[width=9cm]{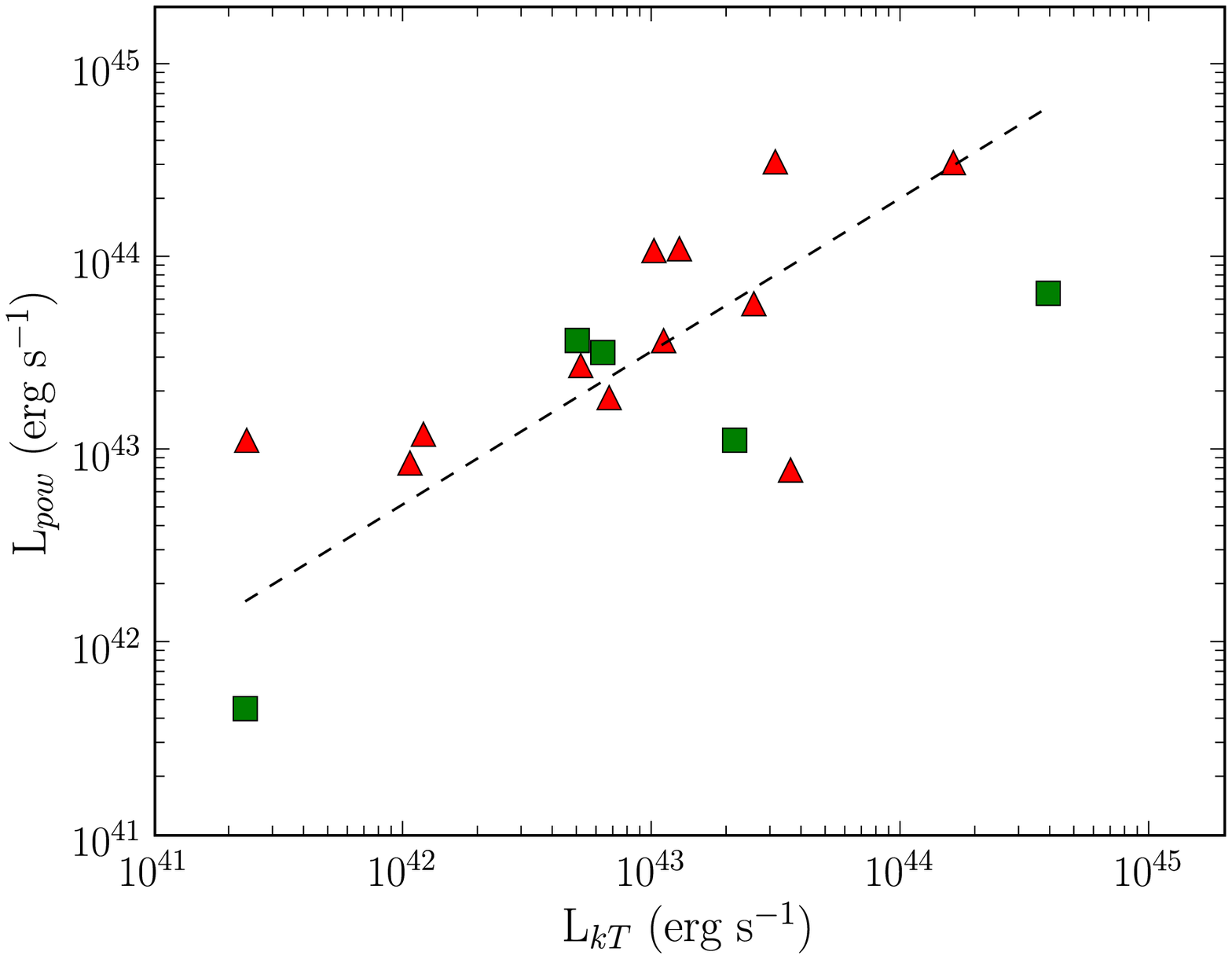}
\caption{In the top row, we plot the blackbody temperature for the 19 simple model sources requiring a soft excess model versus L$^{corr}_{2- 10 keV}$/L$_{Edd}$ and black hole mass, respectively.  There is no correlation between these parameters.  Similarly, we find no correlation between the power law index and blackbody temperature (bottom left).  However, there is a correlation between the unabsorbed luminosity in the power law component and that in the blackbody component (bottom right).  Fitting a line to the data, we find that the relationship is nearly linear (L$_{pow} \propto$\,L$_{kT}$).  Here, triangles represent Sy 1--1.2 sources and squares indicate Sy 1.5--1.9 sources.
\label{fig-simple}}
\end{figure}

\begin{figure}
\includegraphics[width=9cm]{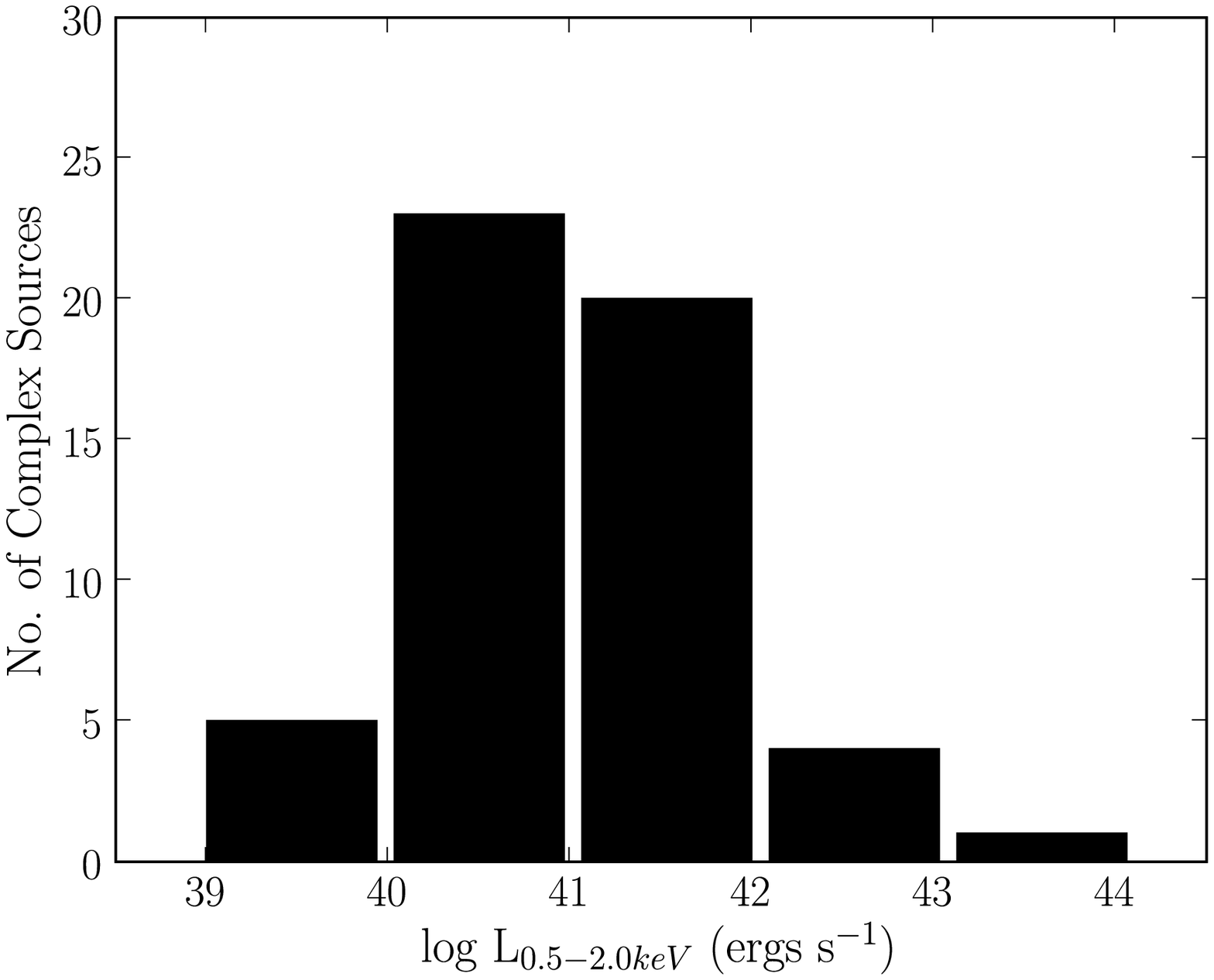}
\includegraphics[width=9cm]{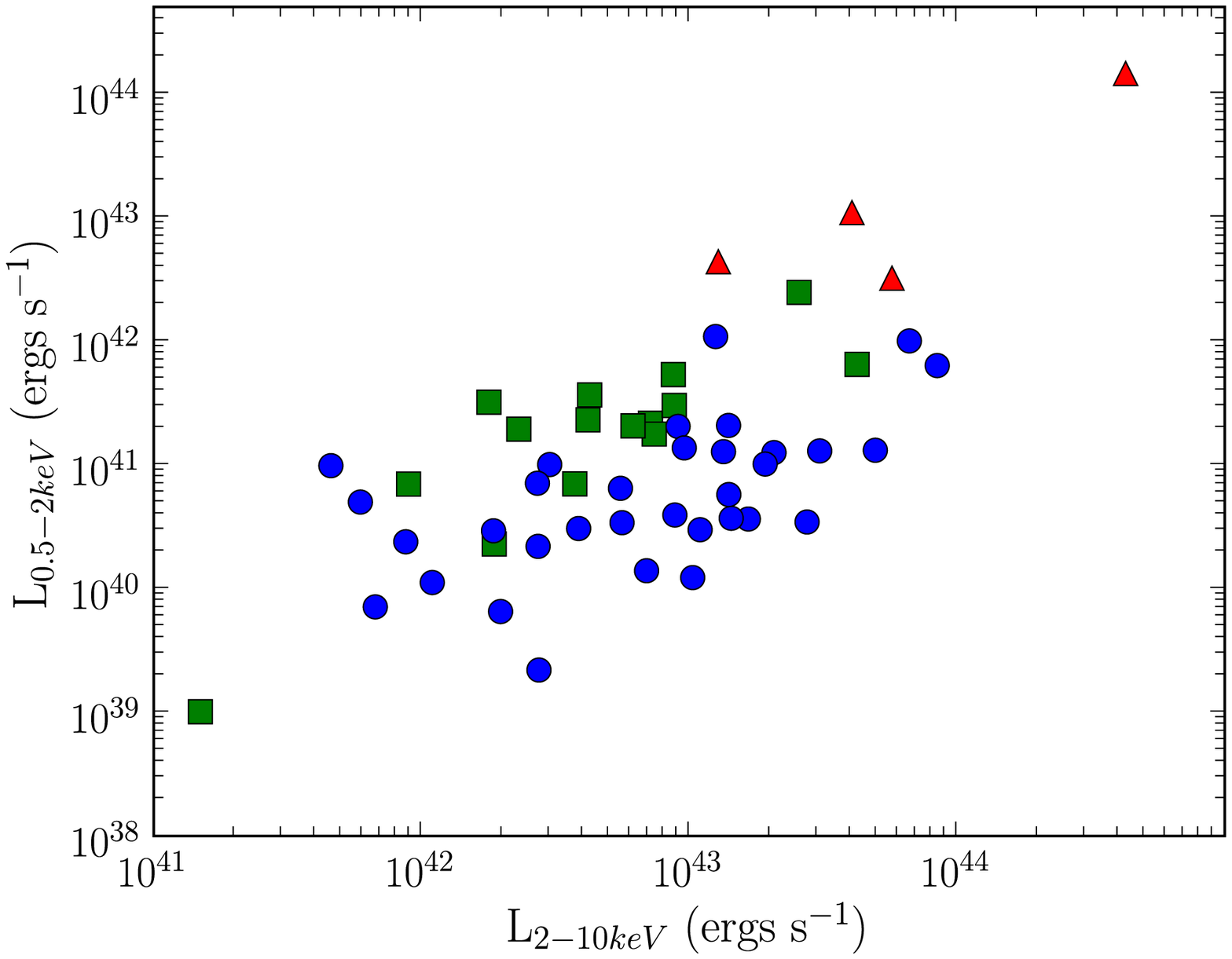}
\caption[L${0.5-2.0 keV}$ for Complex Sources]{Plotted at left is the distribution of soft band X-ray luminosity for sources with spectra best fit by a complex model.  The mean value is $\log L_{0.5-2.0 keV} = 41$ with $\sigma = 0.94$.  This shows that for more than half of the sources the luminosities are low enough that  we can not rule out the idea that the soft emission is from X-ray binaries/star formation/diffuse gas in the host galaxy.  Plotted at right, we show the soft X-ray observed luminosity versus the hard X-ray observed luminosity.  A strong correlation is not seen, indicated by $R^2 = 0.51$.  This further suggests that we can not explain the soft emission as a simple extension of the hard power law emission.  The symbols and color coding are the same as in Figure~\ref{fig-LLEddnh}.
\label{fig-lsoft}}
\end{figure}

\begin{figure}
\includegraphics[width=9cm]{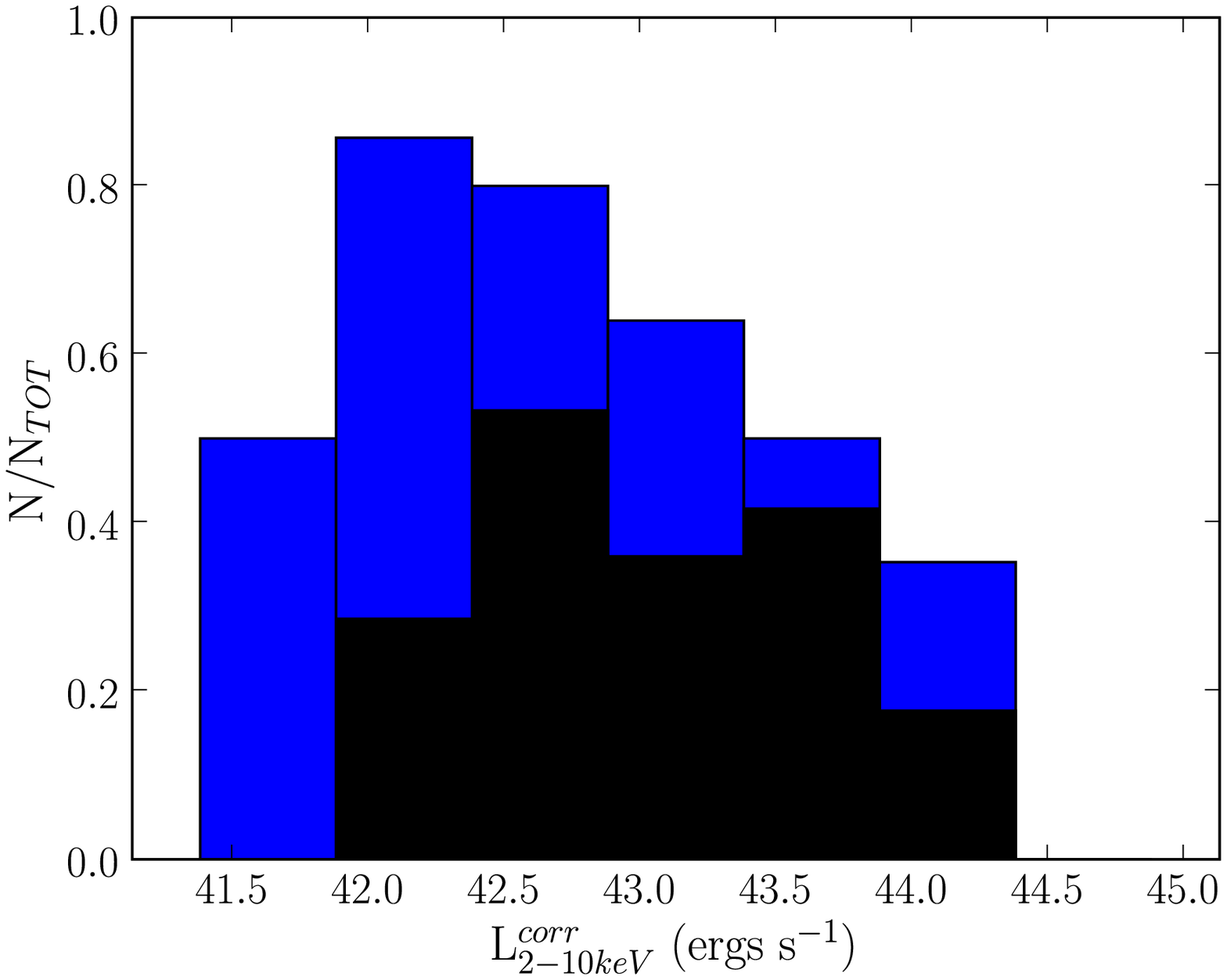}
\includegraphics[width=9cm]{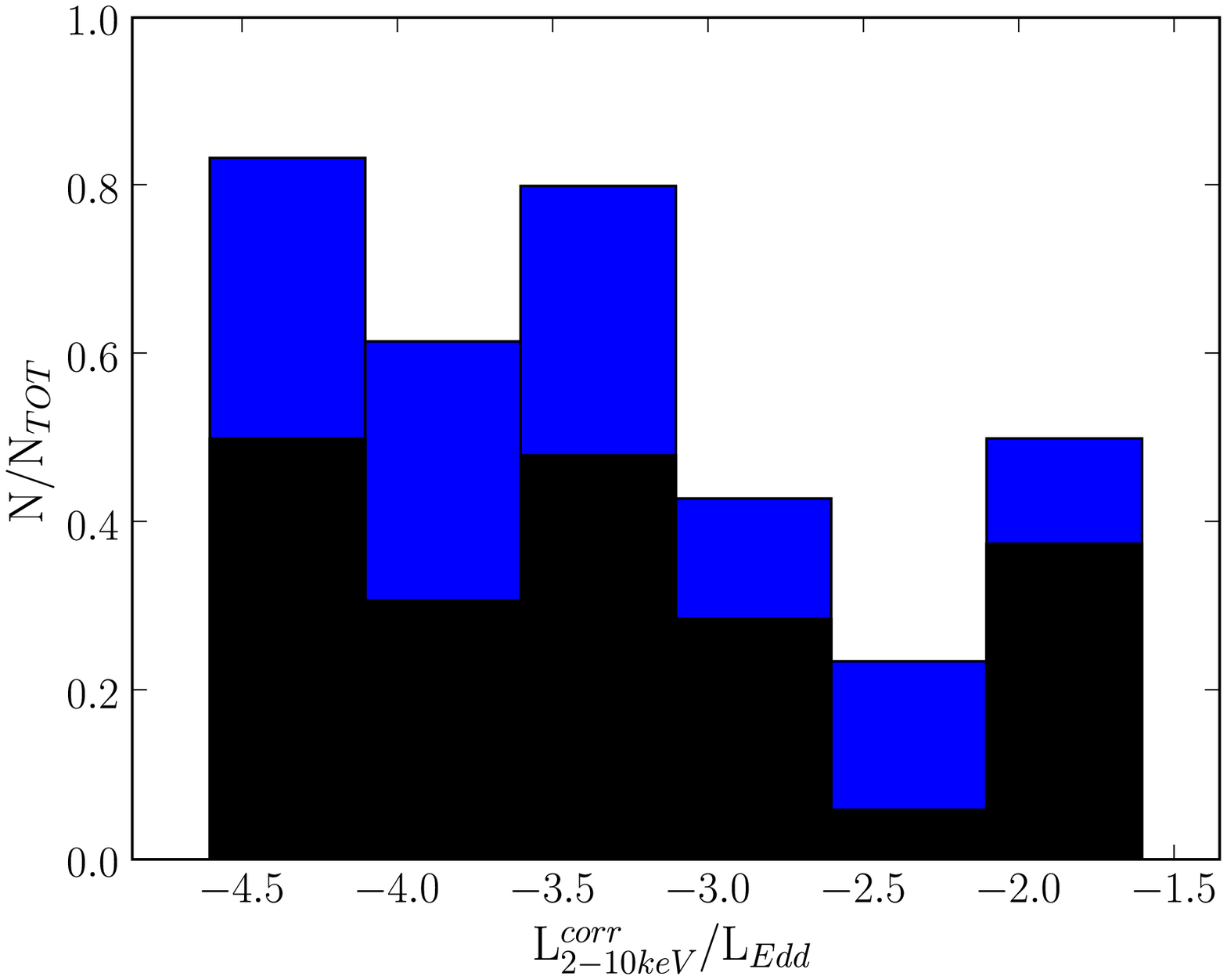}
\caption[Fraction of Obscured Sources]{These plots show the fraction of absorbed sources in a given 2--10\,keV luminosity bin (left) and accretion rate bin (right).  The black bins show the fraction of sources with $\log N_H \ge 23$.  These are a subset of the sources with $\log N_H \ge 22$, pictured in blue.  There are no high column density sources in the highest 2--10\,keV luminosity (unabsorbed) bin.  Further, the fraction of obscured sources is lower at high luminosities.  In the second plot, it is not clear that the fraction of obscured sources increases with lower accretion rate (L$^{corr}_{2- 10 keV}$/L$_{Edd}$).  We find that 50\% of the sources in the highest luminosity bin are absorbed, however, with only 8 sources in this bin, this could be the result of poor sampling.
\label{fig-fraction}}
\end{figure}

\begin{figure}
\centering
\includegraphics[width=15cm]{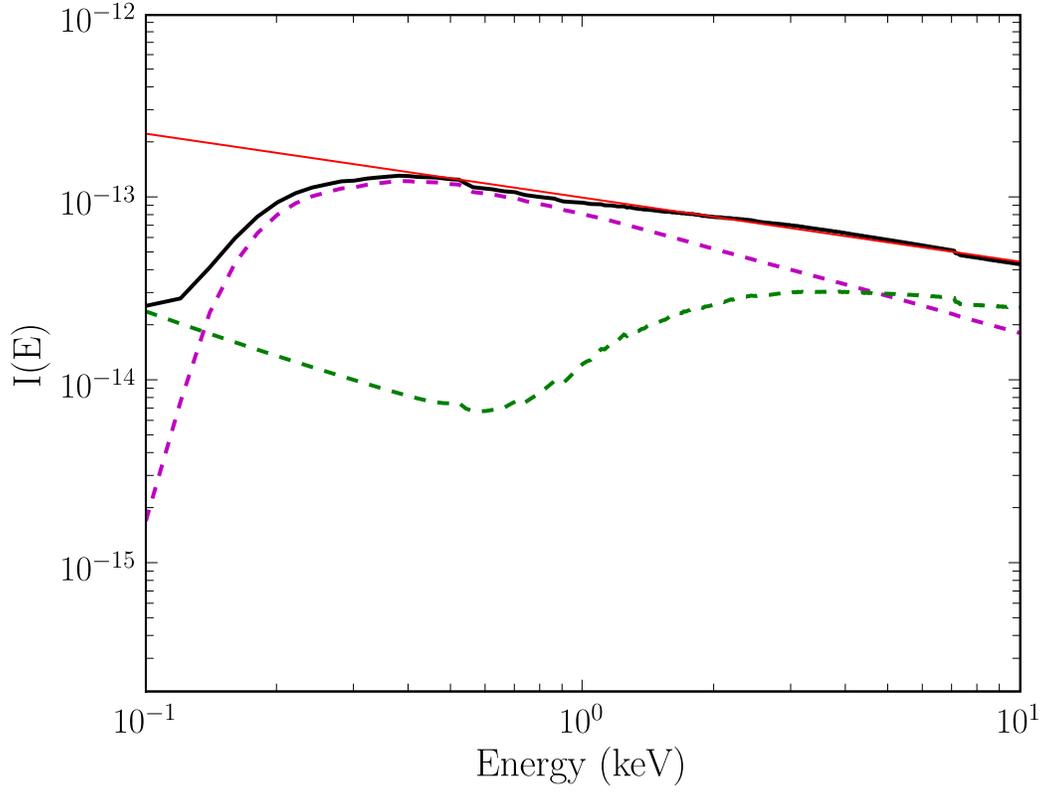}
\caption[Average 0.1--10\,keV Spectrum]{Shown here is the average spectrum constructed from the X-ray fits to our uniform sample of 102 AGN.  The solid black curve is the average spectrum of the AGN sources.  The dashed lines show the contribution from the simple absorption sources (magenta) and the complex absorption sources (green).  We also show a line fit to the average spectrum from 0.6 -- 10\,keV.  The slope of the line, $\Gamma = 1.369 \pm 0.004$, is consistent with the modeled CXB slope.
The total flux from our sources corresponds to only 0.29\% of the entire 2--10\,keV CXB.
\label{fig-avg}}
\end{figure}

\begin{figure}
\begin{center}
\includegraphics[width=15cm]{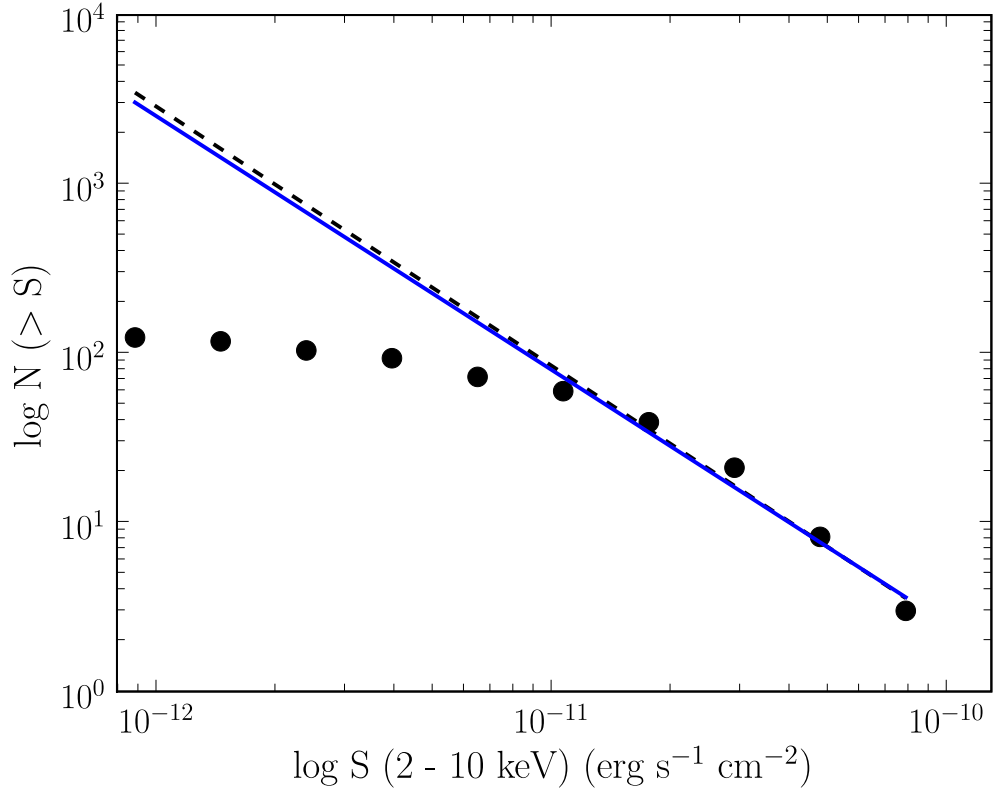}
\end{center}
\caption[Log N--Log S in the 2--10\,keV Band]{Here we plot the relation of $\log N$-$\log S$ for our entire uniform sample ($|b| \ge 15^{\circ}$ or 74\% of the sky).  The value $\log N$ corresponds to the number of AGN with 2 -- 10\,keV fluxes above the indicated $\log S$, corrected for sky coverage using the BAT sky coverage maps and 14--195\,keV fluxes.  The dashed line represents a fit to the points with $\log S \ge -11$.  The slope of this line ($-1.53 \pm 0.12$) is consistent with the cumulative distribution of a uniform density of objects (-1.5, shown as the solid line).  This plot further suggests that we are missing many sources at fluxes below $\log S \approx -11$.  
\label{fig-lognlogs}}
\end{figure}

\begin{figure}
\begin{center}
\includegraphics[width=12cm]{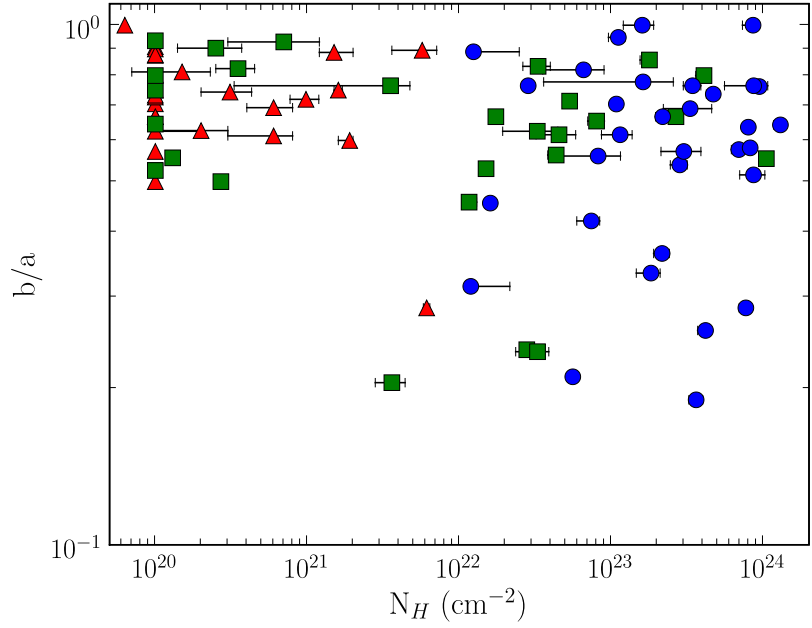}
\end{center}
\caption[Host Inclination vs. X-ray Column Density]{We plot the measured host galaxy minor axis/major axis ($b/a$) from NED versus the X-ray measured column density.  As shown, only one Seyfert 1 (IC 4329A, a Sy 1.2) is in a highly inclined galaxy.  However, this source has a higher X-ray column than most other Seyfert 1s.  The hosts of Seyfert 2s span the entire range of b/a. Note that for sources requiring no additional absorption over the Galactic, we used a standard value of $N_H = 10^{20}$\,cm$^{-2}$.  Here, as in the previous figures, triangles represent Sy 1--1.2 sources, squares indicate Sy 1.5--1.9 sources, and circles indicate Sy2s.\label{fig-ba}}
\end{figure}
\clearpage

\begin{figure}
\includegraphics[width=9cm]{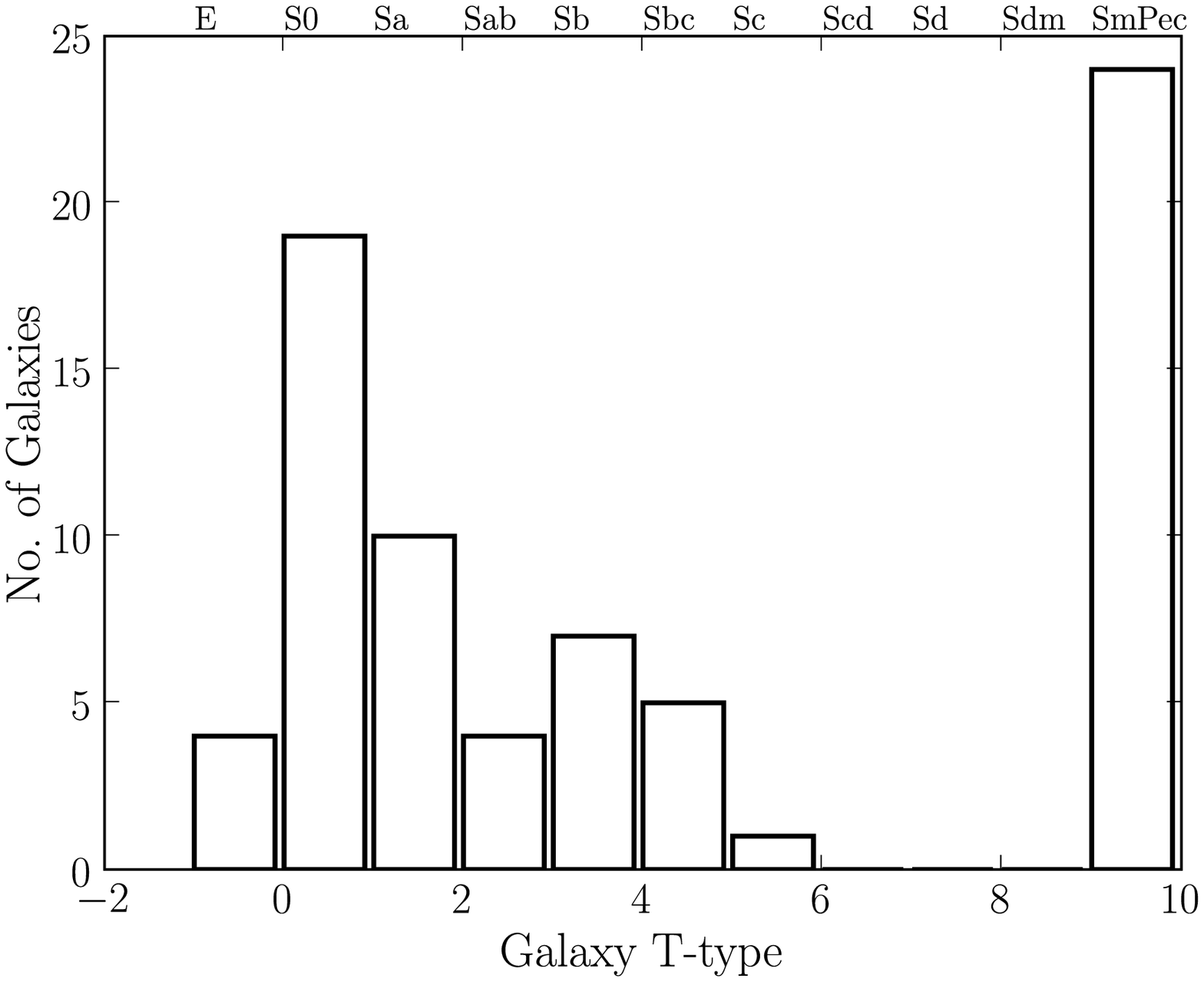}
\includegraphics[width=9cm]{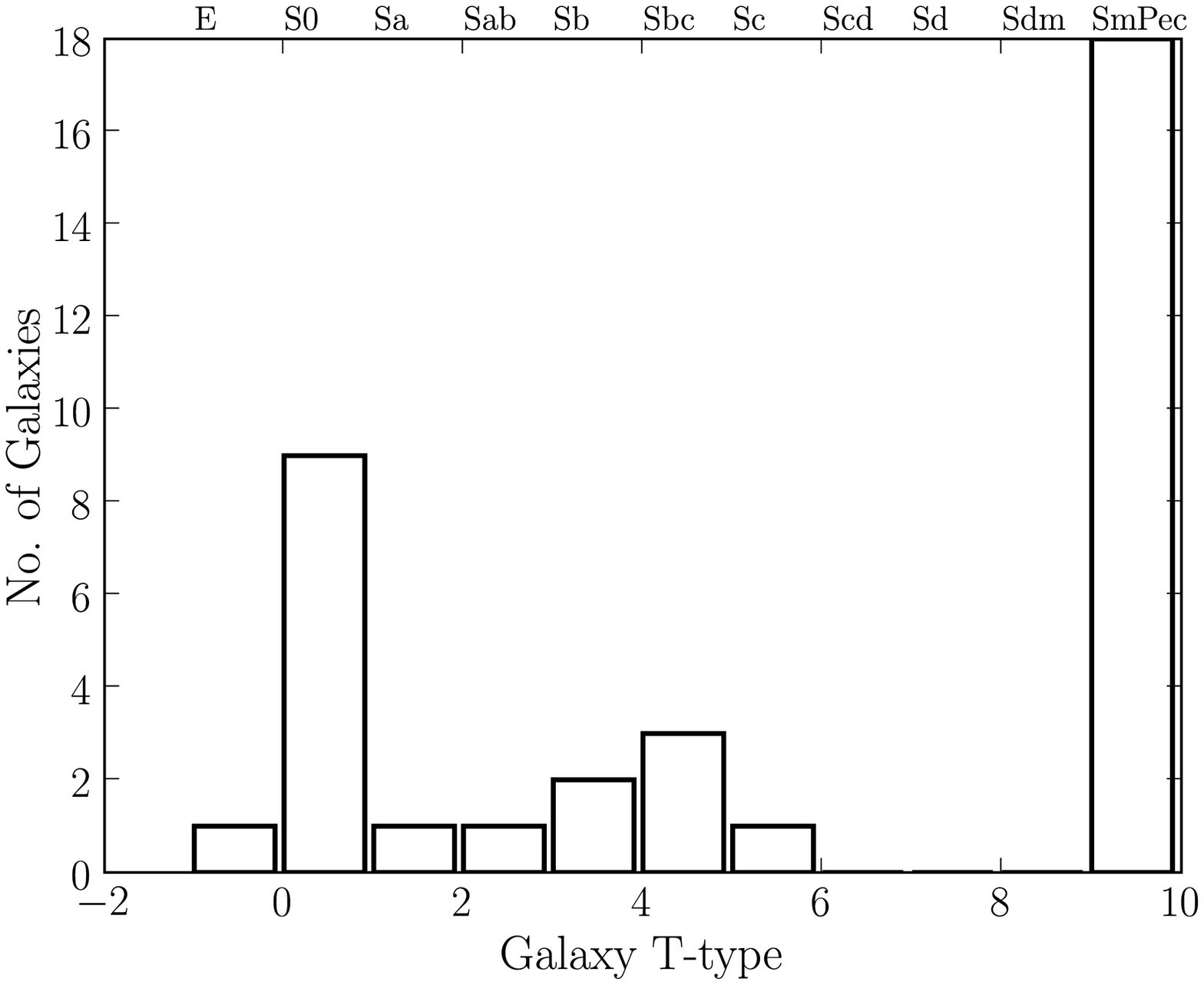}
\caption{We plot the distribution of host galaxy type, from NED, for the uniform sample of BAT-detected AGNs (left).  Also, we plot the distribution of host galaxy type for `interacting' sources in our uniform sample (right).  Clearly, there is no significant difference in the host type for these sources.
\label{fig-hosts}}
\end{figure}
\clearpage

\begin{figure}
\includegraphics[width=10cm]{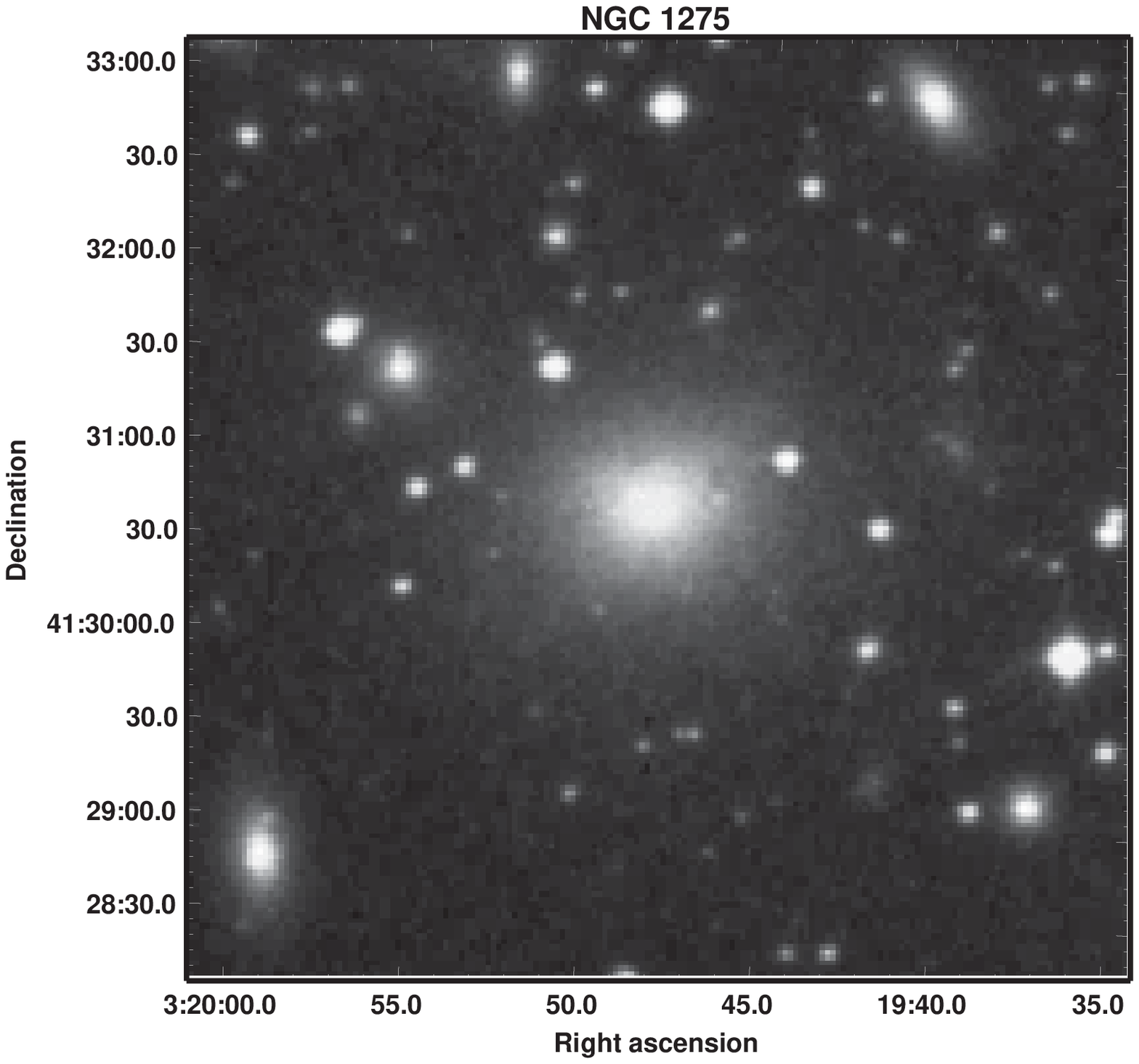}
\hspace{-1cm}
\includegraphics[width=10cm]{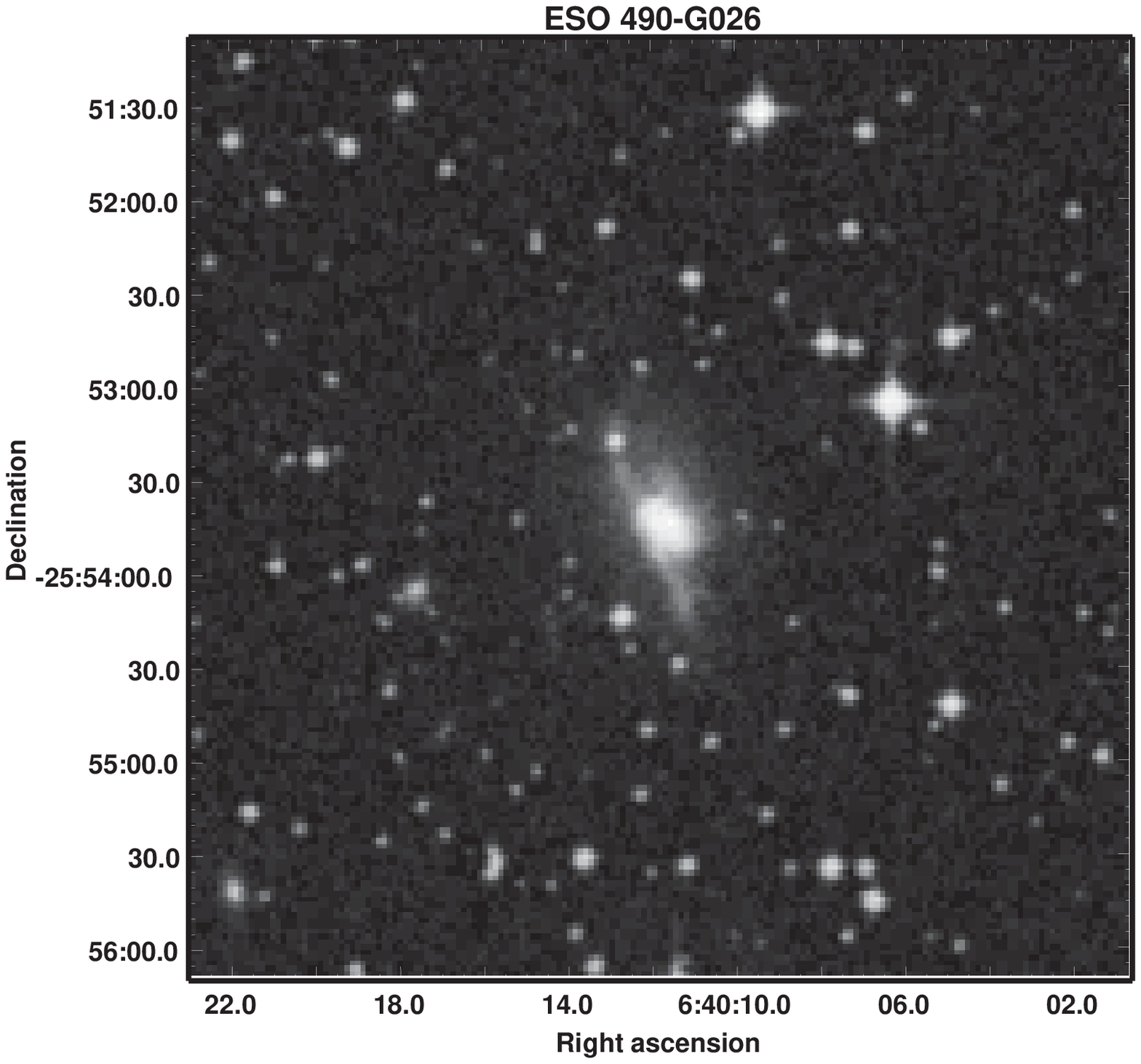} \\
\includegraphics[width=10cm]{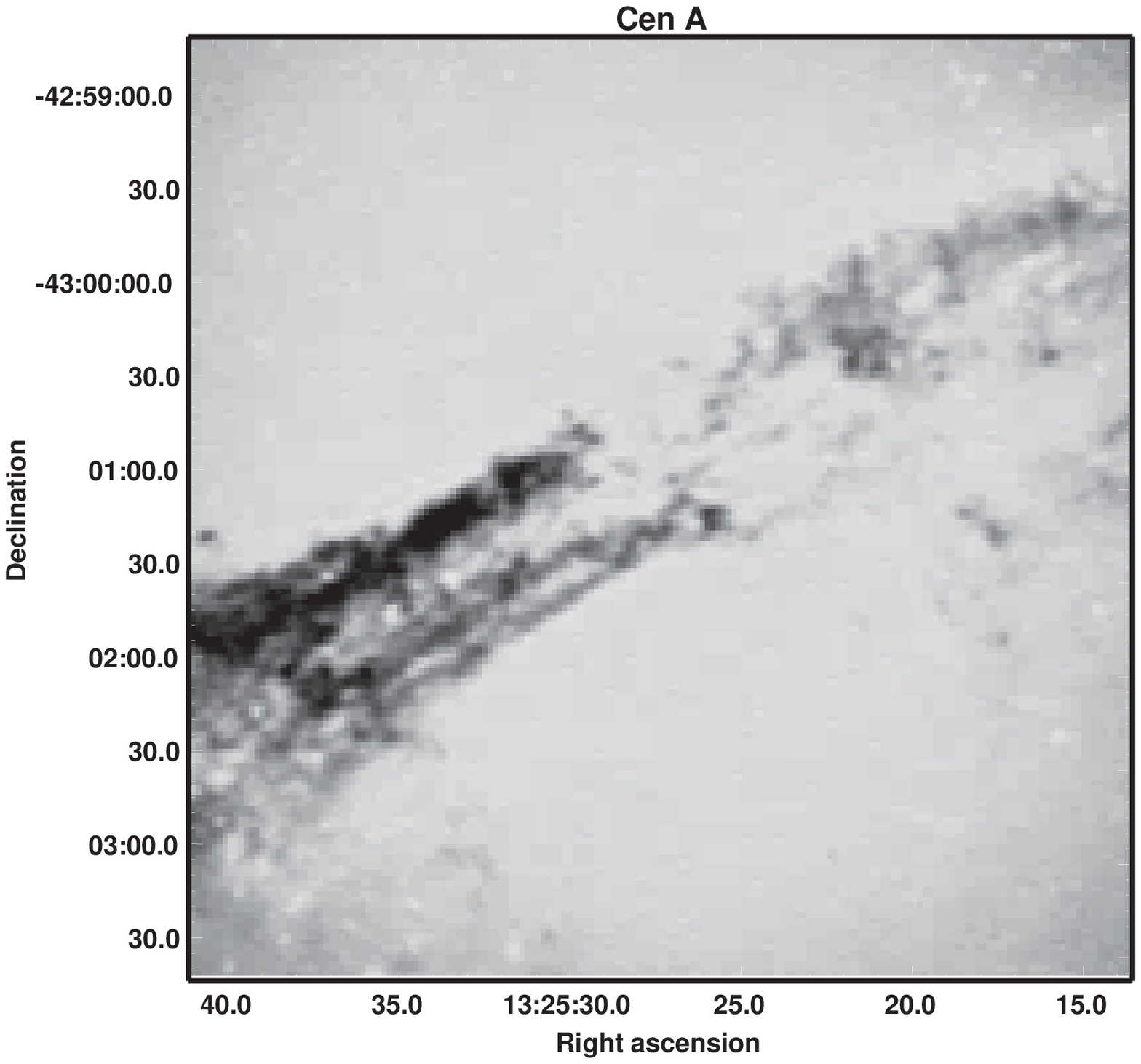}
\hspace{-1cm}
\includegraphics[width=10cm]{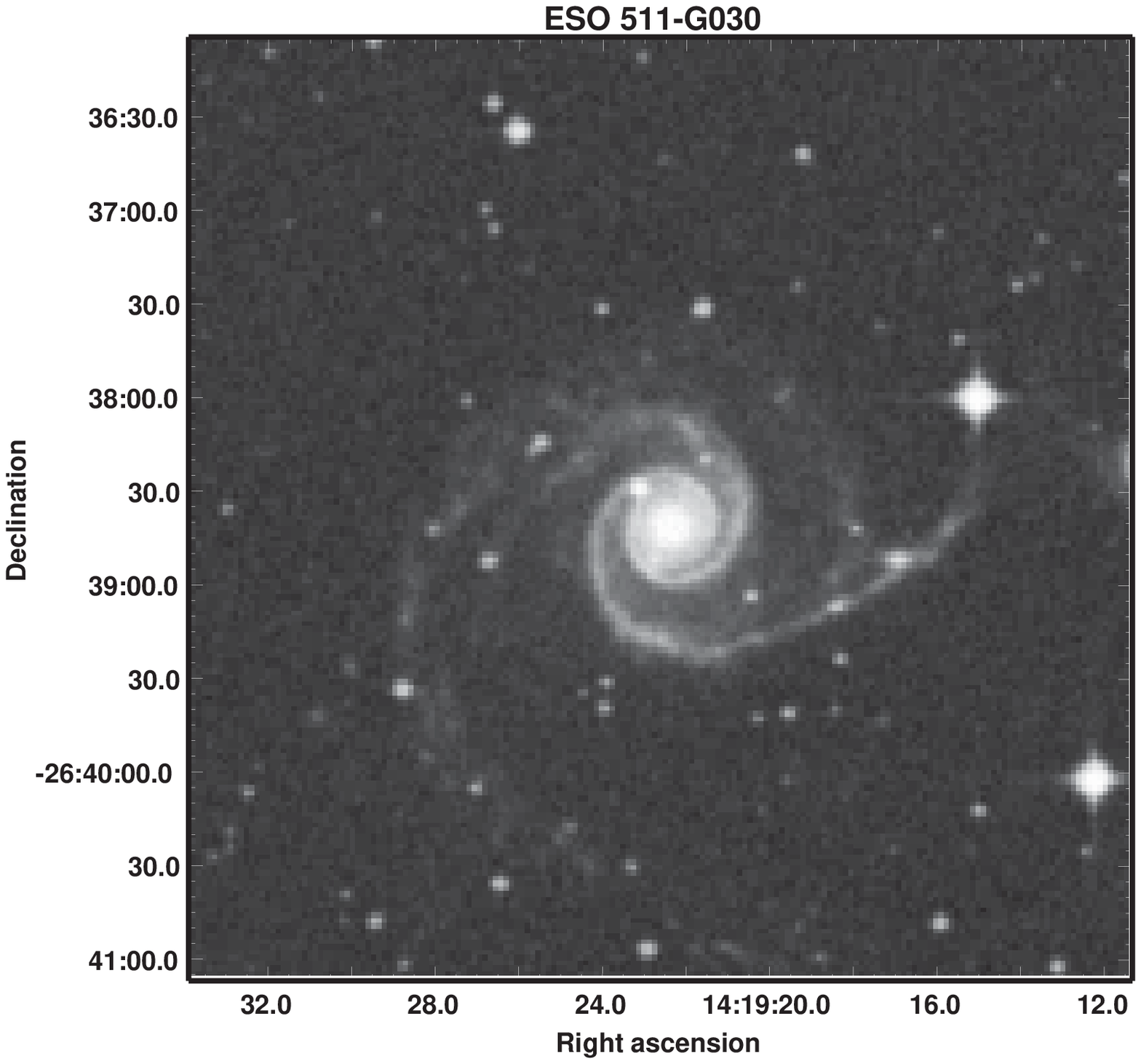}
\caption{We show DSS images of 4 representative peculiar/irregular host galaxies.  For all images, we plot a 5\arcmin$\times$5\arcmin region.  Both NGC 1275 and Cen A have available HST images which reveal in detail large dust lanes.  However, even in the lower resolution DSS image, the dust lane in Cen A is clearly visible.  Of the other sources, ESO 490-G026 is really two colliding galaxies and ESO 511-G030 is a spiral with a small nearby companion galaxy (outside the image). 
\label{fig-peculiar}}
\end{figure}
\clearpage

\begin{figure}
\includegraphics[width=9cm]{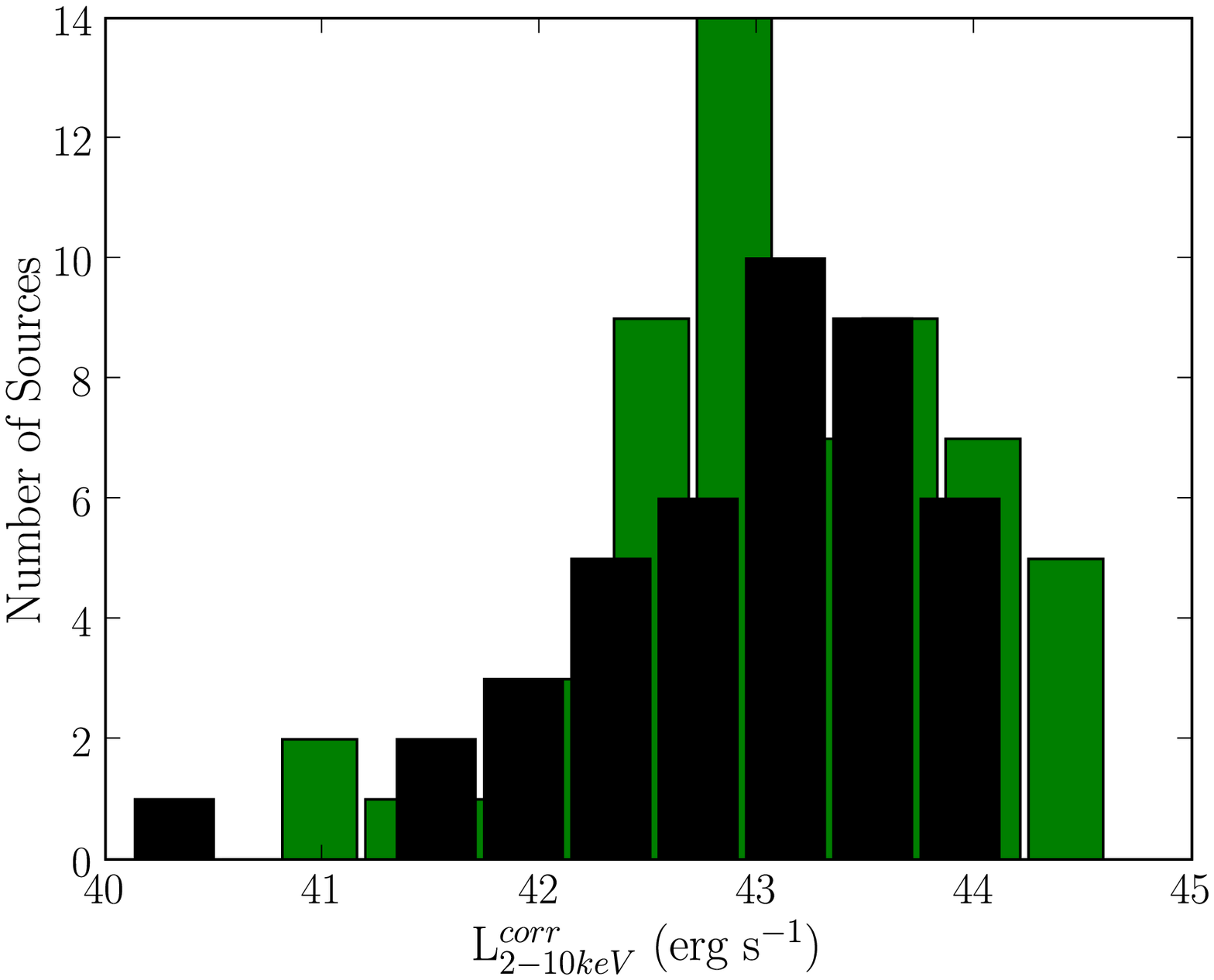}
\includegraphics[width=9cm]{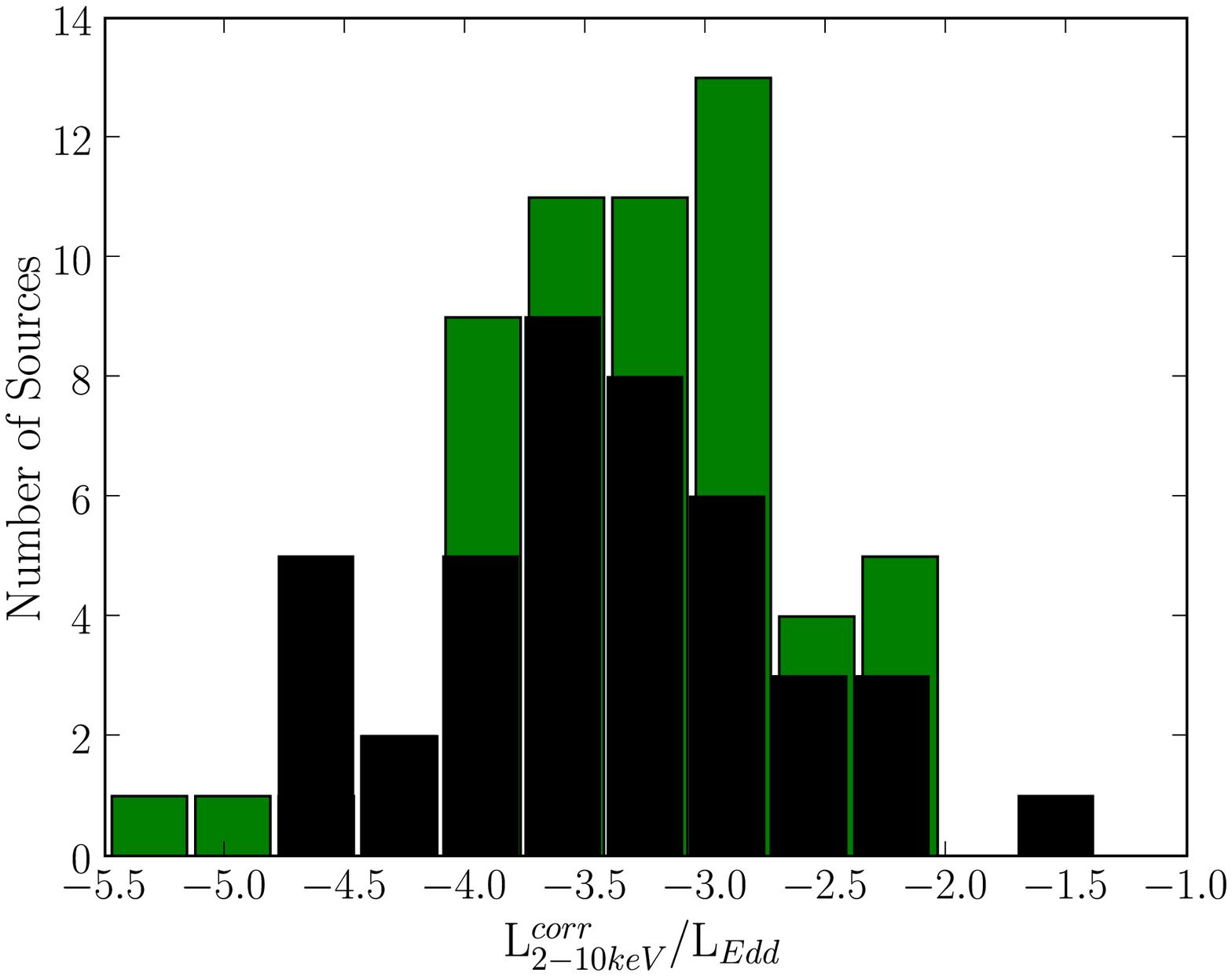}
\caption{Here we plot the distributions of unabsorbed 2--10\,keV luminosity and our Eddington ratio proxy for interacting (black) and non-interacting (green) systems.  There is little difference between the two distributions, showing that AGN in galaxies with a close companion or which underwent a recent merger have the same luminosities and accretion rates as AGN in non-interacting galaxies.
\label{fig-int_lum}}
\end{figure}

\end{document}